%% file: main.tex
\documentclass[11pt,a4paper]{article}
\usepackage[utf8]{inputenc}
\usepackage{amsmath}
\numberwithin{equation}{section}
\usepackage[colorinlistoftodos]{todonotes}

\input{0-title.tex}

\begin{document}
\maketitle
\input{0-abstract.tex}
\tableofcontents
\input{1-introduction.tex}
\input{2-directCPV.tex}
\input{3-Polarization_and_CP_observable.tex}
\input{4-angular-TPA}
\input{5-discussions.tex}
\input{6-summary.tex}
\input{7-appendix.tex}
\input{8-bibliography.tex}
\end{document}

%% file: 0-title.tex
\def\papertitle{\bf CP violation observables in baryon decays} 

\usepackage{authblk}
\title{\papertitle}
\author[1]{Jian-Peng Wang\footnote{Email: wangjp20@lzu.edu.cn, corresponding author}}
\author[2]{Qin Qin\footnote{Email: qqin@hust.edu.cn, corresponding author}}
\author[1]{Fu-Sheng Yu\footnote{Email: yufsh@lzu.edu.cn, corresponding author}}
\affil[1]{Frontiers Science Center for Rare Isotopes, and School of Nuclear Science and Technology, Lanzhou University, Lanzhou 730000, China}
\affil[2]{School of Physics, Huazhong University of Science and Technology, Wuhan 430074, China}

\date{\today}                   
\usepackage[top=1in, bottom=1.25in, left=1in, right=1in]{geometry}
\usepackage{microtype}
\usepackage{lineno}
\usepackage{xspace}
\usepackage{booktabs}
\usepackage{comment}
\usepackage{physics}
\usepackage{caption} 

\usepackage{graphicx}  
\usepackage{color}
\usepackage{colortbl}
\DeclareGraphicsExtensions{.pdf,.PDF,png,.PNG}

\usepackage{amsmath} 
\usepackage{amssymb}
\usepackage{yhmath}
\usepackage{braket}
\usepackage{amsfonts}
\usepackage{amsmath,amsfonts,amssymb,amsthm,bm,upgreek}
\usepackage{upgreek} 

\usepackage{hyperxmp}
\usepackage{hyperref}
\usepackage{hypcap} 
\hypersetup{hidelinks} 

\def\CP      {{\ensuremath{C\!P}}\xspace}

\def\pion   {{\ensuremath{\pi}}\xspace}

\def\pip    {{\ensuremath{\pion^+}}\xspace}
\def\pim    {{\ensuremath{\pion^-}}\xspace}

\def\kaon   {{\ensuremath{K}}\xspace}
\def\Kbar    {{\kern 0.2em\overline{\kern -0.2em \kaon}{}}\xspace}

\def\KorKbar    {\kern 0.18em\optbar{\kern -0.18em K}{}\xspace}

\def\Dbar    {{\kern 0.2em\overline{\kern -0.2em D}{}}\xspace}

\def\DorDbar    {\kern 0.18em\optbar{\kern -0.18em D}{}\xspace}

\def\Xires       {{\ensuremath{\Xi}}\xspace}

\def\Lz          {{\ensuremath{\Lambda^0}}\xspace}
\def\Lbar        {{\ensuremath{\kern 0.1em\overline{\kern -0.1em\Lambda}}}\xspace}
\def\LorLbar    {\kern 0.18em\optbar{\kern -0.18em \PLambda}{}\xspace}

\def\Lb      {{\ensuremath{\Lambda^0_b}}\xspace}

\def\Lc      {{\ensuremath{\Lambda^+_c}}\xspace}

\def\Xibz    {{\ensuremath{\Xires^0_b}}\xspace}

\def\Xic     {{\ensuremath{\Xires_c}}\xspace}

\newcommand{\tev}{\ensuremath{\mathrm{\,Te\kern -0.1em V}}\xspace}
\newcommand{\gev}{\ensuremath{\mathrm{\,Ge\kern -0.1em V}}\xspace}
\newcommand{\mev}{\ensuremath{\mathrm{\,Me\kern -0.1em V}}\xspace}
\newcommand{\kev}{\ensuremath{\mathrm{\,ke\kern -0.1em V}}\xspace}
\newcommand{\gevc}{\ensuremath{{\mathrm{\,Ge\kern -0.1em V\!/}c}}\xspace}
\newcommand{\mevc}{\ensuremath{{\mathrm{\,Me\kern -0.1em V\!/}c}}\xspace}
\newcommand{\gevcc}{\ensuremath{{\mathrm{\,Ge\kern -0.1em V\!/}c^2}}\xspace}
\newcommand{\gevgevcccc}{\ensuremath{{\mathrm{\,Ge\kern -0.1em V^2\!/}c^4}}\xspace}
\newcommand{\mevcc}{\ensuremath{{\mathrm{\,Me\kern -0.1em V\!/}c^2}}\xspace}

\usepackage{cite} 
\usepackage{mciteplus}

%% file: 0-abstract.tex
\begin{abstract}
The era of baryon physics is on the horizon with the accumulation of increasing data by collaborations such as LHCb, Belle II, and BESIII. Despite the wealth of data, one of the critical issues in flavor physics, namely CP violation in baryon decays, still awaits experimental confirmation. It is evident that the development of formulas and phenomenological analyses will play a pivotal role in advancing theoretical investigations and experimental measurements in this domain. In this work, we discuss the fundamental definitions and properties of polarization, asymmetry parameters and their associated CP violating observables that may arise in baryon decays. Additionally, we provide a detailed analysis of b-baryon quasi-two and -three body decays based on angular distributions. We highlight the significance of CP asymmetries defined by spin and momentum correlations, as non-trivial spin distributions are expected in baryon decays due to their non-zero polarization in productions and decays. Of particular importance is a complementary relation can be establised for two different types of observables, that are $\mathcal{T}$-even and -odd CP asymmetries respectively. The complementarity indicates that one observable exhibits the sine dependence on strong phase $\Delta\delta$, while another exhibits as cosine. This gives us an distinctive opportunity to avoid the potential large suppression to CP asymmetries from small phase $\Delta\delta$. Two points (\textbf{i}) the exact proof of strong phase dependence behaviour, and (\textbf{ii}) the criterion for complementary observables are clarified. Based on these arguments from theoretical aside, some brief discussions and suggestions on experimental measurements are offered in the final discussions. Lastly, we present a pedagogical introduction to angular distribution techniques based on helicity descriptions for both two and three body decays.  \newpage
\end{abstract}

%% file: 1-introduction.tex
\section{Introduction}
\subsection{Motivations and current status}
The $\mathcal{CP}$ violation (CPV), being a crucial phenomenon in the realm of flavor physics within the Standard Model (SM), is a fundamental requirement for achieving a dynamic explanation of the baryon-antibaryon asymmetry observed in the present universe \cite{Sakharov:1967dj,Planck:2015fie}. In the SM, it is implemented through the famous quark mixing picture that was firstly suggested by Kobayashi-Maskawa (KM) mechanism, and specified by one weak phase in the Cabibbo-Kobayashi-Maskawa (CKM) matrix \cite{Cabibbo:1963yz,Kobayashi:1973fv}. 
However, the CPV in the SM is almost nine orders of magnitude smaller than the requirement of the matter-antimatter asymmetry in the Universe \cite{Planck:2015fie}, implying that new CPV sources are needed. Besides, the CPV phase in the CKM matrix is the unique phase parameter in the SM. It is typically more sensitive than all the other parameters, offering a pathway to explore various aspects of complex strong dynamics methods developed over the past decades. This sensitivity allows for precise testing of the fundamental flavor structure of the SM and the search for potential new physics (NP) beyond the SM \cite{Belle:2015qfa,Belle:2016dyj,Belle:2009zue,Belle:2016fev,LHCb:2014vgu,LHCb:2017avl,LHCb:2015gmp,LHCb:2015svh,LHCb:2021trn}. 
While the KM mechanism has been well-established in mesonic mixing and decays, including the $K$-, $B$- and $D$-meson systems  \cite{Christenson:1964fg,BaBar:2001pki,Belle:2001zzw,LHCb:2019hro,PDG}, CPV in baryons has yet to be observed. Therefore, investigations into baryonic CPV are crucial to provide guidance for experimental measurements and to establish a complementary platform for testing the KM mechanism and probing new CPV mechanisms.

The field of baryon physics is currently experiencing a renaissance, driven by the wealth of data collected by prominent colliders such as BESIII, Belle II, and LHCb. Recent years have seen significant advancements in baryon physics, exemplified by the observation of the double-charm baryon $\Xi_{cc}^{++}$ by LHCb \cite{LHCb:2017iph,LHCb:2018pcs}, with theoretical implications discussed in works such as \cite{Yu:2017zst}. While experimental efforts related to baryon $\mathcal{CP}$ violation (CPV) have been extensive, it is important to acknowledge that the results obtained are not yet definitive.
BESIII has made remarkable strides in this field, achieving high-precision measurements of hyperon polarization and \CP asymmetries induced by the asymmetry parameter $\alpha(\Lz\to p\pim)$ at the level of $\mathcal{O}(10^{-3})$. However, these measurements still exhibit discrepancies of approximately one to two orders of magnitude compared to theoretical predictions \cite{BESIII:2018cnd,BESIII:2021ypr,Donoghue:1986hh}. Similarly, investigations into CPV, as defined by the ratio of $\beta+\bar{\beta}$ and $\alpha-\bar{\alpha}$, have been conducted in the $\Xi^{-}\to\Lambda \pi^{-}$ decay utilizing a comprehensive angular analysis approach \cite{BESIII:2021ypr}. Such an observable may exhibit a more significant impact compared to direct $\mathcal{CP}$ violation, as it is influenced by the cosine dependence on the strong phase. We will revisit this topic in Section~\ref{sec.3}. Most of weak decay modes of charm baryons \Lc and \Xic have been extensively studied by experiments like BESIII, Belle, and LHCb, with a particular focus on decay parameters and associated $\mathcal{CP}$ violation effects \cite{BESIII:2019odb,BESIII:2022udq,Belle:2021crz}. Notably, BESIII has achieved a milestone by measuring direct $\mathcal{CP}$ violation in the inclusive decay $\Lambda^{+}_{c}\to \Lambda X$ for the first time \cite{BESIII:2018ciw}. In comparison to BESIII and Belle II, LHCb stands out as a prominent factory for $b$-baryons, since the production of $b$-baryons is only one half of $B^{0,+}$ mesons and even more than $B_s^0$ due to the fragmentation factors. LHCb currently produces around $10^{12}$ \Lb baryons, with the yield of $\Xi_b$  being an order of magnitude smaller \cite{Jiang:2018iqa}. CPV in $\Lb\to p\pip\pim\pim$ was reported by LHCb Collaboration with the evidence of $3.3\sigma$ based on the triple product asymmetry \cite{LHCb:2016yco}, and the direct CPV measurements in two body decays $\Lambda^{0}_{b}\to p\pi^{-}$ and $pK^{-}$ have reached to the precision of $10^{-2}$ \cite{LHCb:2018fly}. Meanwhile, it is noteworthy that many three and four body decay modes of $\Lambda^{0}_{b}$ and $\Xi^{0}_{b}$ are investigated in the experiments for baryonic CPV \cite{LHCb:2014yin,LHCb:2014nhe,LHCb:2018fpt,LHCb:2019oke,LHCb:2019jyj}. These studies contribute significantly to our understanding of $\mathcal{CP}$ violation in the baryon sector and pave the way for further discoveries in this intriguing field. 

On the theoretical front, $\mathcal{CP}$ violation in non-leptonic baryon decays has been extensively studied using various methods. The generalized factorization approach has been a popular choice, incorporating improvements such as NLO corrected effective Hamiltonians and introducing the parameter $N_c$ to account for non-factorizable contributions \cite{Hsiao:2014mua,Geng:2016kjv,Hsiao:2017tif}. Final state interactions have also been explored, with a focus on complete hadronic loop integrals to capture the dynamics \cite{Jia:2024pyb}. QCD factorization with the diquark hypothesis has been another avenue of investigation in understanding $\mathcal{CP}$ violation in baryon decays \cite{Zhu:2018jet,Zhu:2016bra}. Perturbative QCD based on $k_T$ factorization has been utilized in several studies, with advancements made to incorporate high twist effects for improved accuracy \cite{Lu:2009cm,Chou:2001bn,Zhang:2022iun, Rui:2022sdc,Rui:2022jff,Yu:2024cjd}. Theoretical predictions based on exact SU(3) flavor symmetry have led to the formulation of relationships between $\mathcal{CP}$ asymmetries and branching ratios \cite{Roy:2019cky,He:2015fsa,He:2015fwa}. In a seminal work by Cheng et al. \cite{Cheng:1996cs}, Cabibbo-allowed two-body hadronic weak decays of bottom baryons were analyzed using the naive factorization approach, employing form factors within a nonrelativistic quark model framework. These theoretical frameworks provide valuable insights into the dynamics of $\mathcal{CP}$ violation in baryon decays and offer predictions that can be tested against experimental data.

On the phenomenological side, The non-zero spin of baryons introduces rich polarization effects, partial waves, and helicity structures that play a crucial role in the search for baryonic $\mathcal{CP}$ violation. The polarization of hyperons such as $\Lambda$ and $\Xi$ produced in $e^+e^-$ collisions or weak decays can provide valuable information for probing asymmetry parameters associated with their decays and exploring $\mathcal{CP}$ violation through these parameters~\cite{BESIII:2018cnd,BESIII:2021ypr}. Recent proposals have outlined strategies to investigate hyperon $\mathcal{CP}$ violation through the decays of charm baryons like \Lc~\cite{Wang:2022tcm,Belle:2022uod,Salone:2022lpt}, and to determine weak phases and strong CP-induced electron dipole moments in $e^+e^-$ collisions \cite{Salone:2022lpt,Fu:2023ose,He:2022jjc,He:1992ng}. The detailed investigation of multibody decaying differential distributions of $b$-baryons has been carried out based on the quasi-two-body picture using the Jacob-Wick helicity formalism~\cite{Durieux:2016nqr}. The $\mathcal{CP}$ asymmetries arising from interference effects between different intermediate resonances or decaying paths have also been taken into account in various studies~\cite{Zhao:2024ren,Zhang:2022iye,Zhang:2022emj,Shen:2023eln}. Intermediate states from weak decays of $b$-baryons are expected to be polarized and correlated in spin space, leading to non-trivial distributions of momentum for their secondary decay products in phase space. Triple product correlations, defined as rotational invariants using spin and momentum vectors, have been proposed to search for $\mathcal{CP}$ violation in baryon decays \cite{Durieux:2015zwa,Bevan:2014nva,Gronau:2015gha,Wang:2022fih,Geng:2021sxe}. The helicity scheme is particularly useful for polarization analysis, as it allows for clear resolution of polarization in the helicity representation, simplifying the decomposition and recombination of angular momentum wave functions \cite{Jacob:1959at}. Therefore, helicity amplitudes are employed in this work to facilitate the analysis of polarization effects in baryon decays.

In mesonic processes, it has been well established that direct $\mathcal{CP}$ violation in bottom decays can be significant, reaching $\mathcal{O}(10\%)$. For instance, observables such as $C_{\pi^+\pi^-}=-0.32\pm0.04$, $\mathcal{A}_{\overline B^0\to K^-\pi^+}=-0.084\pm0.004$, $\mathcal{A}_{\overline B_s^0\to K^+\pi^-}=0.213\pm0.017$ have been measured with large $\mathcal{CP}$ violation effects~\cite{PDG}. Furthermore, substantial $\mathcal{CP}$ violation has been observed in specific regions of phase space in three-body decays like $B^+\to K^+K^-K^+$, $\pi^+\pi^-K^+$ and $\pi^+\pi^-\pi^+$~\cite{PDG}. However, many measurements of direct CPV $a^{dir}_{CP}$ of $b$-baryon decays have generally yielded smaller values. This observation suggests that the strong phases involved in $b$-baryon decays might be significantly smaller than those in $B$ decays, or that cancellations between various partial wave contributions lead to an overall small global $\mathcal{CP}$ violation when integrated over the entire phase space. In light of this, it is essential to consider observables beyond direct $\mathcal{CP}$ asymmetry in the study of $b$-baryon decays. Some of these observables may exhibit distinct dependencies on strong phases compared to $a^{dir}_{CP}$ while others may involve single partial wave contributions or interference terms, offering alternative avenues to overcome suppression in cases where detecting $a^{dir}_{CP}$ directly proves challenging.

\subsection{Strategy to search for baryonic CPV}
Compared to hyperons and charmed baryons, $b$-baryons exhibit the greatest potential for observing baryonic CPV analogous to the observations made in mesons. 
In charmed mesons, $\mathcal{CP}$ asymmetries are typically on the order of $10^{-3}$ or even smaller based on rough estimates, a trend that has been confirmed by model calculations such as the factorization-assisted topological-amplitude (FAT) approach~\cite{Li:2019hho,Li:2012cfa,Qin:2014nxa}, with results consistent with experimental data. Recent studies have conducted a comprehensive analysis of the decays of \Lc to a light baryon and vector meson within the framework of FSIs, incorporating a thorough calculation of hadronic loop integrals. The predicted CPV effects in these decays are all expected to be on the order of $10^{-4}$. Similarly, $\mathcal{CP}$ asymmetries in hyperons are also anticipated to occur at the level of $10^{-4} - 10^{-5}$ within the SM~\cite{Donoghue:1986hh,Deshpande:1994vp}. Given the relatively smaller CPV effects observed in hyperons and charm baryons, $b$-baryons emerge as promising candidates for the initial detection of significant baryonic CPV.

Conventionally, many studies of $\mathcal{CP}$ asymmetry focus on comparing the decay widths of a process with its $\mathcal{CP}$-conjugate process, specially the direct $\mathcal{CP}$ asymmetry $a_{CP}^{dir}=(\Gamma-\bar{\Gamma})/(\Gamma+\bar{\Gamma})$. In a general decay scenario, the corresponding matrix element denoted as $\bra{f}\mathcal{H}\ket{i}$ is expected to be equal to $\bra{\left[CP\right]f}\mathcal{H}\ket{\left[CP\right]i}$ provided $\mathcal{CP}$ is conserved, where $i,f,\mathcal{H}$ are initial, final states and interaction Hamiltonian, respectively. Therefore, $\mathcal{CP}$ violation implies that 
\begin{equation}
	\begin{aligned}
\Gamma\left[i(p_{i},\lambda_{i})\to f_{1}(p_{1},\lambda_{1})+...f_{m}(p_{m},\lambda_{m})\right]\neq \Gamma\left[i(-p_{i},-\lambda_{i})\to f_{1}(-p_{1},-\lambda_{1})+...f_{m}(-p_{m},-\lambda_{m})\right] \; .
	\end{aligned}
\end{equation}
However, the direct $\mathcal{CP}$ asymmetry is typically determined by integrating over the phase space, resulting in the loss of crucial information regarding momentum direction and helicity correlations~\cite{Kayser:1989vw,Nowakowski:1988dx,Nowakowski:1989ju}. This method encounters significant challenges when cancellations occur in different regions of the phase space. Another limitation of the direct $\mathcal{CP}$ asymmetry is its dependence on $\sin\Delta\delta\sin\Delta\phi$, where $\Delta\delta$ and $\Delta\phi$ represent strong and weak phase differences, respectively. This dependence implies that the asymmetry may be too small to observe if the strong phase $\Delta\delta$ is negligible, which is often the case in real-world scenarios such as the decay of hyperons $\Lambda,\Sigma,\Xi$ where typical $\Delta\delta$ values are on the order of $10^{-1}\sim10^{-2}$~\cite{Donoghue:1986hh}. Estimating the strong phase accurately poses a significant challenge, as it is not easily obtained through dynamical methods. Some processes can utilize experimental data from $\pi\pi$ and $N\pi$ scatterings or perturbative QCD methods based on various factorization theorems to estimate the strong phase~\cite{Lu:2000em,Keum:2000wi,Beneke:2003zv,Beneke:1999br,Beneke:2000ry,Beneke:2001ev,LHCb:2019jta,LHCb:2022fpg,AlvarengaNogueira:2015wpj,Cheng:2020ipp,Garrote:2022uub,Bediaga:2022sxw,Wang:2024oyi}. However, predicting the behavior of many decays remains challenging. An alternative approach to address this issue is to consider $\mathcal{CP}$-violating observables that either allow for the extraction of $\mathcal{CP}$ violation from specific regions of the phase space, thereby isolating it from particular partial wave interference terms, or are proportional to $\cos\Delta\delta$, such as $(\beta+\bar \beta)/(\alpha-\bar \alpha)$ and $\mathcal{CP}$ asymmetries arising from $\mathcal{T}$-odd correlations. The benefits of these novel $\mathcal{CP}$-violating observables can be outlined as follows: 
\begin{itemize}
    \item These observables, such as asymmetry parameters and $\mathcal{T}$-odd correlations, are intricately connected to the helicity structures and polarizations in the decays, thereby enabling the comprehensive utilization of spin and valuable partial wave information. As an example, the construction of $\mathcal{T}$-odd correlations can involve the utilization of the $b$ quark spin in quark-level decays. Their realization at the hadronic level~\cite{Bensalem:2000hq}, however, becomes unfeasible when considering $B$ meson decays. In contrast to mesons, the decays of $b$-baryons present an opportunity to investigate these $\mathcal{T}$-odd observables, as they carry non-zero spins.
    \item These observables are always in accordance with the phase space distributions, enabling them to distinguish the contributions from distinct partial waves or resonances, particularly when the associated resonances have different $J^{P}$ quantum numbers leading to disparate momentum distributions of the final products. Consequently, they might provide a way to disentangle various contributions, thereby avoid possible cancellations of them, and finally give some insightful implications for our understanding.
    \item Some of the asymmetry parameters and $\mathcal{T}$-odd correlations serve as valuable tools for capturing the effects of FSIs and higher-order perturbative corrections \cite{Cronin:1963zb,Overseth:1967zz,Cleland:1972fa,Lee:1957qq,Lee:1957he,Hikasa:1985qi,Dharmaratna:1996xd}, thereby offering a robust way to testing various dynamical methods. For instance, the decay asymmetry measured in $\Lambda^{+}_{c}$ decays imposes significant constraints on the associated dynamical models, providing some insights into the underlying processes \cite{Jia:2024pyb,LHCb:2022sck,Belle:2022bsi,Belle:2022uod,BESIII:2019odb,BESIII:2022udq,BESIII:2023wrw}.
    \item The $\mathcal{T}$-odd correlations offer a distinct class of $\mathcal{CP}$ observables that is typically found to be proportional to $\cos\Delta\delta$, thereby providing a pathway to investigate CPV effects without suffering possible large suppression from potential small strong phases \cite{Wang:2022fih,Valencia:1988it,Datta:2003mj,Durieux:2015zwa,Donoghue:1986hh}. Moreover, it is possible to identify a pair of CPV observables depending on $\sin\Delta\delta$ and $\cos\Delta\delta$, that are induced by $\mathcal{T}-$even and -odd asymmetries, respectively. Their complementarity is anticipated to be valuable for exploring CPV in baryons in the future.
    \item Small $\mathcal{T}-$odd asymmetries are also valuable as they provide a new path for investigating potential new physics beyond the SM~\cite{Bensalem:2002ys,Bensalem:2000hq,Bensalem:2002pz,Rui:2022jff,LHCb:2016hwg}. Numerous models of new physics predict the existence of heavier particles at higher energy scales, thereby they could proceed decays and scatterings of SM particles through loop effects with complex interaction vertex, and hence reasonably shift SM predictions for $\mathcal{T}-$odd asymmetries. Finally, they serve as a compelling signal of potential new physics when they are small in SM.
\end{itemize}
In this work, we will conduct a comprehensive review of asymmetry parameters and their associated properties about $\mathcal{CP}$ and $\mathcal{T}$ violations. An exact complementary relation between the CPV observables depending on $\Delta\delta$ as $\sin\Delta\delta$ and those as $\cos\Delta\delta$ is established by exploring the general features of $\mathcal{T}$-even and -odd correlations. In order to reach the aim, we will prove the general conclusion that any $\mathcal{CP}$ asymmetries induced by $\mathcal{T}$-odd correlations depend on cosine of $\Delta\delta$ after incorporating two conditions, and then the criterion for complementarity is discussed based on this exact proof. The realization of these observables in experimental measurements could be arrived by applying angular analysis, hence the polarization, differential distribution and CPV observables are all essential in this works.

The rest of this paper is organized as follows. In Sec.\ref{sec.2}, we firstly provide the general form of amplitudes for many baryon decay modes used in this work. A concise introduction to direct CPV and time reversal violation ($\mathcal{T}$ violation) is given as a start point. Especially, some general comments on $\mathcal{T}$ violation are emphasized owing to their importance for our understandings. Next, in Sec.\ref{sec.3}, a complete review is provided for the polarization and it's helicity representation in weak decays and strong productions including, e.g. the asymmetry parameters $\alpha,\beta$ in baryon decays firstly proposed by Lee and Yang for Parity violation, and the normal polarization of $\Lambda^{0}_{b}$ produced in the proton-proton collisions based on the predictions of perturbative QCD and heavy quark symmetry.
In Sec.\ref{sec.5}, we establish a general complementary relation
between $\mathcal{T}$-even and -odd correlations. In order to connect them to experimental measurements, the angular analysis for quasi-two-body and three-body decays of b-baryons are performed. In Sec.\ref{sec.6}, we show some comments on CPV in b-baryon decays under the picture of topological diagrammatic amplitudes and suggestions on experimental searches. In App.\ref{App.}, we comprehensively review the helicity approach to the scattering and decaying amplitudes of particles with spin. The reason why helicity representation and how to construct helicity state, to calculate decay angular distribution for both of quasi-two-body and three-body are showed pedagogically.

%% file: 2-directCPV.tex
\section{A brief review of direct $\mathcal{CP}$ violation and $\mathcal{T}$ violation}\label{sec.2}
In this section, we will review some discussions on conventional direct CPV and direct-like CPV, e.g. one observable proposed recently named as {\it Partial wave CP asymmetry}~\cite{Zhang:2021fdd}. Additionally, we will incorporate a discussion on $\mathcal{T}$ violation to provide a comprehensive foundation for subsequent chapters. The primary objective of this section is to establish and present fundamental formulas and derivations that will serve as essential references throughout this paper, thereby enhancing the clarity and coherence of our subsequent discussions.

\subsection{Decay amplitudes}\label{sec: decay amplitude}
In general, the decay amplitudes of a $J^P={1\over 2}^+$ baryon $\mathbf{B}_i$ decaying into a $J^P={1\over 2}^\pm$ or a ${3\over2}^\pm$ baryon $\mathbf{B}_f$ and a pseudoscalar meson $P$ or a vector meson $V$ can be expressed as \cite{Cheng:1996cs}
\begin{align}\label{eq:amplitudes}
\mathcal{A}\left(\mathbf{B}_i({1/2}^+)\to \mathbf{B}_f({1/2}^\pm)P\right)
&=\bar u_f\left(A^\pm+B^\pm\gamma_5\right)u_i,
\\
\mathcal{A}\left(\mathbf{B}_i({1/2}^+)\to \mathbf{B}_f({1/2}^\pm)V\right)
&=\epsilon^{*\mu}\bar u_f \left[\left(A_1^\pm\gamma_\mu+A_2^\pm {p_\mu\over M_i}\right)\gamma_5+\left(B_1^\pm\gamma_\mu+B_2^\pm {p_\mu\over M_i}\right)\right]u_i,
\\
\mathcal{A}\left(\mathbf{B}_i({1/2}^+)\to \mathbf{B}_f({3/2}^\pm)P\right)
&=iq_\mu\bar u_f^\mu\left(C^\pm+D^\pm\gamma_5\right)u_i,
\\
\mathcal{A}\left(\mathbf{B}_i({1/2}^+)\to \mathbf{B}_f({3/2}^\pm)V\right)
&=\epsilon^{*\mu}\bar u_f^\nu \bigg[\left(C_1^\pm g_{\mu\nu}+C_2^\pm {p_{1\nu}\over M_i}\gamma_\mu+C_3^\pm {p_{1\nu}p_{2\mu}\over M_i^2}\right)
\nonumber\\
&~~~~~~~~~+\left(D_1^\pm g_{\mu\nu}+D_2^\pm {p_{1\nu}\over M_i}\gamma_\mu+D_3^\pm {p_{1\nu}p_{2\mu}\over M_i^2}\right)\gamma_5 \bigg] u_i,
\end{align}
where the momenta and spin indices of the initial- and final-state spinors $u_{i,f}^{(\mu)}$ have been hidden for convenience, $M_i$ is the initial-satate baryon mass. In the second amplitude, $p$ is the momentum of the involved final-state baryon, and in the last one, $p_{1,2}$ are the momenta of the involved initial- and final-state baryon, $\epsilon^{\mu}$ is the polarization vector of the vector meson, and $A,B,C,D$ are the amplitudes correpsonding to different Lorentz structures.

\subsection{Direct $\mathcal{CP}$ violation}
The direct $\mathcal{CP}$ asymmetry $a^{dir}_{CP}=(\Gamma-\bar{\Gamma})/(\Gamma+\bar{\Gamma})$, being the most straightforward observable, has been extensively studied in baryon decays. It is particularly crucial in two-body non-leptonic b-baryon decays, as achieving polarization in high-energy proton collisions to observe angular distributions such as $\Lambda^{0}_{b}\to p\pi^{-}$ is challenging. In such cases, the direct $\mathcal{CP}$ asymmetry remains as the sole surviving observable. By the way, it shoule be noted that a theory without Charge Conjugation symmetry can not give any difference between $\Gamma$ and $\bar{\Gamma}$ provided CP is conserved since decay width requires the summation of all of helicity components of amplitudes, and vice versa.While other intriguing observables arising from multi-body decays will be discussed later as essential aspects, let us initially concentrate on $a^{dir}_{CP}$. It would be beneficial to provide a clear demonstration of direct CPV in b-baryon decays such as $\Lambda^{0}_{b}\to p\pi^{-},pK^{-}$, which has been experimentally measured~\cite{LHCb:2018fly,CDF:2014pzb}.

Consider now the partial wave amplitudes, $S,P$ waves corresponding to $\mathcal{A}(\Lambda^{0}_{b}\to p\pi^{-},pK^{-})=\bar{u}_{p}(S+P\gamma_{5})u_{\Lambda_{b}}$, and $\bar{S},\bar{P}$ the $\mathcal CP$-conjugate amplitudes. The decay widths $\Gamma$ and $\bar{\Gamma}$ are given by
\begin{equation}
	\begin{aligned}
     \Gamma=\frac{|\vec{p}|}{8\pi M^{2}} \left(|S|^{2}+|P|^{2}\right),~~~
     \bar{\Gamma}=\frac{|\vec{p}|}{8\pi M^{2}}\left(|\bar{S}|^{2}+|\bar{P}|^{2}\right)
	\end{aligned}
\end{equation}
where $|\vec{p}|$ is the final proton momentum in the $\Lambda_b^0$ rest frame, and $M$ is initial $\Lambda^{0}_{b}$ mass. Each partial wave amplitude contains the contributions from tree and penguin diagrams with different weak and strong phases, 
\begin{equation}\label{eq:S and P wave1}
	\begin{aligned}
		S&=|S_{t}|e^{i\delta_{s,t}}e^{i\phi_{t}}+|S_{p}|e^{i\delta_{s,p}}e^{i\phi_{p}}\\
		P&=|P_{t}|e^{i\delta_{p,t}}e^{i\phi_{t}}+|P_{p}|e^{i\delta_{p,p}}e^{i\phi_{p}}\\
		\bar{S}&=-\left\{|S_{t}|e^{i\delta_{s,t}}e^{-i\phi_{t}}+|S_{p}|e^{i\delta_{s,p}}e^{-i\phi_{p}}\right\}\\
		\bar{P}&=|P_{t}|e^{i\delta_{p,t}}e^{-i\phi_{t}}+|P_{p}|e^{i\delta_{p,p}}e^{-i\phi_{p}}\\
	\end{aligned}
\end{equation}
where the subscript in $(S/P)_{t,p}$ represent tree and penguin components of $S/P$ wave, the first subscript in $\delta_{s/p,t/p}$ signs which partial wave is, while the second one denote tree or penguin we are focused on. $\delta,\phi$ are strong and weak phases respectively. 
the subscripts in $(S/P)_{t,p}$ represent the tree and penguin components of the $S/P$ wave, the first subscript in $\delta_{s/p,t/p}$ indicates the specific partial wave, while the second one denotes whether we are focusing on the tree or penguin contribution. The phases $\delta,\phi$ correspond to the strong and weak phases, respectively. Finally, the direct $\mathcal CP$ asymmetry
\begin{equation}\label{eq.dirCP}
	\begin{aligned}
      a_{CP}^{dir}=\frac{\Gamma-\bar{\Gamma}}{\Gamma+\bar{\Gamma}}&=\frac{|S|^{2}+|P|^{2}-|\bar{S}|^{2}-|\bar{P}|^{2}}{|S|^{2}+|P|^{2}+|\bar{S}|^{2}+|\bar{P}|^{2}}\\
      &= -\frac{\sin(\delta_{s,t}-\delta_{s,p})+r\sin(\delta_{p,t}-\delta_{p,p})}{K+\left[\cos(\delta_{s,t}-\delta_{s,p})+r\cos(\delta_{p,t}-\delta_{p,p})\right]\cos\Delta\phi}\sin\Delta\phi
	\end{aligned}
\end{equation}
is proportional to the difference in weak and strong phases, represented in the form of a sine function. The parameters $r=(|P_{t}||P_{p}|)/(|S_{t}||S_{p}|)$, and $K=(|S_{t}|^{2}+|S_{p}|^{2}+|P_{t}|^{2}+|P_{p}|^{2})/(2|S_{t}||S_{p}|)$ are involved in the expression.. Furthermore, the direct $\mathcal CP$ violation in multi-body decays such as $\Lambda_b\to N^*(1/2)K^*\to p\pi K\pi$ and $\Lambda_{b}\to pa_{1}\to p\pi\pi\pi$ can be described in a similar manner, incorporating more partial wave components.

The direct CPV formula in~\eqref{eq.dirCP}, which involves the difference between $\Gamma$ and $\bar{\Gamma}$, is conceptually straightforward. However, it may encounter challenges due to significant cancellations between contributions from different partial waves, as highlighted in a recent study~\cite{Yu:2024cjd}. To mitigate this issue, alternative CP asymmetry observables can be introduced. There are many possibilities for achieving this goal, and one promising approach is the utilization of the {\it Partial Wave CP Asymmetry}, which is based on the Legendre decomposition of $|\mathcal{M}|^{2}$ instead of the amplitude $\mathcal{M}$ \cite{Zhang:2021fdd}. Let us delve deeper into this concept.
Taking the decay $\Lambda_{b}\to N^*(3/2)K^*\to p\pi K\pi$ as an example, we integrate out the entire phase space except for $\cos\theta$ where $\theta$ is the relative angle between the momenta of $N^{*}$ in the $\Lambda^{0}_{b}$ rest frame and that of proton in the $N^{*}$ rest frame, 
    \begin{equation}
	\begin{aligned}
     d\Gamma\propto \langle|\mathcal{M}|^{2}\rangle d\cos\theta
	\end{aligned}
\end{equation}
with $\langle|\mathcal{M}|^{2}\rangle$ representing the average over the whole phase space except for the $\cos\theta$, is expanded as Legendre polynomial series
    \begin{equation}
	\begin{aligned}
     \langle|\mathcal{M}|^{2}\rangle=\sum_{j}\langle\omega^{j}\rangle P_{j}(\cos\theta) \; . 
	\end{aligned}
\end{equation}
Based on this, a set of CPV observables named as {\it Partial wave CP asymmetry} are defined as
\begin{equation}\label{eq:defination}
	\begin{aligned}
		A_{CP}^{j}=\frac{\langle\omega^{j}\rangle-\langle\bar{\omega}^{j}\rangle}{\langle\omega^{j}\rangle+\langle\bar{\omega}^{j}\rangle}
	\end{aligned}
\end{equation}
Moreover, $\langle\omega^{j}\rangle$ are extracted through convoluting $\langle|\mathcal{M}|^{2}\rangle$ with additional weight function $P_{j}(\cos\theta)$ by unitizing the orthogonality of Legendre polynomials. In other words, it could be measured in the experiment by taking bibs by bins with different weight varying with $P_{j}(\cos\theta)$
\begin{equation}
	\begin{aligned}
	 \langle\omega^{j}\rangle = \int_{-1}^{+1} \langle|\mathcal{M}|\rangle^{2}P_{j}(\cos\theta)d\cos\theta \; .
	\end{aligned}
\end{equation}
Explicitly, for the example decay channel $\Lambda_{b}\to N^*(3/2^{-})K^*\to p\pi K\pi$,
\begin{equation}
	\begin{aligned}
		\langle \frac{d\Gamma}{d \cos\theta}\rangle&\propto(\mathcal{A}_{2}+\mathcal{A}_{3}-\mathcal{A}_{1})P_{2}(\cos\theta)+(1+\mathcal{A}_{1}+\frac{\mathcal{A}_{2}+\mathcal{A}_{3}}{2})P_{0}(\cos\theta)
	\end{aligned}
\end{equation}
where $\mathcal{A}_{1},\mathcal{A}_{2},\mathcal{A}_{3}$ are expressed in the helicity amplitudes as 
\begin{equation}
	\begin{aligned}
		\mathcal{A}_{1}&=\left|\mathcal{H}_{+1,+\frac{3}{2}}\right|^{2}+\left|\mathcal{H}_{-1,-\frac{3}{2}}\right|^{2}\; ,\\
		\mathcal{A}_{2}&=\left|\mathcal{H}_{+1,+\frac{1}{2}}\right|^{2}+\left|\mathcal{H}_{-1,-\frac{1}{2}}\right|^{2} \; ,\\
		\mathcal{A}_{3}&=\left|\mathcal{H}_{0,-\frac{1}{2}}\right|^{2}+\left|\mathcal{H}_{0,+\frac{1}{2}}\right|^{2}  \; .
	\end{aligned}
\end{equation}
Two CPV observables appear according to the definition in~\eqref{eq:defination}, 
\begin{equation}
	\begin{aligned}
		A_{CP}^{0}=\frac{\langle\omega^{0}\rangle-\langle\bar{\omega}^{0}\rangle}{\langle\omega^{0}\rangle+\langle\bar{\omega}^{0}\rangle} \; ,~		A_{CP}^{2}=\frac{\langle\omega^{2}\rangle-\langle\bar{\omega}^{2}\rangle}{\langle\omega^{2}\rangle+\langle\bar{\omega}^{2}\rangle} \; .
	\end{aligned}
\end{equation}
The dependence of $A_{CP}^{0}$ and $A_{CP}^{2}$ on strong phases exhibits a similarity to direct CP violation, characterized by a sine function. However, a crucial distinction lies in their capability to discern the contributions from distinct helicity amplitudes $\mathcal{H}_{i,j}$, whereas direct CP violation combines all helicity amplitude squares. From a phenomenological viewpoint, these quantities hold significant experimental promise, as specific $A^{j}_{CP}$ values may experience amplification due to interference effects stemming from diverse intermediate resonances in multi-body decays, thereby yielding substantial strong phases. An illustration of this phenomenon can be observed in the investigation of interference effects between $\Delta(1232)(3/2^+)$ and $N(1440)(1/2^+)$ in the $\Lambda^{0}_{b}$ decay, as elaborated in~\cite{Zhang:2021fdd}.

\subsection{$\mathcal{T}$ violation}\label{sec: T violation}
Time reversal, as a fundamental symmetry at the microscopic level, has long been anticipated as a consequence of CPV, given that $\mathcal{CPT}$ is considered a robust symmetry of nature within the framework of modern quantum field theory. Furthermore, in modern quantum field theory, the conservation of $\mathcal{CPT}$ is established as a rigorous theorem that arises from any local Hermitian and Lorentz invariant quantum field theory, as elucidated by Schwinger in~\cite{Schwinger:1953tb}. Experimentally, $\mathcal{T}$ violation was initially observed in the neutron-K meson system through a comparison of the transformation probabilities of $K^{0}$ transforming into $\bar{K}^{0}$ and $\bar{K}^{0}$ into $K^{0}$~\cite{CPLEAR:1998dvs}. In general, testing time reversal symmetry can be achieved by comparing the cross section of a scattering reaction with its inverse evolution. However, this test is not feasible in decay processes since there is no direct counterpart to reactions like $p\pi^{-}\to \Lambda$ for $\Lambda\to p\pi^{-}$. Furthermore, even if one could produce the inverse reaction of a decay in experiments, it remains a significant challenge as the time reversal operation involves not only a reversal of momentum and spin but also quantum phases in wave functions~\cite{Lee:1981mf}. As a result, it is argued that $\mathcal{T}$-odd observables may offer insights to confirm whether time reversal symmetry is respected in decays. Numerous studies have been conducted in this regard, including investigations in $b$-baryon decays \cite{Leitner:2006sc,Leitner:2006nb,Geng:2021sxe,Wang:2022fih,Valencia:1988it,Datta:2003mj,Durieux:2015zwa,Donoghue:1986hh,Dutta:2021del,
Geng:2021lrc,
Rui:2022jff}.

A type of $\mathcal{T}$-odd observables can be constructed using momentum, polarization and spin vectors. The simplest case is the triple product correlation 
\begin{equation}
	\begin{aligned}
	  \mathcal{O}_{-}=(\vec{v}_{i}\times\vec{v}_{j})\cdot\vec{v}_{k} \; ,
	\end{aligned}
\end{equation}
where $\vec{v}_{i}$ is the spin or momentum of the $i$th particle, and the subscript $-$ in $\mathcal{O}_{-}$ implies its time reversal transformation phase factor. Hence, an asymmetry parameter is defined naturally through the expectation value of operator $\mathcal{O}_{-}$
\begin{equation}\label{def: T odd quantity}
	\begin{aligned}
\Delta_{t}=\bra{f}\mathcal{O}_{-}\ket{f}= \sum_{t}\left[t \; \Gamma(t)\right] /\Gamma \; , 
	\end{aligned}
\end{equation}
where we take $t$ to represent the eigenvalue of $\mathcal{O}_{-}$, $\ket{f}$ denotes the final state vector, and $\Gamma(t)/\Gamma$ signifies the probability of the final state having an eigenvalue of $t$. A non-vanishing asymmetry $\Delta_{t}$ can arise even if both the theory and the final state are $\mathcal{T}$-invariant, as a result of unavoidable final state interactions~\cite{Lee:1981mf}. The statement that non-zero $\Delta_{t}$ implies $\mathcal{T}$ violation is actually valid only in the absence of final state interaction like decay $\pi^{-}\to\mu^{-}\nu_{\mu}$. The real physics world, however, is occupied by many reactions with products of final hadrons that scatter each other before evolving to states that are observed in the experiments. Taking an example to illustrate~\cite{Bigi:2000yz,Lee:1981mf}, we consider a weak decay $i\to f$ receiving contributions from different partial waves. We temporarily assume that the weak Hamiltonian $\mathcal{H}_{W}$ conserves $\mathcal{T}$, or in the language of commutation relation $\mathcal{T}\mathcal{H}_{W}=\mathcal{H}_{W}\mathcal{T}$. Under the leading order for the weak interaction and all orders for the strong interaction, picking these scattering effects up, one has the transition amplitude
\begin{equation}
	\begin{aligned}
	  \mathcal{A}_{i\to f}=\bra{f}U(+\infty,0)\mathcal{H}_{W}\ket{i} \; .
	\end{aligned}
\end{equation}
Decompose the final states $\ket{f}$ in the angular momentum representation, we arrive at
    \begin{equation}\label{2.23}
        	\begin{aligned}
        		\ket{f}=\sum_{\omega,J,m}\ket{\omega,J,m}\bra{\omega,J,m}U(\infty,0)\mathcal{H}_{W}\ket{i_{m}}
        	\end{aligned}
    \end{equation}
where $\ket{\omega,J,m}$ is an angular momentum eigenstate with specific parity $\omega$. Assumption of parity conservation for scattering operator $U(\infty,0)$ can be imposed reasonably, since the considered final interactions are always strong and electromagnetic ones. Further, we only take elastic scattering as an illustration since it is enough to demonstrate our point. Hence,
   \begin{equation}\label{2.24}
    	\begin{aligned}
    		U(\infty,0)\ket{\omega,J,m}=exp\left\{-i\delta_{\omega,J,m}\right\}\ket{\omega,J,m}\; .
    	\end{aligned}
    \end{equation}
Substituting both of (\ref{2.23}) and (\ref{2.24}) into the expectation value of $\mathcal{O}_{-}$, we arrive at
\begin{equation}
	\begin{aligned}
		\bra{f}\mathcal{O}_{-}\ket{f}=\sum_{\omega^{\prime},J^{\prime},m^{\prime},\omega,J,m}&\bra{\omega^{\prime},J^{\prime},m^{\prime}}\mathcal{O}_{-}\ket{\omega,J,m}
		exp\left\{i(\delta_{\omega^{\prime},J^{\prime},m^{\prime}}-\delta_{\omega,J,m})\right\}\\&\mathcal{H}^{*}_{W}(\omega^{\prime},J^{\prime},m^{\prime})
		\mathcal{H}_{W}(\omega,J,m) \; , \\
	\end{aligned}
\end{equation}
where $\mathcal{H}_{W}(\omega,J,m)$ defined by $\bra{\omega,J,m}\mathcal{H}_{W}\ket{i_{m}}$. We perform time reversal operator $\mathcal{T}$ on sub-elements $\bra{\omega^{\prime},J^{\prime},m^{\prime}}\mathcal{O}_{-}\ket{\omega,J,m}$ with the constraints from spatial-rotation symmetry and hermiticity
\begin{equation}\label{eq2.26}
	\begin{aligned}
	\bra{\omega^{\prime},J^{\prime},m^{\prime}}\mathcal{O}_{-}\ket{\omega,J,m}=-\bra{\omega^{\prime},J^{\prime},m^{\prime}}\mathcal{O}_{-}\ket{\omega,J,m}^{*}=-\bra{\omega,J,m}\mathcal{O}_{-}\ket{\omega^{\prime},J^{\prime},m^{\prime}}
	\end{aligned}
\end{equation}
Replacing the $\bra{\omega^{\prime},J^{\prime},m^{\prime}}\mathcal{O}_{-}\ket{\omega,J,m}$ by (\ref{eq2.26}) and interchanging $\omega,\omega^{\prime}$ since they are dummy indices, we have
\begin{equation}
		\begin{aligned}
			\bra{f}\mathcal{O}_{-}\ket{f}\propto\sum_{\omega^{\prime},J^{\prime},m^{\prime},\omega,J,m} \mathcal{H}_{W}(\omega^{\prime},J^{\prime},m^{\prime})
			\mathcal{H}_{W}(\omega,J,m)\sin(\delta_{\omega^{\prime},J^{\prime},m^{\prime}}-\delta_{\omega,J,m} +\phi)
		\end{aligned}
	\end{equation}
where the phase difference $\delta_{\omega^{\prime},J^{\prime},m^{\prime}}-\delta_{\omega,J,m}$ and $\mathcal{H}_{W}(\omega,J,m)$ are real since we have assumed $\mathcal{T}$ is conserved for the weak $\mathcal{H}_{W}$, hence the associated weak phase $\phi$ is vanishing. It is apparent that the non-trivial final state interactions provide a non-zero phase shift which varies with different paths, while $\Delta_{t}$ is proportional to this interference effect. Hence, the expectation $\bra{f}\mathcal{O}_{-}\ket{f}$ is not necessarily vanishing even though decaying interaction respects time reversal symmetry. Only if the underlying effects of strong scatterings can be precisely determined, the part of the asymmetry $\Delta_{t}$ independent of final state interactions can reflect whether weak Hamiltonian conserve $\mathcal{T}$ or not. However, extracting the contamination from final state interactions in a clear and unambiguous manner is often fraught with challenges in most cases. The most promising case is no pollution from final state interactions, thus the associated expectation value $\bra{f}Q_{-}\ket{f}$ is largely simplified as
	\begin{equation}
		\begin{aligned}
			\bra{f}\mathcal{O}_{-}\ket{f}&=(\mathcal{T}\mathcal{H}_{W}i_{m},\mathcal{T}\mathcal{O}_{-}\mathcal{H}_{W}i_{m})^{*}
			=(\mathcal{T}\mathcal{H}_{W}\mathcal{T}^{-1}\mathcal{T}i_{m},\mathcal{T}\mathcal{O}_{-}\mathcal{T}^{-1}\mathcal{T}\mathcal{H}_{W}\mathcal{T}^{-1}\mathcal{T}i_{m})^{*}\\
			&=(\mathcal{H}_{W}i_{-m},- \mathcal{O}_{-}\mathcal{H}_{W}i_{-m})^{*}
			=(\mathcal{H}_{W}i_{+m},- \mathcal{O}_{-}\mathcal{H}_{W}i_{+m})
			=-\bra{f}\mathcal{O}_{-}\ket{f} \; , 
		\end{aligned}
	\end{equation}
where $i_{m}$ represents the initial state with spin projection $m$, and the second line of the expression applies the rotation invariance and hermiticity. An illustrative example is the the polarization of $\mu^{-}$ perpendicular to the decay plane in the process $K^{-}\to \pi\mu^{-}\nu_{\mu}$. This observable is anticipated to serve as a manifestation of time reversal violation or CP violation in the decay of $K^{-}$ decay since the associated final state interaction can be disregarded with confidence, as discussed in~\cite{KEK-E246:2004xuu}.

A well-defined CP or $\mathcal{T}$ violation can be established by canceling the impact of final state interactions through the comparison of a decay process with its CP-conjugate counterpart
\begin{align}
a^{\mathcal{T}-odd}_{CP}=\frac{1}{2}\left[\langle \mathcal{O}_{-}\rangle -\langle({CP}) \mathcal{O}_{-}({CP})^\dag
\rangle\right] \; . 
\end{align}
The above observable is typically simplified applying the parity property of the object $\mathcal{O}_{-}$
\begin{align}\label{2.26}
a^{\mathcal{T}-odd}_{CP,1}=\frac{1}{2}\left[\langle \mathcal{O}_{-}\rangle -\langle({C}) \mathcal{O}_{-}({C})^\dag
\rangle\right]
\end{align}
\begin{align}\label{2.27}
a^{\mathcal{T}-odd}_{CP,2}=\frac{1}{2}\left[\langle \mathcal{O}_{-}\rangle +\langle({C})\mathcal{O}_{-}({C})^\dag
\rangle\right]
\end{align}
Correspondingly, the operator $\mathcal{O}_{-}$ exhibits parity properties even and odd in the definitions $a^{\mathcal{T}-odd}_{CP,1}$ or $a^{\mathcal{T}-odd}_{CP,2}$ respectively. Here,  It is assumed that the expectation value $\langle \mathcal{O}_{-}\rangle$ has been normalized by the total decay width, rendering it dimensionless. Consequently, the associated CP 
$\mathcal{T}$ violation is defined as half of their summation or difference. It is emphasised to note that there is no necessity to define $a^{\mathcal{T}-odd}_{CP,1}$ as $\left[\langle \mathcal{O}_{-}\rangle -\langle({C}) \mathcal{O}_{-}({C})^\dag
\rangle\right]/\left[\langle \mathcal{O}_{-}\rangle +\langle({C}) \mathcal{O}_{-}({C})^\dag
\rangle\right]$, as there is no inherent principle ensuring that this value must be constrained to be less than $1$. As a well-defined CP asymmetry, it is expected to fall within the range of $-1$ to $1$. However, definitions such as \eqref{2.26} generally satisfy the requirement of falling within the range of $-1$ to $1$. In this work, we will commonly refer to these forms of CP violations as $\mathcal{T}$-odd correlation-induced violations.

%% file: 3-Polarization_and_CP_observable.tex
\section{Polarization and asymmetry parameters}\label{sec.3}
In this section, we will provide a comprehensive overview of polarizations, which are aften useful to reflect the underlying dynamics in decays and scatterings. Moreover, some new observables associated with CP violation in phenomenological analyses are closely connected to certain polarization components. The polarization components, such as those along the transverse or longitudinal directions in a specific frame, have distinct properties under Parity and Time Reversal transformations \cite{Leitner:2006nb}. In principle, one can construct more CP-odd observables when the spin of the initial state is aligned or when the final polarization is measured. Therefore, it is reasonable to suggest that CP violation in the baryon sector may be more readily confirmed through polarization consideration compared to the meson sector, given that baryons carry non-zero spin. Here, we will discuss the polarization in baryon productions and decays in detail, especially, their associtaed properties under CP and $\mathcal{T}$ transformation.
\subsection{Polarization components and asymmetry parameters}\label{sec:Polarization component}
Let us begin by examining a simple decay process, such as $\Lambda\to p\pi$, with the initial polarization of $\Lambda$ denoted as $\vec{S}_{i}$. The final spin of the proton, $\vec{S}_{f}$, can be expressed in a specific frame $\vec{e}_{L,N,T}$
\begin{equation}
	\begin{aligned}
      \vec{S}_{f}=P_{L}\vec{e}_{L}+P_{N}\vec{e}_{N}+P_{T}\vec{e}_{T}
	\end{aligned}
\end{equation}
with three unit directions $\vec{e}_{L},\vec{e}_{N},\vec{e}_{T}$ defined by
\begin{equation}
	\begin{aligned}
     \vec{e}_{L}=\hat{p},~~\vec{e}_{N}=\frac{\vec{S}_{i}\times\hat{p}}{\left|\vec{S}_{i}\right|}=\hat{S}_{i}\times\hat{p},~~\vec{e}_{T}=\vec{e}_{N}\times\vec{e}_{L}
	\end{aligned}
\end{equation}
where $\hat{p}$ is proton unit momentum. We will refer to these directions as the longitudinal, normal, and transverse directions, respectively. By applying both Parity and Time reversal transformations to them, the outcomes are trivial and summarized in Table \ref{Tab.1}. Using this example as a basis, we will give a detailed discussion of the initial and final polarizations.
	\begin{table}[ht] 
		\centering
		\caption{\label{tab:test} Vector-polarization under Parity and TR operations} 
		\begin{tabular}{lll} 
			\toprule 
			Vector-polarization   &  ~~~Parity &  ~~~Time reversal   \\ 
			\midrule 
			~~~~~~~~~~~~$ \vec{S}_{i} $ & ~~~~even &   ~~~~~~~~~odd \\ 
			~~~~~~~~~~~~$ \vec{S}_{f} $ & ~~~~even &   ~~~~~~~~~odd \\
			~~~~~~~~~~~~$ \hat{p} $ & ~~~~odd &   ~~~~~~~~~odd \\
			~~~~~~~~~~$ \vec{S}_{i}\times\hat{p} $ & ~~~~odd &   ~~~~~~~~~even \\
			~~~~~~~~$ (\vec{S}_{i}\times\hat{p})\times\hat{p} $ & ~~~~even &   ~~~~~~~~~odd \\
			~~~~~~~~~~~~$ P_{L} $ & ~~~~odd &   ~~~~~~~~~even \\
			~~~~~~~~~~~~$ P_{N} $ & ~~~~odd &   ~~~~~~~~~odd \\
			~~~~~~~~~~~~$ P_{T} $ & ~~~~even &   ~~~~~~~~~even \\
			\bottomrule 
		\end{tabular} \label{Tab.1}
	\end{table}

The direction and magnitude of initial polarization $\vec{S}_{i}$ is always determined through it's production. There are generally two production mechanisms: one involves Parity-violating weak interactions, while the other entails Parity-conserved strong or electromagnetic interactions. For example, in the weak decay of $\Lambda^{+}_{c}\to \Lambda$, a longitudinally polarized $\Lambda$ is produced \cite{Wang:2022tcm}, which is crucial for measuring $\Lambda$ asymmetry parameters and associated CP asymmetries. This production mechanism can be intuitively understood through the picture of Parity violation, where the probability of generating a left-handed final $\Lambda$ is not equal to that of a right-handed one, leading naturally to a non-zero polarization along its direction of motion. The polarization provided by weak decays has been extensively utilized to determine asymmetry parameters in various decaying channels for charm baryons and hyperons, as well as triple product correlations in $B\to VV$ processes in experimental measurements. These applications have facilitated precise investigations of the underlying interactions in these decay processes.

\begin{figure}[h!]
\centering
\includegraphics[width=0.9\textwidth]{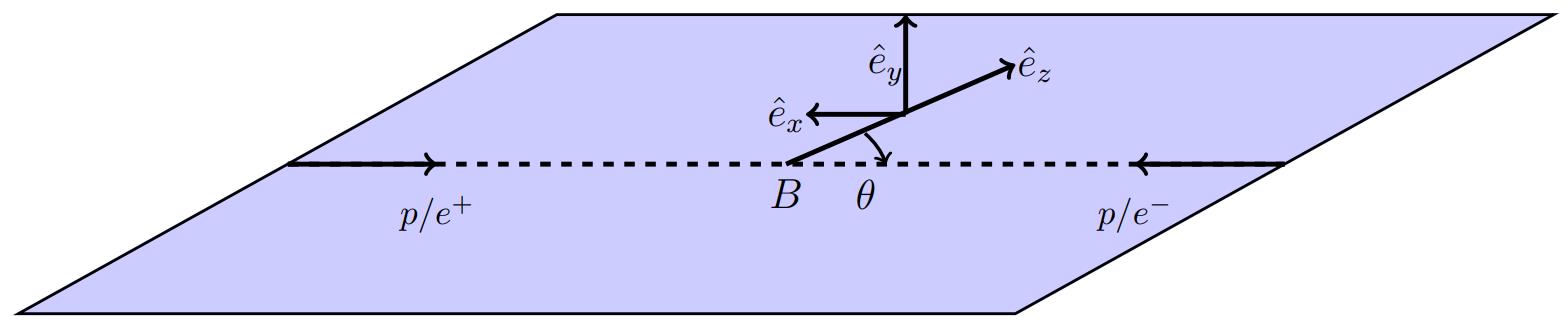}
\caption{In general, a non-zero projection of polarization along the normal vector $\hat{e}_{y}$ is present in $pp$ or $e^+e^-$ collisions. This does not violate parity conservation, as $\vec{\sigma}\cdot\hat{e}_{y}$ remains invariant under parity transformation. However, the other projections, namely $\vec{\sigma}\cdot\hat{p}$ and $\vec{\sigma}\cdot(\hat{e}_{y}\times\hat{p})$, vanish in accordance with the requirement of parity conservation.}\label{fig-T Polarization1}
\end{figure}

More interestingly, there exists another mechanism for generating polarization in strong and electromagnetic scatterings, while parity is conserved for both types interactions. To be precised, a non-zero polarization normal to the reaction plane, as illustrated by $\hat{e}_{y}$ in Fig. \ref{fig-T Polarization1}, can be produced in processes such as the strong scattering $\pi+p\to \Lambda+K$ \cite{Lee:1957he} and electromagnetic production $e^{+}e^{-}\to\Lambda\bar{\Lambda}$. Notably, this normal polarization serves as a $\mathcal{T}$-odd quantity, but its non-zero value does not violate $\mathcal{T}$ symmetry, as previously discussed. Conversely, the transverse $\vec{e}_{x}$ and longitudinal $\vec{e}_{z}$ projections of polarization are prohibited due to the constraints imposed by parity symmetry. This normal polarization can be attributed to helicity flip effects in the reaction and is typically proportional to the imaginary part of interference terms arising from different helicity amplitudes. Extensive experimental investigations have been conducted on this phenomenon. For example, the normal polarization of the hyperon $\Lambda$ produced in $e^{+}e^{-}$ collisions has been accurately measured at BESIII by observing the channel $e^{+}e^{-}\to J/\psi\to \Lambda(\to p\pi^{-})\bar{\Lambda}(\to \bar{p}\pi^{+})$, leading to the precise determination of the asymmetry parameter $\alpha$ in the decay $\Lambda\to p\pi$ with increased precision \cite{BESIII:2018cnd}. Moreover, the production polarization of hyperons at higher energy regimes has been explored by various experimental collaborations such as ALICE, Belle, and STAR. Additionally, the normal polarization of $\Lambda^{0}_{b}$ in $pp$ collisions was measured by LHCb \cite{LHCb:2013hzx,LHCb:2020iux} and CMS \cite{CMS:2018wjk} through angular analyses of $\Lambda^{0}_{b}\to \Lambda J/\Psi\to p\pi^{-}\mu^{+}\mu^{-}$, with the results consistently indicating zero polarization. Theoretical predictions suggest a polarization of single b-quarks at the order of $10\%$ in high-energy $pp$ collisions within the framework of perturbative QCD subprocesses \cite{Dharmaratna:1996xd}, implying a potential non-zero polarization for $\Lambda^{0}_{b}$ after b-quark hadronization. Furthermore, based on the heavy quark effective theory (HQET), a significant normal polarization of $\Lambda_{b}^{0}$ is anticipated \cite{Falk:1993rf}.
Before moving on, it is crucial to emphasize the implications of the polarization of $\Lambda^{0}_{b}$, particularly in the context of b-baryon CP violation phenomenology. The presence of this polarization is essential for probing the decay asymmetry parameter $\alpha$ in the two-body decay of $\Lambda_{b}\to p\pi^{-}/K^{-}$ since a non-trivial differential distribution is emergent with the help of it
\begin{equation}\label{eq:transverseLb}
\begin{aligned}
\frac{d\Gamma}{d\Omega}\propto 1+P\alpha \cos\theta
\end{aligned}
\end{equation}
with $\theta$ the angle between the direction of the $\Lambda^{0}_{b}$ polarization and final proton momentum. CP violated parameter $a^{\alpha}_{CP}$ induced by $\alpha$ thus can be extracted from the non-trivial distribution of the decay products. This observable is anticipated to provide a complementary perspective if the CP asymmetries from the two waves cancel out in the total rate of CP violation. We will discuss it in following sections.

\begin{figure}[h!]
\centering
\includegraphics[width=0.30\textwidth]{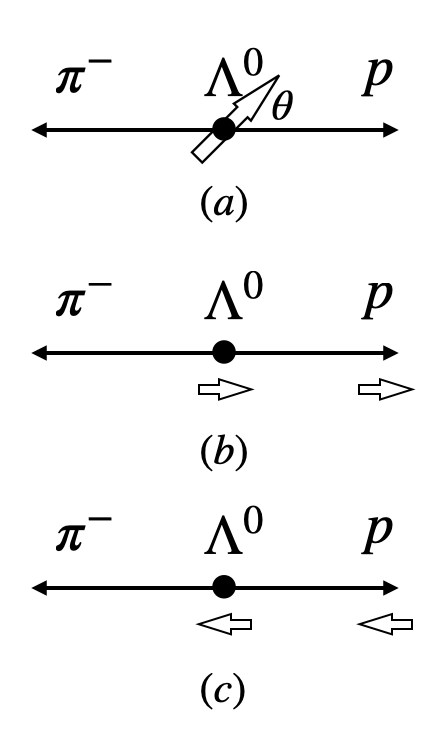}
\caption{(a) A general case of a polarized hyperon decay with $\theta$ as the angle between the polarization of the hyperon and the momentum of the proton. (b) A specific case of $\theta=0$, corresponding to the helicity amplitude of  $\mathcal{H}_{\lambda_\Lambda=+{1\over2},\lambda_p=+{1\over2}}$. (c) A specific case of $\theta=\pi$, corresponding to the helicity amplitude of  $\mathcal{H}_{\lambda_\Lambda=-{1\over2},\lambda_p=-{1\over2}}$.  }\label{fig-LambdaPolarization}
\end{figure}

Following the discussion on the production of polarization in the decaying initial particles, we now introduce the polarization of the final proton in the example of $\Lambda\to p\pi^{-}$, which is initially proposed by Lee and Yang to explore parity symmetry in reactions without neutrinos after the confirmation of parity violation in $\beta$ decay \cite{Wu:1957my}. These polarization components, also called asymmetry parameters, have significant implications for understanding the underlying dynamics of weak decays in charm and beauty baryons up to now. With the improvement of data samples, an increasing number of experimental measurements are becoming feasible. For simplicity, considering a completely polarized $\Lambda$, the differential distribution simplifies to $1+\alpha \cos\theta$, as illustrated in Fig. \ref{fig-LambdaPolarization}(a). Since $\cos\theta$ is odd under parity transformation, the parameter $\alpha$ quantifies the parity violation in the $\Lambda\to p\pi^-$ decay process, as seen by
\begin{align}\label{4.1.2}
    {\alpha}={2\left[\int_0^{1}\Gamma(\cos\theta)d\cos\theta-\int_{-1}^0\Gamma(\cos\theta)d\cos\theta\right]\over \int_{-1}^{1}\Gamma(\cos\theta)d\cos\theta}.
\end{align}
From Fig.\ref{fig-LambdaPolarization} (b) and (c), the cases of $\theta=0$ and $\theta=\pi$ correspond to the helicity amplitudes of 
\begin{equation}
	\begin{aligned}
    \left|\mathcal{H}_{\lambda_\Lambda=+{1\over2},\lambda_p=+{1\over2}}\right|^2&\propto 1+\alpha, \\
    \left|\mathcal{H}_{\lambda_\Lambda=-{1\over2},\lambda_p=-{1\over2}}\right|^2&\propto 1-\alpha, 
	\end{aligned}
\end{equation}
respectively. Due to the zero spin of pion and the conservation of angular momentum, the helicity symbols of hyperon and proton must be same. For simplicity, we will use $\mathcal{H}_{+{1\over2}}$ and $\mathcal{H}_{-{1\over2}}$ to replace the above helicity amplitudes. Then we get 
\begin{equation}\label{eq:alpha-helicity}
	\begin{aligned}
     \alpha=\frac{\left|\mathcal{H}_{+\frac{1}{2}}\right|^2-\left|\mathcal{H}_{-\frac{1}{2}}\right|^2}{\left|\mathcal{H}_{+\frac{1}{2}}\right|^2+\left|\mathcal{H}_{-\frac{1}{2}}\right|^2}.
	\end{aligned}
\end{equation}
Under the parity transformation, the helicity amplitudes of $\mathcal{H}_{+{1\over2}}$ and $\mathcal{H}_{-{1\over2}}$ interchange with each other. 
It is clear again that $\alpha$ is a parity-odd (P-odd) quantity, which can be used to reflect the parity violation. From Eq.(\ref{eq:alpha-helicity}), 
anther physical meaning of $\alpha$ can be seen that it counts how large the proton is polarized longitudinally in the case that the hyperon is unpolarized.

A more comprehensive description is implemented by the decay amplitude directly
\begin{align}
    \mathcal{A}(\Lambda^0\to p\pi^-)=\bar u_p(S+P\gamma_5)u_\Lambda.
\end{align}
Transforming it to Pauli spinor form, one arrives at
\begin{equation}
	\begin{aligned}		\mathcal{A}(\Lambda^0\to p\pi^-)=i\chi^{\dagger}_{p}(S+P\vec{\sigma}\cdot\hat{p})\chi_{\Lambda}
	\end{aligned}
\end{equation}
with $\hat{p}$ unit momentum of the proton. The helicity amplitude is related to partial wave as
\begin{equation}\label{HP}
	\begin{aligned}
    \mathcal{H}_{+{1\over2}}&=\frac{1}{\sqrt{2}} (S+P),\\
    \mathcal{H}_{-{1\over2}}&=\frac{1}{\sqrt{2}} (S-P).\\
	\end{aligned}
\end{equation}
Then we have 
\begin{equation}
	\begin{aligned}
		\alpha&=\frac{2\mathcal{R}e(S^{*}P)}{|S|^{2}+|P|^{2}}.
	\end{aligned}
\end{equation}
Under the parity transformation, $\hat P\alpha \hat P^\dag=-2\mathcal{R}e(S^{*}P)/|S|^{2}+|P|^{2}$, which can be obtained directly from Eq.(\ref{eq:alpha-helicity}). The other two asymmetry parameters $\beta,\gamma$ reflecting the polarization along $\vec{e}_{N},\vec{e}_{T}$ of final proton are given by squaring the amplitude without spin indices summation
\begin{equation}\label{eq: trace and square amplitude}
	\begin{aligned}
		\mathcal{A}\mathcal{A}^{\dagger}&=Tr\left(\left\{\chi^{\dagger}_{p}(S+P\vec{\sigma}\cdot\hat{p})\chi_{\Lambda}\right\}\left\{\chi_{\Lambda}^{\dagger}(S^{*}+P^{*}\vec{\sigma}\cdot\hat{p})\chi_{p}\right\}\right)\\
		&=\frac{1}{4}Tr\left((1+\vec{\xi}\cdot\vec{\sigma})(S+P\vec{\sigma}\cdot\hat{p})(1+\vec{\eta}\cdot\vec{\sigma})(S^{*}+P^{*}\vec{\sigma}\cdot\hat{p})\right)
	\end{aligned}
\end{equation}
where $\vec{\xi}$ and $\vec{\eta}$ are polarization vectors of proton and hyperon respectively. To get above expression, the following equations are used
\begin{equation}
	\begin{aligned}
	 \chi_{\Lambda}\chi_{\Lambda}^{\dagger}&=\frac{1}{2}(1+\vec{\eta}\cdot\vec{\sigma})\\
	 \chi_{B}\chi_{B}^{\dagger}&=\frac{1}{2}(1+\vec{\xi}\cdot\vec{\sigma})\\
	\end{aligned}
\end{equation}
Consequently, the distribution function of final nucleon momentum and spin, that is proportional to $\mathcal{A}\mathcal{A}^{\dagger}$, is derived explicitly employing the variables $\vec{\xi},\vec{\eta},\hat{p}$  
\begin{equation}
	\begin{aligned}
	 d\Gamma(\vec{\xi},\vec{\eta},\hat{p})\propto\left\{|S|^{2}+|P|^2\right\}&\left\{1+\alpha(\vec{\eta}\cdot\hat{p}+\vec{\xi}\cdot\hat{p})+\beta \vec{\xi}\cdot(\vec{\eta}\times\hat{p})\right.\\&\left.+\gamma \vec{\eta}\cdot\vec{\xi}+(1-\gamma)(\vec{\eta}\cdot\hat{p})(\vec{\xi}\cdot\hat{p})\right\}
	\end{aligned}\label{proton polarization}
\end{equation}
with the parameters $\beta,\gamma$ defined as
\begin{equation}
	\begin{aligned}
		\beta=\frac{2\mathcal{I}m(S^{*}P)}{|S|^{2}+|P|^{2}}~,~~
		\gamma=\frac{|S|^{2}-|P|^{2}}{|S|^{2}+|P|^{2}}
	\end{aligned}
\end{equation}
or equivalently, by the helicity amplitude as in \ref{HP}
\begin{equation}\label{eq:beta and gamma in helicity}
	\begin{aligned}
	\beta=\frac{2\mathcal{I}m(\mathcal{H}_{+1/2}\mathcal{H}^{*}_{-1/2})}{\left|\mathcal{H}_{+1/2}\right|^{2}+\left|\mathcal{H}_{-1/2}\right|^{2}},~~
	\gamma=\frac{2\mathcal{R}e(\mathcal{H}_{+1/2}\mathcal{H}^{*}_{-1/2})}{\left|\mathcal{H}_{+1/2}\right|^{2}+\left|\mathcal{H}_{-1/2}\right|^{2}}
	\end{aligned}
\end{equation}
As we know, the decay $B\to B^{\prime}P$, with $B,B^{\prime}$ and $P$ are spin one-half baryons and pseudo-scalar meson respectively, is completely described employing $S,P$ wave amplitudes. Alternatively, the magnitudes and relative phase of $S,P$ are all embedded in another set of three real parameters $\alpha,\beta,\gamma$ as utilized here. As a reminder, we should note that the normal and transverse polarizations indeed vanish in the case of no initial polarization. This is because the directions $N$ and $T$ are dependent on the presence of the initial polarization $\vec{S}_{i}$. Meanwhile, it is noted that $\alpha,\beta,\gamma$ are nothing but the normalized polarization projection of final proton. Hence
\begin{equation}
	\begin{aligned}
	 \alpha^2+\beta^2+\gamma^2=1
	\end{aligned}
\end{equation}
if the final nucleon polarization is not measured
\begin{equation}
	\begin{aligned}
	 d\Gamma(\vec{\xi},\vec{\eta},\vec{p})\propto\left\{|S|^{2}+|P|^2\right\}&\left(1+\alpha\vec{\eta}\cdot\vec{p}\right)d\Omega
	\end{aligned}
\end{equation}
which is exactly consistent with earlier discussions. Therefore, in a cascade weak decay scenario such as $\Omega^{0}_{c}\to \Omega\pi$ with $\Omega\to \Lambda\pi\to p\pi\pi$, the $\Omega$ baryon will carry polarization with a magnitude of $\alpha(\Omega^{0}_{c}\to \Omega\pi)$, even if the initial $\Omega^{0}_{c}$ baryon is not polarized. Consequently, the polarization of the $\Lambda$ baryon will follow the pattern described in Eq.(\ref{proton polarization}), allowing the asymmetry parameters $\beta$ and $\gamma$ of the reaction $\Omega\to \Lambda\pi$ to be extracted through the angular distribution of the final proton. This approach has proven to be highly effective in measuring these asymmetry parameters in experiments such as BESIII \cite{BESIII:2019odb,Chen:2019hqi} and Belle II \cite{Belle:2021crz,Belle:2021zsy}. These discussions are equally applicable to the case of $\Lambda^{0}_{b}$ decays, although it remains challenging to obtain a sufficient number of polarized b-baryons in experiments to reach precise measurements. Moreover, analogous to the decay $\Lambda\to p\pi$, decaying asymmetries in the case of $B\to B^{\prime}V$, where $V$ denotes a vector meson, can also be formulated using helicity amplitudes, as discussed in the next part.
Finally, we conclude that although the measurements of the Lee-Yang parameters and  corresponding CP asymmetries require polarized $b$-baryons which is difficult to be satisfied in the current experiments, the polarized $b$-baryons could be produced abundantly in  the electron-ion colliders proved in the US and proposed in China could perform collisions with polarized electron and polarized proton \cite{Accardi:2012qut} in the future. Hence, the CPV induced by the Lee-Yang parameters could be measured at EIC and EicC then.

\subsection{Helicity description}
In this section, we will focus on the question how to express the polarization components in the helicity scheme explicitly. This connection will play a vital role as a bridge to relate three parts of total paper as a whole object that are polarization, CPV observable and angular distribution respectively, since all of them are easily resolved in the helicity representation. A pedagogical appendix is provided in the App.\ref{App.} as a detailed basis for the following discussion.

It is well known that polarization is defined as the expectation value of spin, of course which can be formulated in a specific representation. Comparing with conventional partial wave method, the helicity approach to deal with polarization firstly developed by Jacob and Wick\cite{Jacob:1959at}, is convenient since complex couplings and decomposition of orbital and spin angular momentum are avoided naturally. Nowadays, it has been generally accepted in the study of decays and scatterings. Under the helicity representation,  a general state with spin is given as a superposition
\begin{equation}\label{eq: superposition}
   	\begin{aligned}
   	\ket{\psi}=\sum_{j}a_{j}\ket{j}
   	\end{aligned}
\end{equation}
with the coefficient $a_{j}$ associated to helicity state $\ket{j}$. Thus, the projection of polarization along direction $i$ is
\begin{equation}\label{eq:Expectation}
   	\begin{aligned}
   	P_{i}=\bra{\psi}\hat{S}_{i}\ket{\psi}
   	\end{aligned}
\end{equation}
Substituting (\ref{eq: superposition}) into (\ref{eq:Expectation}), we arrive at
\begin{equation}\label{eq: formula of polarization}
   	\begin{aligned}
   	P_{i}=\sum_{m,n}a^{*}_{m}a_{n}\bra{m}\hat{S}_{i}\ket{n}
   	\end{aligned}
\end{equation}
It becomes straightforward when we express $\hat{S}_{i}$ in the helicity representation. Alternatively, the expression above (\ref{eq: formula of polarization}) can be equivalently formulated using a density matrix $\rho_{n,m}=\bra{n}\ket{\psi}\bra{\psi}\ket{m}$
\begin{equation}\label{eq:trace}
	\begin{aligned}
	P_{i}&=\sum_{m,n}\bra{\psi}\ket{m}\bra{m}\hat{S}_{i}\ket{n}\bra{n}\ket{\psi}\\
 &=\sum_{m,n}\bra{m}\hat{S}_{i}\ket{n}\bra{n}\ket{\psi}\bra{\psi}\ket{m}\\
	&=\sum_{m,n}S_{m,n}\rho_{n,m}=Tr(\rho \hat{S}_{i})
	\end{aligned}
\end{equation}
To illustrate this point, let's consider the polarization of the proton in the decays $\Lambda\to p\pi$ and $\Lambda^{0}_{b}\to p\rho^{-}$ as examples. In the context of a specific decay process, the coefficients $a_{j}$ and $\rho_{n,m}$ mentioned earlier are all connected to transition amplitudes. For instance, when the initial particle is unpolarized, let's take the decay $\Lambda\to p\pi$ firstly. In this case, the proton state can be decomposed into two components in the helicity representation
\begin{equation}
	\begin{aligned}
	  \ket{\psi}=\mathcal{N}\left\{h_{+}\ket{+\frac{1}{2}}+h_{-}\ket{-\frac{1}{2}}\right\}
	\end{aligned}
\end{equation}
where the superposition coefficients $h_{+}$ and $h_{-}$ represent the decaying helicity amplitudes, the subscripts $+$ and $-$ are the helicity signs of the proton. The normalization factor $N$ is defined as 
\begin{equation}
	\begin{aligned}
	 \mathcal{N}=\frac{1}{\left|h_{+}\right|^{2}+\left|h_{-}\right|^{2}}
	\end{aligned}
\end{equation}
The direction of the proton momenta is set to align with the $z$-axis, leading to the natural polarization projected along the $z$-axis
\begin{equation}
	\begin{aligned}
	  P_{L}=\bra{\psi}\sigma_{z}\ket{\psi}=\mathcal{N}\left\{\left|h_{+}\right|^{2}-\left|h_{-}\right|^{2}\right\}
	\end{aligned}
\end{equation}
and along the other two directions
\begin{equation}
	\begin{aligned}
	  P_{N}=\mathcal{N}(h^{*}_{+},h^{*}_{-})\begin{pmatrix} 0 & -i\\ i & 0\end{pmatrix} \begin{pmatrix} h_{+}\\ h_{-} \end{pmatrix}=2\mathcal{N}\mathcal{I}m(h^{*}_{+}h_{-})
	\end{aligned}
\end{equation}
\begin{equation}
	\begin{aligned}
	  P_{T}=\mathcal{N}(h^{*}_{+},h^{*}_{-})\begin{pmatrix} 0 & 1\\ 1 & 0\end{pmatrix} \begin{pmatrix} h_{+}\\ h_{-} \end{pmatrix}=2\mathcal{N}\mathcal{R}e(h^{*}_{+}h_{-})
	\end{aligned}
\end{equation}
The three directions are defined to be self-consistent with the discussions in Sec.\ref{sec:Polarization component}. Let us now confirm the above conclusion using the density matrix approach, while also preparing for the subsequent discussion on $\Lambda^{0}_{b}\to pV$. The spin density matrix of the proton in the $\Lambda\to p\pi$ decay is given by
\begin{equation}
	\begin{aligned}
    \rho_{n,m}=\bra{n}\ket{\psi}\bra{\psi}\ket{m}=\mathcal{N}\begin{pmatrix} \left|h_{+1/2}\right|^{2}& h_{+1/2}h^{*}_{-1/2}\\ h^{*}_{+1/2}h_{-1/2} & \left|h_{-1/2}\right|^{2}  \end{pmatrix}
	\end{aligned}
\end{equation}
Therefore, the three polarization components can be obtained by taking the trace as per Eq.(\ref{eq:trace}).
\begin{equation}
	\begin{aligned}
	  P_{L}&=Tr(\hat{\sigma}_{z}\rho)=\mathcal{N}\left\{\left|h_{+}\right|^{2}-\left|h_{-}\right|^{2}\right\}\\
	  P_{N}&=Tr(\hat{\sigma}_{y}\rho)=2\mathcal{N}\mathcal{I}m(h^{*}_{+}h_{-})\\	  P_{T}&=Tr(\hat{\sigma}_{x}\rho)=2\mathcal{N}\mathcal{R}e(h^{*}_{+}h_{-})\\
	\end{aligned}
\end{equation}
If the initial polarization is considered, the final spin density matrix turns to
\begin{equation}
	\begin{aligned}
	 \rho_{n,m}&=\bra{n}\ket{\psi}\bra{\psi}\ket{m}=\bra{n}\mathcal{H}_{eff}\ket{\psi^{I}}\bra{\psi^{I}}\mathcal{H}_{eff}\ket{m}\\
	 &=\sum_{i,j}\bra{n}\mathcal{H}_{eff}\ket{i}\bra{i}\ket{\psi^{I}}\bra{\psi^{I}}\ket{j}\bra{j}\mathcal{H}_{eff}\ket{m}\\
	 &=\sum_{i,j}\rho^{I}_{i,j}h_{i,n}h^{*}_{j,m}\delta_{i,n}\delta_{j,m}\\
	\end{aligned}
\end{equation}
where $h_{i,n}$ represent decaying helicity amplitudes, $\ket{\psi^{I}}$ is initial helicity state, and $\rho^{I}_{i,j}$ is initial density matrix given by
\begin{equation}
	\begin{aligned}
		\rho^{I}=\frac{1}{2}\begin{pmatrix} 1+P\cos\theta & P\sin\theta \\ P\sin\theta & 1-P\cos\theta\end{pmatrix}
	\end{aligned}
\end{equation}
where $P$ is the magnitude of the initial polarization, $\theta$ is the polar angle between $\vec{S}_{i}$ and the moving direction of the final proton. Thus, the final polarizations for the case of a polarized $\Lambda$, denoted as $P^{\prime}_{L}$, $P^{\prime}_{N}$, and $P^{\prime}_{T}$, are given
\begin{equation}
	\begin{aligned}
	  P^{\prime}_{L}&=Tr(\hat{\sigma}_{z}\rho)=\mathcal{N}\left\{P_{L}+P\cos\theta\right\}\\
	  P^{\prime}_{N}&=Tr(\hat{\sigma}_{y}\rho)=2\mathcal{N}\mathcal{I}m(h^{*}_{+}h_{-})P\sin\theta\\	  P^{\prime}_{T}&=Tr(\hat{\sigma}_{x}\rho)=2\mathcal{N}\mathcal{R}e(h^{*}_{+}h_{-})P\sin\theta\\
	\end{aligned}
\end{equation}
Besides, the normalization factor $\mathcal{N}$ now becomes the total angular distribution in this case since it is obtained as the trace of $\rho_{n,m}$.

For the decay of $\Lambda^{0}_{b}\to pV^{-}$, some difference arise due to the emergent of vector meson which  require that a two-particle state must be taken into account when we investigate the spin of system. Unlike the $p\pi$, only proton is considered because $\pi$ is spinless. However, there are only four independent states due to the conservation of helicity. The corresponding spin density matrix is obtained by direct product of that of proton and vector meson. We can trace out the spin of proton or vector meson in order to extract one of them separately. For example, the spin density matrix of vector meson is 
\begin{equation}
	\begin{aligned}
		\rho^{V}_{m,m^{\prime}}=\sum_{\lambda,\alpha,\beta}\rho^{I}_{\alpha,\beta}h_{\lambda,m}h^{*}_{\lambda,m^{\prime}}\delta_{\lambda-m,\alpha}\delta_{\lambda-m^{\prime},\beta}
	\end{aligned}
\end{equation}
where subscript $\lambda$ is the helicity symbol of proton. The spin operators of vector meson are $SO(3)$ generators, in the helicity representation which are expressed as \cite{Georgi:2000vve}
\begin{equation}
	\begin{aligned}
	S_{z}=\begin{pmatrix} 1 & 0 & 0 \\ 0 & 0 & 0 \\0 & 0 & -1 \end{pmatrix},~S_{x}=\frac{1}{\sqrt{2}}\begin{pmatrix} 0 & 1 & 0 \\ 1 & 0 &1 \\0 & 1 & 0 \end{pmatrix},~S_{y}=\frac{i}{\sqrt{2}}\begin{pmatrix} 0 & -1 & 0 \\ 1 & 0 & -1 \\0 & 1 & 0 \end{pmatrix}
	\end{aligned}
\end{equation}
Therefore, the normalized polarizations $P^{V}_{L},P^{V}_{N},P^{V}_{T}$ of vector meson are given\cite{Leitner:2006nb}
\begin{equation}
	\begin{aligned}
	 P^{V}_{L}\frac{d\Gamma}{d\Omega}&=Tr(S_{z}\rho^{V})= \gamma_{-}\left|h_{+1/2,+1}\right|^{2}-\gamma_{+}\left|h_{-1/2,-1}\right|^{2}\\
	 P^{V}_{T}\frac{d\Gamma}{d\Omega}&=Tr(S_{x}\rho^{V})=\frac{1}{\sqrt{2}}\left\{\mathcal{R}e(h_{+1/2,+1}h^{*}_{+1/2,0})+\mathcal{R}e(h^{*}_{-1/2,-1}h_{-1/2,0})\right\}P\sin\theta\\
	 P^{V}_{N}\frac{d\Gamma}{d\Omega}&=Tr(S_{y}\rho^{V})=-\frac{1}{\sqrt{2}}\left\{\mathcal{I}m(h_{+1/2,+1}h^{*}_{+1/2,0})+\mathcal{I}m(h^{*}_{-1/2,-1}h_{-1/2,0})\right\}P\sin\theta\\
	\end{aligned}
\end{equation}
with $\gamma_{\pm}$ 
\begin{equation}
	\begin{aligned}
	 \gamma_{+}=\frac{1}{2}(1+P\cos\theta),~\gamma_{-}=\frac{1}{2}(1-P\cos\theta)
	\end{aligned}
\end{equation}
Similarly, the polarization of the proton can be derived. In the first step, the density matrix of the proton is
	\begin{equation}
		\begin{aligned}
		\rho^{p}_{m,m^{\prime}}=\sum_{\lambda,\alpha,\beta}\rho^{I}_{\alpha,\beta}\mathcal{H}_{m,\lambda}\mathcal{H}_{m^{\prime},\lambda}\delta_{m-\lambda,\alpha}\delta_{m^{\prime}-\lambda,\beta}
		\end{aligned}
	\end{equation}
Consequently, the polarizations along $\vec{e}_{L},\vec{e}_{T},\vec{e}_{N}$ $P_{L}$ are 
\begin{equation}
	\begin{aligned}
	 P_{L}\frac{d\Gamma}{d\Omega}&=Tr(\sigma_{z}\rho^{p})= \gamma_{-}\left\{\left|h_{+1/2,+1}\right|^{2}-\left|h_{-1/2,0}\right|^{2}\right\}+\gamma_{+}\left\{\left|h_{+1/2,0}\right|^{2}-\left|h_{-1/2,-1}\right|^{2}\right\}\\
	 	\end{aligned}
\end{equation}
\begin{equation}
	\begin{aligned}
	 P_{T}\frac{d\Gamma}{d\Omega}&=Tr(\sigma_{x}\rho^{p})=\mathcal{R}e(h_{+1/2,0}h^{*}_{-1/2,0})P\sin\theta\\	 P_{N}\frac{d\Gamma}{d\Omega}&=Tr(\sigma_{x}\rho^{p})=-\mathcal{I}m(h_{+1/2,0}h^{*}_{-1/2,0})P\sin\theta\\
	\end{aligned}
\end{equation}
The polarization is straightforward if the initial polarization $P$ is zero
\begin{equation}
	\begin{aligned}
	 P^{V}_{L}\frac{d\Gamma}{d\Omega}&=Tr(S_{z}\rho^{V})= \left|h_{+1/2,+1}\right|^{2}-\left|h_{-1/2,-1}\right|^{2}\\	 	 P_{L}\frac{d\Gamma}{d\Omega}&=Tr(\sigma_{z}\rho^{p})= \left|h_{+1/2,+1}\right|^{2}+\left|h_{+1/2,0}\right|^{2}-\left|h_{-1/2,0}\right|^{2}-\left|h_{-1/2,-1}\right|^{2}\\
	\end{aligned}
\end{equation}
From a phenomenological perspective, the phase space distribution of final product momenta associated with longitudinal and the other two polarizations are significantly different due to the fundamental property of rotational symmetry. Specifically, the longitudinal direction alone is not sufficient to completely break the $SO(3)$ symmetry. Therefore, the final distribution function is always dependent on only one polar angle $\theta$ when only longitudinal polarization is considered. However, if an additional physical direction is utilized to define transverse and normal directions, it becomes evident that more angular variables are necessary to characterize the distribution functions. By employing the Wigner–Eckart theorem, the decay amplitudes of polarized particles can be separated into dynamical and geometrical parts \cite{Jacob:1959at}.
\begin{equation}
	\begin{aligned}
\bra{f,\lambda_{f}}\mathcal{H}_{eff}\ket{i,\lambda_{i}}\propto \mathcal{A}_{\lambda_{i},\lambda_{f}}D^{*j}_{\lambda_{i},\lambda_{f}}(\phi,\theta,0)
	\end{aligned}
\end{equation}
where $\mathcal{A}_{\lambda_{i},\lambda_{f}}$, representing the magnitude of the transition amplitude, is angle-independent due to rotational symmetry. The spatial dependencies are all factorized into rotation functions $D^{*j}_{\lambda_{i},\lambda_{f}}(\phi,\theta,0)$. In this work, we will take the Jackson convention to define the reference frame and angle variables. Consequently, the angular distribution is given
\begin{equation}
	\begin{aligned}
	\frac{d\Gamma}{d\Omega}&\propto \sum_{\lambda_{i},\lambda_{f}}\left|\bra{F,\lambda_{f}}\mathcal{H}_{eff}\ket{P,\lambda_{i}}\right|^{2}\\
	&=\sum_{\lambda_{i},\lambda^{\prime}_{i},\lambda_{f}}\rho_{\lambda_{i},\lambda^{\prime}_{i}}\mathcal{A}_{\lambda_{i},\lambda_{f}}\mathcal{A}^{*}_{\lambda^{\prime}_{i},\lambda_{f}}D^{*j}_{\lambda_{i},\lambda_{f}}(\phi,\theta,0)D^{j}_{\lambda^{\prime}_{i},\lambda_{f}}(\phi,\theta,0)
	\end{aligned}
\end{equation}
Apparently, the azimuth $\phi$ dependence is closely related to the off-diagonal elements of density matrix, or equivalently, the transverse and normal polarization components. 

\subsection{Polarization production through three body decays}\label{sec: three body}
\begin{figure}[h!]
\centering
\includegraphics[width=0.5\textwidth]{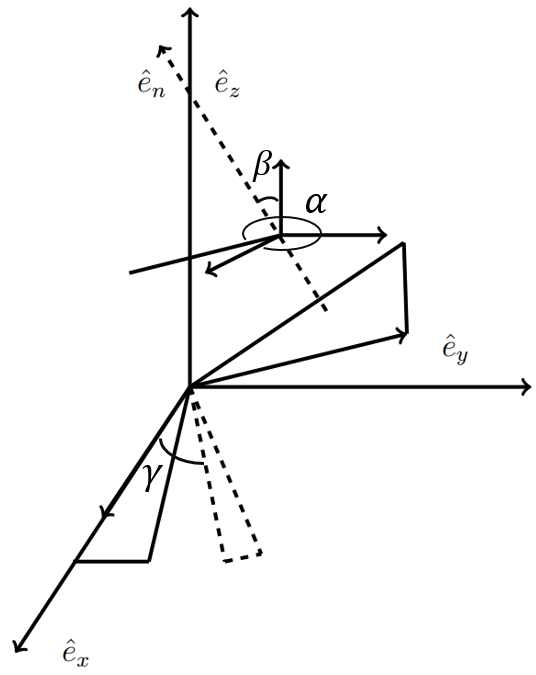}
\includegraphics[width=1.0\textwidth]{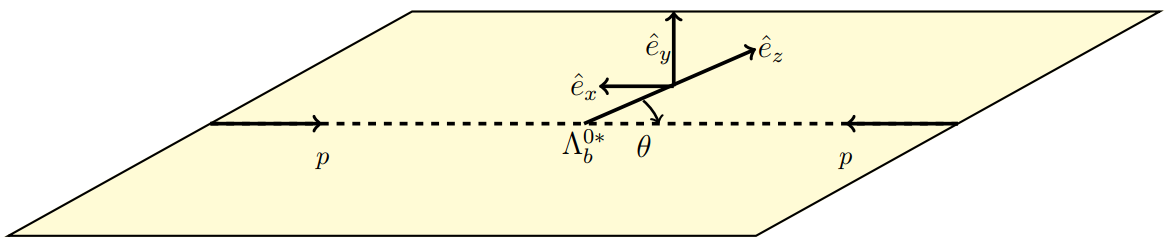}
\caption{The kinematics of three body decay can be represented by a closed triangle in the frame of initial particle due to the conservation of momenta. $\hat{e}_{x},\hat{e}_{y},\hat{e}_{z}$ is a specific coordinate system where initial particle is located in, and it can be fixed through the production of initial state. $\gamma$ denote the rotation about the normal vector of decaying plane, while $\beta,\gamma$ are polar and azimuth angles of normal vector itself which is labeled as $\hat{e}_{n}$. Specifically, the coordinate system $\hat{e}_{xyz}$, of the production and decay of $\Lambda_{b}^{0*}$, is showed in the second diagram. A non-vanishing normal polarization parallel to $\hat{e}_{n}$ is obtained for $\Lambda^{0}_{b}$ in the decay $\Lambda^{0*}_{b}\to \Lambda^{0}_{b}\pi\pi$ although $\Lambda_{b}^{*0}$ is not required to be polarized. }
\label{fig-T Polarization3body}
\end{figure}

In this section, we will discuss strong three-body decays, which are particularly interesting for providing normal polarization. It is well-known that the description of three-body decays requires five variables: two invariant masses $\omega_{1}$ and $\omega_{2}$, and three Euler angles $\alpha$, $\beta$, and $\gamma$, respectively. The Dalitz plot is defined as a distribution with $\omega_{1}$ and $\omega_{2}$, while the angular dependence can be separately depicted based solely on rotational symmetry. In comparison with two-body decays, an additional rotation along the normal vector of the decaying plane is required in three-body decays. The construction of the three-body state and general helicity formulas has been developed in \cite{Berman:1965gi}, with specific derivations provided in Appendix \ref{App.}. Conventionally, the rotation about the normal vector of the decaying plane is described using the Euler angle $\gamma$, where the normal vector is determined by azimuth and polar angles $\alpha$ and $\beta$, as depicted in Fig. \ref{fig-T Polarization3body}. Based on the aforementioned convention, the angular distribution normal to the decaying plane is generally expressed as
\begin{equation}\label{eq:decay width1}
	\begin{aligned}
		\left(\frac{d\Gamma}{d\Omega}\right)=\sum_{M}\sum_{m,m^{\prime}} \rho_{m,m^{\prime}}D^{j*}_{m,M}(\alpha,\beta,\gamma)D^{j}_{m^{\prime},M}(\alpha,\beta,\gamma)\left|\mathcal{R}_{M}\right|^{2}
	\end{aligned}
\end{equation}
with the subscript $m,M$ the projection of total angular momentum $j$ along $z$ axis and the normal direction of decaying plane, respectively. $ \rho_{m,m^{\prime}}$ is density matrix of initial state, and $\left|\mathcal{R}_{M}\right|^{2}$ is the dynamical part of decays, which includes helicity amplitudes and the associated Dalitz distribution integration
\begin{equation}
	\begin{aligned}
	 \left|\mathcal{R}_{M}\right|^{2}=2\pi\sum_{\lambda_{1},\lambda_{2},\lambda_{2}}\int d\omega_{1}d\omega_{2}\left|F_{M}(\omega_{1},\lambda_{1};\omega_{2},\lambda_{2};\omega_{3},\lambda_{3})\right|^{2}
	\end{aligned}
\end{equation}
where $\lambda_{1},\lambda_{2},\lambda_{3}$ are helicity symbol of three final particles. Formula (\ref{eq:decay width1}) can be simplified explicitly
\begin{equation}\label{eq: compact formula}
	\begin{aligned}
		\left(\frac{d\Gamma}{d\Omega}\right)&=\sum_{M\geq0}\sum_{m,m^{\prime}}\left\{\left(\mathcal{R}e\rho_{m,m^{\prime}}\cos(m-m^{\prime})\alpha-\mathcal{I}m\rho_{m,m^{\prime}}\sin(m-m^{\prime})\alpha\right)\right.\\&\left.~~~~~~~~\left[R^{+}_{M}Z_{m,m^{\prime}}^{jM+}(\beta)+R^{-}_{M}Z_{m,m^{\prime}}^{jM-}(\beta)\right]\right\}
	\end{aligned}
\end{equation}
in which $R_{M}^{\pm}$ and $Z_{m,m^{\prime}}^{jM\pm}(\beta)$ are abbreviation of
\begin{equation}
	\begin{aligned}
	 R_{M}^{\pm}&=\frac{1}{2}\left[\left|R_{M}\right|^{2}\pm\left|R_{-M}\right|^{2}\right]\\
	 Z_{m,m^{\prime}}^{jM\pm}(\beta)&=d^{j}_{m,M}(\beta)d^{j}_{m^{\prime},M}(\beta)\pm d^{j}_{m,-M}(\beta)d^{j}_{m^{\prime},-M}(\beta)
	\end{aligned}
\end{equation}
It is emphasised that, for the decay such as $\Lambda^{0*}_{b}(5920)/\Lambda^{0*}_{b}(5912)\to \Lambda_{b}\pi^{+}\pi^{-}$, a non-zero polarization normal to decaying plane arises for the final $\Lambda^{0}_{b}$. This polarization can be described within the helicity framework as following below. The production frame of $\Lambda_{b}^{0*}$ can be constructed using the momenta of $\Lambda_{b}^{0*}$ and proton beam, in which $z,y$ axis replace the momentum of $\Lambda_{b}^{*}$ and the normal vector of production plane, respectively.
The angular distributions normal to decaying plane are given using formula (\ref{eq: compact formula}) for two specific decays 
\begin{itemize}
    \item (a) for $\Lambda^{0*}_{b}(5912)\to \Lambda_{b}\pi\pi$
\begin{equation}
	\begin{aligned}
		\frac{d\Gamma}{d\Omega}=C_{2}\left[R^{+}_{\frac{1}{2}}+R^{-}_{\frac{1}{2}}\cos\beta\right]
	\end{aligned}
\end{equation}
\item (b) for $\Lambda^{0*}_{b}(5920)\to \Lambda_{b}\pi\pi$
\begin{equation}
	\begin{aligned}
		\frac{d\Gamma}{d\Omega}&=\frac{1}{4}C_{1}\left[R^{+}_{\frac{3}{2}}(3\cos^{2}\beta+1)+3R^{+}_{\frac{1}{2}}\sin^{2}\beta\right]+\frac{1}{4}C_{2}\left[3R^{+}_{\frac{3}{2}}\sin^{2}\beta+R^{+}_{\frac{1}{2}}(3\cos^{2}\beta+1)\right]\\
		&+\frac{1}{4}C_{1}^{\prime}\left[R^{-}_{\frac{3}{2}}\cos\beta(3+\cos^{2}\beta)+3R^{-}_{\frac{1}{2}}\sin^{2}\beta\cos\beta\right]\\
		&+\frac{1}{4}C_{2}^{\prime}\left[3R^{-}_{\frac{3}{2}}\sin^{2}\beta\cos\beta-R^{-}_{\frac{1}{2}}\cos\beta(9\cos^{2}\beta-5)\right]
	\end{aligned}
\end{equation}
\end{itemize}
where parameters $C_{1},C_{2},C^{\prime}_{1},C^{\prime}_{2}$ are 
\begin{equation}
	\begin{aligned}
	 C_{1}&=\rho_{\frac{3}{2},\frac{3}{2}}+\rho_{-\frac{3}{2},-\frac{3}{2}},~
	 C_{2}=\rho_{\frac{1}{2},\frac{1}{2}}+\rho_{-\frac{1}{2},-\frac{1}{2}}\\
	 C^{\prime}_{1}&=\rho_{\frac{3}{2},\frac{3}{2}}-\rho_{-\frac{3}{2},-\frac{3}{2}},~
	 C^{\prime}_{2}=\rho_{\frac{1}{2},\frac{1}{2}}-\rho_{-\frac{1}{2},-\frac{1}{2}}\\
	\end{aligned}
\end{equation}
We assume that the $\Lambda_{b}^{0*}$ is unpolarized in its strong production. Hence, only the diagonal elements of $\rho_{m,m^{\prime}}$ contribute to the decay distribution. The strong decay $\Lambda^{0*}_{b}(5912)/\Lambda^{0*}_{b}(5920)\to \Lambda_{b}\pi\pi$ respects parity such that the final state is located at the parity eigenstate
\begin{equation}
	\begin{aligned}
	\frac{1}{\sqrt{2}}\left(\ket{\omega_{1},\lambda=+\frac{1}{2},\omega_{2},0,\omega_{3},0;jmM}+\epsilon(-1)^{M}\ket{\omega_{1},\lambda=-\frac{1}{2},\omega_{2},0,\omega_{3},0;jmM}\right)
	\end{aligned}
\end{equation}
where $\epsilon$ is the intrinsic parity of initial $\Lambda^{*0}_{b}$. It is easily to check that this state satisfies the eigen-equation of $\sigma_{y}$ with eigenvalue $\epsilon(-1)^{M+\frac{1}{2}}$. Correspondingly, the normalized normal polarization $\langle S_{y}\rangle$ of $\Lambda_{b}$ is obtained as
\begin{itemize}
    \item (a) for $\Lambda_{b}(5912)\to \Lambda_{b}\pi\pi$
\begin{equation}\label{polarization 1}
	\begin{aligned}
		\mathcal{P}_{N}\left(\frac{d\Gamma}{d\Omega}\right)&=\sum_{M}\sum_{m}\epsilon(-1)^{M+\frac{1}{2}}d^{j}_{m,M}(\beta)d^{j}_{m,M}(\beta)\left|\mathcal{R}_{M}\right|^{2}\\
		&=\sum_{M}\epsilon(-1)^{M+\frac{1}{2}}\left|\mathcal{R}_{M}\right|^{2}=\epsilon\left(\left|\mathcal{R}_{-\frac{1}{2}}\right|^{2}-\left|\mathcal{R}_{\frac{1}{2}}\right|^{2}\right)=-\epsilon\mathcal{R}_{\frac{1}{2}}^{-}
	\end{aligned}
\end{equation}
\item (b) for $\Lambda_{b}(5920)\to \Lambda_{b}\pi\pi$
\begin{equation}\label{polarization 2}
	\begin{aligned}
		\mathcal{P}_{N}\left(\frac{d\Gamma}{d\Omega}\right)&=\sum_{M=\pm\frac{1}{2},\pm\frac{3}{2}}\epsilon(-1)^{M+\frac{1}{2}}\left|\mathcal{R}_{M}\right|^{2}=\epsilon\left(\left|\mathcal{R}_{-\frac{1}{2}}\right|^{2}-\left|\mathcal{R}_{\frac{1}{2}}\right|^{2}+\left|\mathcal{R}_{\frac{3}{2}}\right|^{2}-\left|\mathcal{R}_{-\frac{3}{2}}\right|^{2}\right)\\
		&=\epsilon\left(\mathcal{R}^{-}_{\frac{3}{2}}-\mathcal{R}^{-}_{\frac{1}{2}}\right)
	\end{aligned}
\end{equation}
\end{itemize}
The transverse polarization mentioned above can also be derived using the spin density matrix, with detailed discussions provided in Appendix \ref{App.}. The numerical value of this polarization depends on the specific strong dynamics, making it extremely difficult to control due to the non-perturbative nature of low-energy QCD. By utilizing this normal polarization, the asymmetry parameter $\alpha$ and the associated CP asymmetry $a^{\alpha}_{CP}$ of the two-body decays $\Lambda_{b}\to p\pi/K$ could be measured through angular analysis, with the unknown normal polarization $P_{N}$ being involved. If we assume that the polarization $P_{N}$ is equal to its charge conjugation counterpart $\bar{P}_{N}$, then its effect can be factorized as an overall factor in the definition of $a^{\alpha}_{CP}$, indicating that it can be used to determine whether CP is violated in $\Lambda_{b}\to p\pi/K$ or not, even in the absence of precise information about $P_{N}$.

At the experimental side, the signal events of $\Lambda_b^{**}(6070)\to\Lambda_b\pi\pi$ with $\Lambda_b^0\to p J/\Psi K^-$ are $550\pm80$ using the LHCb data of 9 fb$^{-1}$ \cite{LHCb:2020lzx}. Considering the branching fractions of $Br(\Lambda_b\to p J/\Psi K^-)Br(J/\Psi\to \mu^+\mu^-)=(3.2^{+0.6}_{-0.5})\times10^{-4}\times6\%\sim2\times10^{-5}$ and $Br(\Lambda_b\to pK^-)=(5.4\pm1.0)\times10^{-6}$, the signal events of $\Lambda_b^{**}(6070)\to\Lambda_b\pi\pi$ with $\Lambda_b^0\to p K^-$ would be at the order of one hundred. Combining more resonances, the signal events could be even larger, such as $N(\Lambda_b^{*}(5912)\to\Lambda_b\pi\pi)=57\pm9$, $N(\Lambda_b^{*}(5920)\to\Lambda_b\pi\pi)=204\pm17$, $N(\Lambda_b^{*}(6146)\to\Lambda_b\pi\pi)=103\pm22$ and $N(\Lambda_b^{*}(6152)\to\Lambda_b\pi\pi)=90\pm21
$, with $\Lambda_b^0\to p J/\Psi K^-$ \cite{LHCb:2020lzx}. Therefore, we conclude that confirming $a^{\alpha}_{CP}(\Lambda_{b}\to p\pi/K)$ in experiments based on this approach is challenging due to the very small associated branching ratios. However, this polarization might be useful for exploring other $\Lambda^{0}_{b}$ decays with larger decay rates in the future.

\subsection{$\Lambda^{0}_{b}$ production polarization}
The precise extraction of $\alpha(\Lambda_{b}\to p\pi^{-}/K^{-})$ requires the definite initial polarization $\vec{P}$. Since there is no definitive experimental evidence to verify the theoretical prediction on $\Lambda^{0}_{b}$ production polarization up to now, here a concise scheme is proposed to explore it in experiments. Specifically, by combining two decaying channels $\Lambda^{0}_{b}\to \Lambda^{+}_{c}\pi^{-}\to \Lambda\pi^{+}\pi^{-}$ and $\Lambda^{0}_{b}\to \Lambda^{+}_{c}\pi^{-}$, both the polarization of $\Lambda_b^0$ and the decaying asymmetry $\alpha(\Lambda^{0}_{b}\to \Lambda^{+}_{c}\pi^{-})$ can be extracted simultaneously. The first step, assuming we have integrated out the initial polarization effect, the angular distribution of the first decaying chain $\Lambda^{0}_{b}\to \Lambda^{+}_{c}\pi^{-}\to \Lambda\pi^{+}\pi^{-}$ is given by
\begin{equation}\label{eq:3.5}
	\begin{aligned}
	 \frac{d\Gamma}{d\Omega}\propto 1+\alpha(\Lambda^{0}_{b}\to \Lambda^{+}_{c}\pi^{-})\times\alpha(\Lambda^{+}_{c}\to \Lambda\pi^{+})\cos\theta \; ,
	\end{aligned}
\end{equation}
where $\theta$ is the polar angle between $\Lambda^{+}_{c}$ momentum in $\Lambda^{0}_{b}$ frame and that of $\Lambda$ in $\Lambda^{+}_{c}$ frame. Fortunately, the precise measurement of $\alpha(\Lambda^{+}_{c}\to \Lambda\pi^{+})$ as $-0.84\pm0.09$ has been reported \cite{P.A. Zylaet al.}. Meanwhile, it is expected that the asymmetry parameter $\alpha(\Lambda^{0}_{b}\to \Lambda^{+}_{c}\pi^{-})$ is close to $-1$ within the framework of HQET, indicating a significant angular distribution with respect to $\cos\theta$. Its suggests that one can determine $\alpha(\Lambda^{0}_{b}\to \Lambda^{+}_{c}\pi^{-})$ through the linear distribution given in equation \eqref{eq:3.5}. In the next step, once the asymmetry parameter $\alpha(\Lambda^{0}_{b}\to \Lambda^{+}_{c}\pi^{-})$ is established, the polarization $P_{b}$ of $\Lambda^{0}_{b}$ can be measured by analyzing the angular distribution of $\Lambda^{0}_{b}\to \Lambda^{+}_{c}\pi^{-}$ with initially polarized $\Lambda^{0}_{b}$
\begin{equation}\label{eq: the measureement polarization of b baryon}
	\begin{aligned}
	 \frac{d\Gamma}{d\Omega}\propto 1+P_{b}\alpha(\Lambda^{0}_{b}\to \Lambda^{+}_{c}\pi^{-})\cos\theta^{\prime}
	\end{aligned}
\end{equation}
with $\theta^{\prime}$ representing the angle between the momentum of $\Lambda^{+}_{c}$ in the $\Lambda^{0}_{b}$ frame and $\vec{P}_{b}$, where $\vec{P}_{b}$ is set to be normal to its production plane. This channel may offer better prospects due to its significant statistics. Furthermore, the angular analysis mentioned above is more accessible compared to that of $\Lambda^{0}_{b}\to \Lambda J/\Psi\to p\pi l^{+}l^{-}$, which involves numerous variables and terms. Hence, we anticipate that this proposal can be explored in future experiments. In principle, the secondary decay $\Lambda^{+}_{c}\to pK^-\pi^+$ provides a more favorable data statistics. However, there is no single well-defined $\alpha$ for this three-body weak decay due to the presence of multiple possible resonances, and partial wave analysis may be helpful.

\begin{figure}[!]
\centering
\includegraphics[width=0.8\textwidth]{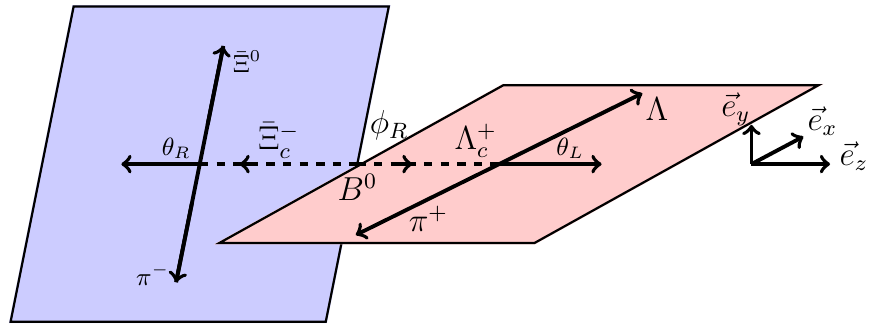}
\caption{The depicted figures of angular distributions of $B^{0}\to \Lambda^{+}_{c}\bar{\Xi}^{-}_{c}$ with $\Lambda^{+}_{c}\to \Lambda\pi^{+},\bar{\Xi}^{-}_{c}\to \bar{\Xi}^{0}\pi^{-}$. The angle $\theta_{L},\theta_{R}$ are defined in the rest frames of $\bar{\Xi}^{-}_{c}$ and $\Lambda^{+}_{c}$, respectively.}
\label{fig}
\end{figure}

Similarly, decay channel $\Lambda^{0}_{b}\to \Lambda^{+}_{c}\pi\to pK_{s}\pi$, as another proposal, is also prospective to determine the polarization $P_{b}$. The asymmetry parameter $\alpha(\Lambda^{+}_{c}\to pK_{s})$, determined to be $0.18\pm0.43\pm0.14$, unfortunately, suffers from significant uncertainty at BESIII due to measurement errors in the polarization of $\Lambda^{+}_{c}$ production in $e^{+}e^{-}$ collisions \cite{BESIII:2019odb}. The issue could potentially be addressed by enhancing the statistics of $\Lambda^{+}_{c}$ at BESIII, or by exploring the $B$ meson decay at Belle and LHCb experiments. It is worth noting that decays such as $B^{0}\to \Lambda^{+}_{c}\bar{p}$ and $B^{0}\to \Lambda^{+}_{c} \bar{\Xi}^{-}_{c}$ are particularly appealing due to their relatively large branching fractions \cite{P.A. Zylaet al.}
\begin{equation}
	\begin{aligned}
	 Br(B^{0}\to \Lambda^{+}_{c}\bar{p})&=(1.54\pm 0.18)\times 10^{-5}\\
	 Br(B^{0}\to \Lambda^{+}_{c}\bar{\Xi}^{-}_{c})&=(1.2\pm 0.8)\times 10^{-3}\\
	\end{aligned}
\end{equation}
In these two decays, the longitudinally polarized $\Lambda^{+}_{c}$ is anticipated to be produced as a result of parity violation. This longitudinal polarization of $\Lambda^{+}_{c}$ can be expressed as the difference between two helicity amplitudes
\begin{equation}\label{Polarization from B decay}
	\begin{aligned}
	 P_{L}=\frac{\left|\mathcal{H}_{+\frac{1}{2}}\right|^{2}-\left|\mathcal{H}_{-\frac{1}{2}}\right|^{2}}{\left|\mathcal{H}_{+\frac{1}{2}}\right|^{2}+\left|\mathcal{H}_{-\frac{1}{2}}\right|^{2}}
	\end{aligned}
\end{equation}
where the subscripts $\pm1/2$ denote the helicity of the final $\Lambda^{+}_{c}$. The presence of this polarization $P_{L}$ offers an opportunity to determine $\alpha(\Lambda^{+}_{c}\to pK_{s})$, even though the precise value of $P_{L}$ is currently unknown. In particular, considering the decay chains $B^{0}\to \Lambda^{+}_{c}\bar{p}$ or $B^{0}\to \Lambda^{+}_{c}\bar{\Xi}^{-}_{c}$ with $\Lambda^{+}_{c}\to pK_{s}$, the angular distribution can be expressed as
\begin{equation}
	\begin{aligned}
	 \frac{d\Gamma}{d\Omega}\propto 1+P_{L}\alpha(\Lambda^{+}_{c}\to pK_{s})\cos\theta
	\end{aligned}
\end{equation}
the definition of $\theta$ is similar to the $\theta_{L}$, as depicted in Figure.\ref{fig}. The combined coefficient $P_{L}\alpha(\Lambda^{+}_{c}\to pK_{s})$ could be determined by measuring the relative asymmetry between $\theta>\pi/2$ and $\theta<\pi/2$, with a normalization to the total decay width $\Gamma$.
\begin{align}
    P_{L}\alpha(\Lambda^{+}_{c}\to pK_{s})={2\left[\int_0^{1}\Gamma(\cos\theta)d\cos\theta-\int_{-1}^0\Gamma(\cos\theta)d\cos\theta\right]\over \int_{-1}^{1}\Gamma(\cos\theta)d\cos\theta}.
\end{align}
Nevertheless, the quantities $P_{L}$ need to be determined independently to accurately extract the true value of $\alpha(\Lambda^{+}_{c}\to pK_{s})$. Once again, we can draw upon the decay $\Lambda^{+}_{c}\to \Lambda\pi$ by using the analogy of \eqref{eq:3.5} and \eqref{eq: the measureement polarization of b baryon} although this may be challenging in experiments due to limited statistics. Besides, one could further explore the decay channel $B^{0}\to \Lambda^{+}_{c}\bar{\Xi}^{-}_{c}$ with $\Lambda_{c}\to \Lambda\pi$ or $pK_{s}$, and $\bar{\Xi}^{-}_{c}\to \bar{\Xi}^{0}\pi^{-}$. The corresponding angular distribution can be constructed based on the Figure.\ref{fig}
	\begin{equation}
		\begin{aligned}     
			\frac{d\Gamma}{d\Omega}&\propto 1	+\alpha_{\Lambda}\alpha_{\bar{\Lambda}}\cos\theta_{R}\cos\theta_{L}+\alpha(\alpha_{\Lambda}\cos\theta_{R}+\alpha_{\bar{\Lambda}}\cos\theta_{L})\\
			&+\beta\alpha_{\Lambda}\alpha_{\bar{\Lambda}}\sin\theta_{R}\cos\theta_{R}\sin\theta_{L}\cos\theta_{L}\sin\phi_{R}\\
			&+\gamma\alpha_{\Lambda}\alpha_{\bar{\Lambda}}\sin\theta_{R}\cos\theta_{R}\sin\theta_{L}\cos\theta_{L}\cos\phi_{R}
		\end{aligned}
	\end{equation}
three asymmetry parameters denoted as $\alpha$, $\beta$, $\gamma$
	\begin{equation}
		\begin{aligned}     
			\alpha&=(\left|\mathcal{H}_{+}\right|^{2}-\left|\mathcal{H}_{-}\right|^{2})/(\left|\mathcal{H}_{+}\right|^{2}+\left|\mathcal{H}_{-}\right|^{2})\\
			\beta&=2\mathcal{I}m(\mathcal{H}_{+}\mathcal{H}^{*}_{-})/(\left|\mathcal{H}_{+}\right|^{2}+\left|\mathcal{H}_{-}\right|^{2})\\
			\gamma&=2\mathcal{R}e(\mathcal{H}_{+}\mathcal{H}^{*}_{-})/(\left|\mathcal{H}_{+}\right|^{2}+\left|\mathcal{H}_{-}\right|^{2})
		\end{aligned}
	\end{equation}
Here, we take the abbreviations $\pm$ for $\pm1/2$. The similar discussion are also completely valid for other decay channels like $B^{0}\to\Lambda\bar{\Lambda}$ with $\Lambda\to p\pi^{-}$ and $\bar{\Lambda}\to \bar{p}\pi^{+}$.

\subsection{CP asymmetries }
One can establish two CP violation observables by contrasting $\alpha,\beta$ with $\bar{\alpha},\bar{\beta}$, where the quantities with a bar denote the asymmetry parameters of charge-conjugated decays. Specifically, a convenient definition of CP asymmetries induced by $\alpha,\beta,\gamma$ is given by \cite{Donoghue:1986hh}
\begin{equation}\label{abc def}
	\begin{aligned}
a^{\alpha}_{CP}&=\frac{\Gamma\alpha+\bar{\Gamma}\bar{\alpha}}{\Gamma\alpha-\bar{\Gamma}\bar{\alpha}}\propto \mathcal{R}e(S^{*}P)+\mathcal{R}e(\bar{S}^{*}\bar{P}) \\
a^{\beta}_{CP}&=\frac{\Gamma\beta+\bar{\Gamma}\bar{\beta}}{\Gamma\beta-\bar{\Gamma}\bar{\beta}}\propto \mathcal{I}m(S^{*}P)+\mathcal{I}m(\bar{S}^{*}\bar{P})\\
a^{\gamma}_{CP}&=\frac{\Gamma\gamma-\bar{\Gamma}\bar{\gamma}}{\Gamma\gamma+\bar{\Gamma}\bar{\gamma}}\propto \left(|S|^{2}-|P|^{2}\right)-\left(|\bar{S}|^{2}-|\bar{P}|^{2}\right)\\
	\end{aligned}
\end{equation}
We will revisit this definition later. For simplicity, let's assume that only two diagrams contribute to the decay amplitudes. Specifically, the partial waves $S$ and $P$ can be parameterized as in equation (\ref{eq:S and P wave1}). By doing so, we can determine the strong phases dependence of the observables $a_{CP}^{\alpha,\beta,\gamma}$. It can then be shown that
\begin{equation}
	\begin{aligned}
a_{CP}^{\alpha}&\propto\left[r_{s}\sin(\delta_{p,p}-\delta_{s,t})-r_{p}\sin(\delta_{p,t}-\delta_{s,p})\right]\sin\Delta\phi\\
a_{CP}^{\beta}&\propto\left[r_{p}\cos(\delta_{p,t}-\delta_{s,p})-r_{s}\cos(\delta_{p,p}-\delta_{s,t})\right]\sin\Delta\phi\\
	a_{CP}^{\gamma}&\propto\left[|S_{t}||S_{p}|\sin(\delta_{s,t}-\delta_{s,p})-|P_{t}||P_{p}|\sin(\delta_{p,t}-\delta_{p,p})\right]\sin\Delta\phi\\
	\end{aligned}
\end{equation}
where $r_{s},r_{p}$ are ratios of $\left|S_{t}\right|/\left|S_{p}\right|,\left|P_{t}\right|/\left|P_{p}\right|$, respectively, $\Delta\phi=\phi_{t}-\phi_{p}$. Here, it is important to note that the CP asymmetries $a_{CP}^{\alpha},a_{CP}^{\beta}$ are determined by the strong phase $\delta_{p,p}-\delta_{s,t},\delta_{p,t}-\delta_{s,p}$  as they reflect the differences between the interference terms of $S,P$ waves and its CP conjugation, while the direct CP asymmetry and $a^{\gamma}_{CP}$ are dependent on variables $\delta_{p,p}-\delta_{p,t},\delta_{s,t}-\delta_{s,p}$ since they provides the difference between $|S|^{2}\pm|P|^{2}$ and that of CP conjugated decays. Furthermore, it is worth noting that $a_{CP}^{\alpha}$ and $a_{CP}^{\beta}$ exhibit different dependencies on the strong phase difference $\Delta\delta$ as sine and cosine functions, respectively. In summary, $a_{CP}^{\alpha}$ is suppressed by small strong phase differences, while $a_{CP}^{\beta}$ is more advantageous for small strong phases. Generally, a pair of such observables might offer a way to reduce the uncertainty arising from $\Delta\delta$, which is often challenging to determine precisely in both experiment and theory, especially for processes involving complex strong interaction dynamics. It is possible for someone to argue that $a_{CP}^{\beta}$ could be small even if the strong phase difference are large, due to a potential cancellation between two terms of it. Nevertheless, this scenario would require a fortuitous cancellation between two large quantities to get a small quantity, which of coincidence rather than a general case. We do not explore this specific case in our work, although it is theoretically possible.

As a supplement, it is necessary to illustrate the minus sign in $\bar{S}$ in equation (\ref{eq:S and P wave1}). Consider a theory governed by a Hamiltonian $\mathcal{H}$ that violates parity but conserves CP symmetry. In this case, one can decompose the Hamiltonian $\mathcal{H}$ into two parts, one being Parity-odd and another Parity-even
\begin{equation}
	\begin{aligned}
	 \mathcal{H}=\mathcal{H}_{1}+\mathcal{H}_{2}
	\end{aligned}
\end{equation}
with 
\begin{equation}
	\begin{aligned}
	 \mathcal{H}_{1}=\mathcal{H}+P\mathcal{H}P^{-1},~~\mathcal{H}_{2}=\mathcal{H}-P\mathcal{H}P^{-1}
	\end{aligned}
\end{equation}
In order to retain the invariance of $\mathcal{H}$ under $CP$ symmetry, we must have
\begin{equation}\label{eq: transformation under Charge}
	\begin{aligned}
	 C\mathcal{H}_{1}C^{-1}=\mathcal{H}_{1},~~C\mathcal{H}_{2}C^{-1}=-\mathcal{H}_{2}
	\end{aligned}
\end{equation}
Specifically, the partial wave $P,S$, with opposite parity, in the decay $B\to B^{\prime}P$ are induced by $\mathcal{H}_{1}$ and $\mathcal{H}_{2}$ respectively according to parity selection rule. Therefore, the minus sign appears under the transformation of charge conjugation. Alternatively, one can choose another convention under which the transformation of $\mathcal{H}_{1}$ and $\mathcal{H}_{2}$ are completely opposite to Eq. (\ref{eq: transformation under Charge}) since the overall phase factor $-1$ has no physical effect, but the relative minus sign between $\mathcal{H}_{1}$ and $\mathcal{H}_{2}$ is always significant.

Another useful asymmetry parameter, $\gamma$, can help distinguish the contributions from the $S$ and $P$ waves. This is achieved by introducing a new parameter, $\gamma^{\prime}$, defined as
\begin{equation}
	\begin{aligned}
		\gamma^{\prime}=\gamma\Gamma=|S|^{2}-|P|^{2},~~
		\bar{\gamma}^{\prime}=\bar{\gamma}\bar{\Gamma}=|\bar{S}|^{2}-|\bar{P}|^{2}
	\end{aligned}
\end{equation}
and obviously, 
	 \begin{equation}
	 \begin{aligned}
	 \Delta_{CP}^{\gamma^{\prime}}\equiv\frac{\gamma^{\prime}-\bar{\gamma}^{\prime}}{\Gamma+\bar{\Gamma}}&=\frac{\left(|S|^{2}-|\bar{S}|^{2}\right)-\left(|P|^{2}-|\bar{P}|^{2}\right)}{\Gamma+\bar{\Gamma}}
	 \end{aligned}
    \end{equation}
\begin{equation}
	 \begin{aligned}
	 a^{dir}_{CP}\equiv\frac{\Gamma-\bar{\Gamma}}{\Gamma+\bar{\Gamma}}&=\frac{\left(|S|^{2}-|\bar{S}|^{2}\right)+\left(|P|^{2}-|\bar{P}|^{2}\right)}{\Gamma+\bar{\Gamma}}
	 \end{aligned}
\end{equation}
Two additional CP observables, which involve only the information of a single partial wave amplitude, can be defined based on the linear combination of $\Delta^{\gamma}_{CP}$ and $a^{dir}_{CP}$
\begin{equation}
	 \begin{aligned}
	 a^{S}_{CP}&=\frac{1}{2}\left(a^{dir}_{CP}+\Delta_{CP}^{\gamma}\right)\\
	 a^{P}_{CP}&=\frac{1}{2}\left(a^{dir}_{CP}-\Delta_{CP}^{\gamma}\right)\\
	 \end{aligned}
\end{equation}
In the context of b-baryon decays, the simple asymmetry parameters mentioned earlier suffer significant restrictions due to the following reasons: (i) the current difficulty in achieving the initial polarization of b-baryons, and (ii) the great majority of $\Lambda_{b},\Xi_{b}$ decaying channels are multi-body rather than two body modes. Despite these challenges, these asymmetry parameters provide valuable insights that prompt us to explore CP-violating observables with different dependencies on strong phases. In the subsequent discussion, we will discuss some CP-violating phenomena similar to those induced by $\alpha$ or $\beta$ in the multi-body decays of b-baryons. We will refer to these as $\alpha$-like or $\beta$-like CP asymmetries, respectively. These observables are, in fact, special cases of $\mathcal{T}$-odd and $\mathcal{T}$-even correlations that will be discussed later.
\begin{figure}[h!]
\centering
\includegraphics[width=0.90\textwidth]{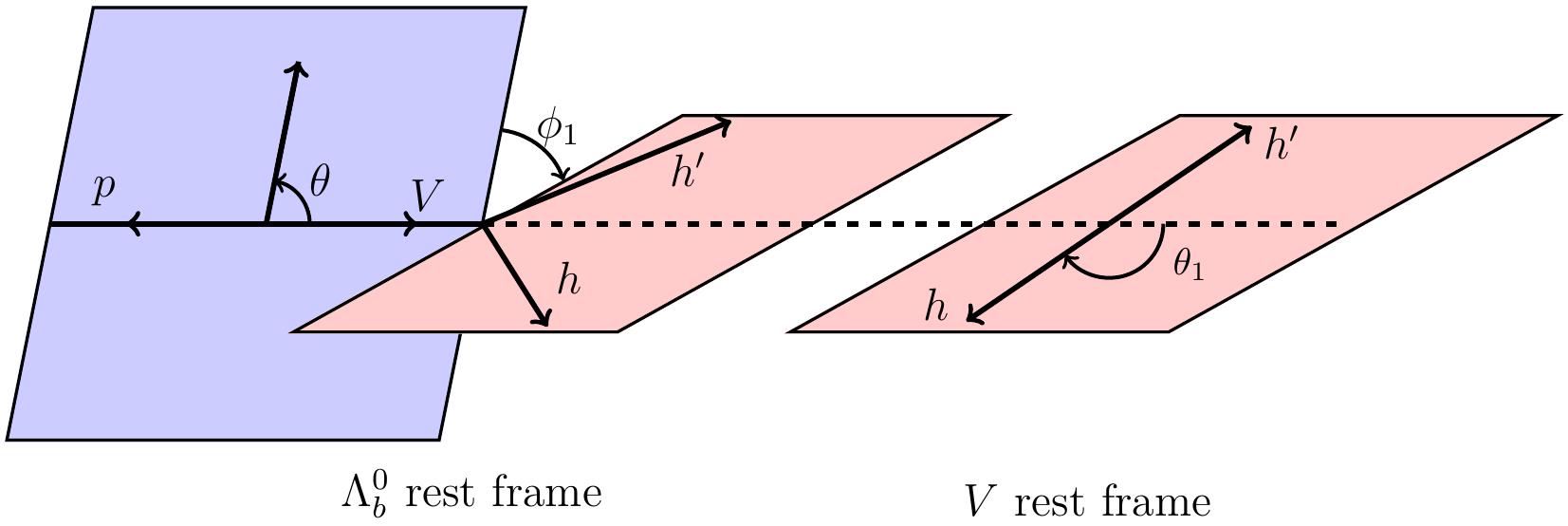}
\caption{The kinematics of three body decay modes of $\Lambda^{0}_{b}\to pV\to phh^{\prime}$, where $\theta$ is the angle between initial polarization and vector meson momentum in the $\Lambda^{0}_{b}$ rest frame, and $\theta_{1},\phi_{1}$ are polar and azimuth angle of $h$ with respect plane consist of $pV$ as blue one in above figure.}\label{Fig-4}
\end{figure}

Firstly, the chain decay $\Lambda^{0}_{b}\to phh^{\prime}$ with resonances $K^{*}/\rho\to hh^{\prime}$ is took into consideration under the condition of polarized $\Lambda^{0}_{b}$ since it provide more constructions of CP observables, seeing Figure.\ref{Fig-4}. The angular distribution in this case can be derived using the helicity representation
\begin{equation}\label{eq: angular distribution three body}
	 \begin{aligned}
	  \mathcal{W}(\theta,\theta_{1},\phi_{1})\propto \sum_{m,m^{\prime}}\rho^{Res}_{m,m^{\prime}}D^{*j=1}_{m,0}(\phi_{1},\theta_{1},0)D^{j=1}_{m^{\prime},0}(\phi_{1},\theta_{1},0)h_{m}h^{*}_{m^{\prime}}
	 \end{aligned}
\end{equation}
where $h_{m}$ represents the helicity amplitude of $K^{*}(\rho)\to hh^{\prime}$, and $\rho^{Res}_{m,m^{\prime}}$ denotes the spin density matrix of $K^{*}(\rho)$, which is given by
\begin{equation}\label{eq: res density matrix}
	 \begin{aligned}
	  \rho^{Res}_{m,m^{\prime}}\propto \sum_{\alpha,\beta,\lambda}\rho^{I}_{\alpha,\beta}\mathcal{H}_{\lambda,m}\mathcal{H}^{*}_{\lambda,m^{\prime}}\delta_{\lambda-m,\alpha}\delta_{\lambda-m^{\prime},\beta}
	 \end{aligned}
\end{equation}
with $\lambda$ the helicity of the proton, taking values of $\pm1/2$, $\mathcal{H}_{\lambda,m}$ denotes the helicity amplitudes of the primary decay $\Lambda^{0}_{b}\to pK^{*}/\rho$, and the spin density matrix of the initial b-baryon, $\rho^{I}_{\alpha,\beta}$, is defined as
\begin{equation}
	\begin{aligned}
		\rho^{I}=\frac{1}{2}\begin{pmatrix} 1+P\cos\theta & P\sin\theta \\ P\sin\theta & 1-P\cos\theta \end{pmatrix}
	\end{aligned}
\end{equation}
The Kronecker $\delta$ in the (\ref{eq: res density matrix}) is the requirement of helicity conservation. Explicitly, $\mathcal{W}(\theta,\theta_{1},\phi_{1})$ is expanded as
\begin{equation}
	 \begin{aligned}
	  \mathcal{W}(\theta,\theta_{1},\phi_{1})&\propto \sin^{2}\theta_{1}(\rho^{Res}_{+1,+1}+\rho^{Res}_{-1,-1})+2\cos^{2}\theta_{1}\rho^{Res}_{0,0}+\sqrt{2}\mathcal{R}e(\rho^{Res}_{-1,0}e^{-i\phi_{1}})\sin2\theta_{1}\\
	  &-\sqrt{2}\mathcal{R}e(\rho^{Res}_{1,0}e^{i\phi_{1}})\sin2\theta_{1}\\
	  &=\sin^{2}\theta_{1}(\rho^{Res}_{+1,+1}+\rho^{Res}_{-1,-1})+2\cos^{2}\theta_{1}\rho^{Res}_{0,0}\\
	  &+\sqrt{2}\sin2\theta_{1}\left\{\mathcal{R}e(\rho^{Res}_{-1,0}-\rho^{Res}_{1,0})\cos\phi_{1}+\mathcal{I}m(\rho^{Res}_{-1,0}+\rho^{Res}_{1,0})\sin\phi_{1}\right\}
	 \end{aligned}
\end{equation}
It is important to note that the angular distribution involves interference terms from different helicity configurations. Two asymmetry parameters can be determined as
\begin{equation}
	 \begin{aligned}
	  \kappa_{1}=\mathcal{R}e(\rho^{Res}_{-1,0}-\rho^{Res}_{1,0}),~~\kappa_{2}=\mathcal{I}m(\rho^{Res}_{-1,0}+\rho^{Res}_{1,0})
	 \end{aligned}
\end{equation}
The matrix elements $\rho^{Res}_{m,m^{\prime}}$ are listed below
\begin{equation}
	 \begin{aligned}
	  &\rho^{Res}_{+1,+1}=\frac{1}{2}(1-P\cos\theta)\left|\mathcal{H}_{+1/2,+1}\right|^{2}\\
	 & \rho^{Res}_{-1,-1}=\frac{1}{2}(1+P\cos\theta)\left|\mathcal{H}_{-1/2,-1}\right|^{2}\\
	 &\rho^{Res}_{+1,0}=\rho^{*Res}_{0,+1}=\mathcal{H}_{+1/2,+1}\mathcal{H}^{*}_{+1/2,0}P\sin\theta\\
	 &\rho^{Res}_{-1,0}=\rho^{*Res}_{0,-1}=\mathcal{H}_{-1/2,-1}\mathcal{H}^{*}_{-1/2,0}P\sin\theta\\
  	 \rho^{Res}_{0,0}=\frac{1}{2}&(1-P\cos\theta)\left|\mathcal{H}_{-1/2,0}\right|^{2}+\frac{1}{2}(1+P\cos\theta)\left|\mathcal{H}_{+1/2,0}\right|^{2}\\
	 \end{aligned}
\end{equation}
Therefore, substituting them into $\kappa_{1,2}$, one obtains
\begin{equation}
	 \begin{aligned}
	  \kappa_{1}&=\mathcal{R}e(\mathcal{H}_{-1/2,-1}\mathcal{H}^{*}_{-1/2,0}-\mathcal{H}_{+1/2,+1}\mathcal{H}^{*}_{+1/2,0})P\sin\theta\\
	  \kappa_{2}&=\mathcal{I}m(\mathcal{H}_{-1/2,-1}\mathcal{H}^{*}_{-1/2,0}+\mathcal{H}_{+1/2,+1}\mathcal{H}^{*}_{+1/2,0})P\sin\theta
	 \end{aligned}
\end{equation}
and we introduce 
\begin{equation}
	 \begin{aligned}
	  \langle\kappa_{1}\rangle&=\mathcal{R}e(\mathcal{H}_{-1/2,-1}\mathcal{H}^{*}_{-1/2,0}-\mathcal{H}_{+1/2,+1}\mathcal{H}^{*}_{+1/2,0})/\Gamma\\
	  \langle\kappa_{2}\rangle&=\mathcal{I}m(\mathcal{H}_{-1/2,-1}\mathcal{H}^{*}_{-1/2,0}+\mathcal{H}_{+1/2,+1}\mathcal{H}^{*}_{+1/2,0})/\Gamma
	 \end{aligned}
\end{equation}
for simplicity. It is evident that $\kappa_{1(2)}$ are odd (even) under Parity transformation and even (odd) under time reversal, respectively. Consequently, if $\kappa_{1}$ is expressed in terms of partial waves, only interference terms arising from waves with opposite parity will persist. Conversely, for $\kappa_{2}$, only interference terms with even parity and potential modulo terms will be present. Referring to Eq.\eqref{2.26}, two observables, $a^{\kappa_{1}}_{CP}$ and $a^{\kappa_{2}}_{CP}$ are defined
\begin{equation}
	\begin{aligned}
a^{\kappa_{1}}_{CP}
=\frac{1}{2}\left[\langle\kappa_{1}\rangle+\langle\bar{\kappa_{1}}\rangle\right],~~~
a^{\kappa_{2}}_{CP}
=\frac{1}{2}\left[\langle\kappa_{2}\rangle-\langle\bar{\kappa_{2}}\rangle\right]\\
	\end{aligned}
\end{equation}
Significantly, $a^{\kappa_{1}}_{CP}$ and $a^{\kappa_{2}}_{CP}$ are respectively $\alpha$-like and $\beta$-like in the dependence of strong phase if we temporarily ignore the direct CPV.

Another decay process, $\Lambda_b^0\to \Lambda^0(\to p\pi^-)h^+h'^-$ with $h=K$ or $\pi$, offers a platform to explore the CP violation in $\Lambda_b\to \Lambda V$ through the sequential decay $\Lambda^{0}\to p\pi^{-}$. The angular analysis in this decay is similar to $\Lambda^{0}_{b}\to pV$ but is more intricate. An insightful discussion involves the angular distribution of the cascade decay $\Lambda^{0}_{b}\to\Lambda^{0} V\to p\pi^{-}h^{+}h^{-}$ with $h=K$ or $\pi$, where $\Lambda^{0}_{b}$ is considered unpolarized for simplicity. Specifically, the angular distribution is given by \cite{Korner:1992wi}
\begin{equation}\label{eq:angularLb2LzKstar}
\begin{aligned}
\frac{d\Gamma}{d\cos\theta_{\Lambda}d\cos\theta_{V} d\varphi}&\propto  \frac{3}{4}\sin^{2}\theta_{K}\left\{\left|\mathcal{H}_{+\frac{1}{2},+1}\right|^{2}(1+\alpha \cos\theta_{\Lambda})+\left|\mathcal{H}_{-\frac{1}{2},-1}\right|^{2}(1-\alpha \cos\theta_{\Lambda})\right\}\\&
+\frac{3}{2}\cos^{2}\theta_{K}\left\{\left|\mathcal{H}_{+\frac{1}{2},0}\right|^{2}(1+\alpha \cos\theta_{\Lambda})+\left|\mathcal{H}_{-\frac{1}{2},0}\right|^{2}(1-\alpha \cos\theta_{\Lambda})\right\}\\
&-\frac{3}{2\sqrt{2}}\alpha \cos\varphi \sin\theta_{\Lambda} \sin2\theta_{K} \mathcal{R}e\left(\mathcal{H}_{+\frac{1}{2},+1}\mathcal{H}^{*}_{-\frac{1}{2},0}-\mathcal{H}_{-\frac{1}{2},-1}\mathcal{H}^{*}_{+\frac{1}{2},0}\right)\\
&-\frac{3}{2\sqrt{2}}\alpha \sin\varphi \sin\theta_{\Lambda} \sin2\theta_{K} \mathcal{I}m\left(\mathcal{H}_{+\frac{1}{2},+1}\mathcal{H}^{*}_{-\frac{1}{2},0}+\mathcal{H}_{-\frac{1}{2},-1}\mathcal{H}^{*}_{+\frac{1}{2},0}\right)  
\end{aligned}
\end{equation}
where $\theta_{\Lambda}$ and $\theta_{K}$ represent the polar angles of the proton in the $\Lambda$ frame and the $K$ meson in the $K^{*}$ frame, respectively, and $\varphi$ is the relative azimuth angle between the two decay planes. The expression can be simplified by integrating out $\theta_{K}$ and $\varphi$
\begin{equation}\label{eq: average lambdab polarization}
{d\Gamma\over d\cos\theta}
\propto
1+\langle \alpha_{\Lambda_{b}}\rangle\alpha_\Lambda\cos\theta
\end{equation}
and average longitudinal polarization of $\Lambda$, $\langle \alpha_{\Lambda_b}\rangle$, is formulated by helicity amplitudes 
\begin{equation}\label{eq: polarization}
	\begin{aligned}
		\langle \alpha_{\Lambda_b}\rangle=\frac{\left|\mathcal{H}_{+\frac{1}{2},0}\right|^{2}+\left|\mathcal{H}_{+\frac{1}{2},+1}\right|^{2}-\left|\mathcal{H}_{-\frac{1}{2},0}\right|^{2}-\left|\mathcal{H}_{-\frac{1}{2},-1}\right|^{2}}{\left|\mathcal{H}_{+\frac{1}{2},0}\right|^{2}+\left|\mathcal{H}_{-\frac{1}{2},-1}\right|^{2}+\left|\mathcal{H}_{-\frac{1}{2},0}\right|^{2}+\left|\mathcal{H}_{+\frac{1}{2},+1}\right|^{2}}
	\end{aligned}
\end{equation}
Given that $\alpha_\Lambda=0.75\pm0.01$ has been accurately measured by BESIII \cite{BESIII:2018cnd}, the value of $\langle \alpha_{\Lambda_{b}}\rangle$ can be determined from the measurement of the angular distributions mentioned above. Additionally, the CP asymmetry parameter $a_{CP}^{\alpha_\Lambda}=(\alpha_\Lambda+\bar \alpha_\Lambda)/(\alpha_\Lambda-\bar \alpha_\Lambda)=-0.006\pm0.014$ was measured by BESIII \cite{BESIII:2018cnd}, with theoretical predictions on the order of $10^{-4}\sim10^{-5}$ \cite{Donoghue:1986hh}. This implies that the CP violation associated with $\alpha_\Lambda$ is negligible in the context of measuring the CP violation in $\Lambda_b$ decays, which is expected to be on the order of percent or even higher \cite{Geng:2021sxe}. Consequently, the CP violation in $\Lambda_b^0$ decays can be extracted effectively.
On the other hand, the interference terms of $\mathcal{H}_{\lambda_{\Lambda},\lambda_{K^{*}}}$ in Eq.(\ref{eq:angularLb2LzKstar}) implies the $\alpha$ and $\beta$-like observable analogy to decay $\Lambda_b^0\to PV$. However, unlike decay modes of $ pV $, this is just achieved in the case of no initial polarization.
Meanwhile, $\Lambda^{0}_{b}\to p\pi h^{+}h^{-}$ distribution with normally polarized $\Lambda^{0}_{b}$, has been extensively investigated in \cite{Geng:2021sxe}, and more independent $\alpha$ and $\beta$-like observable related to $P_{b}$ are constructed. Experimental measurement has also been performed on the triple product of this process, but the CPV is consistent with zero\cite{LHCb:2016hwg}.

Finally, the helicity amplitudes $\mathcal{H}_{\lambda_{1},\lambda_{K^{*}}}$ for $1/2^{+}\to 1/2^{+}V $ can be formulated explicitly based on the equation in the section (\ref{sec: decay amplitude})
\begin{equation}
	\begin{aligned}
	\mathcal{H}_{+\frac{1}{2},+1}(V)&=-\sqrt{2M(E^{\prime}-M^{\prime})}B^{+}_{1}\\	\mathcal{H}_{+\frac{1}{2},+1}(A)&=-\sqrt{2M(E^{\prime}+M^{\prime})}A^{+}_{1}\\
	\end{aligned}
\end{equation}
\begin{equation}
	\begin{aligned}
	\mathcal{H}_{+\frac{1}{2},0}(V)&=\frac{1}{M_{a_{1}}}\sqrt{2M(E^{\prime}-M^{\prime})}\left[(E^{\prime}+M^{\prime})(B^{+}_{1}+B^{+}_{2})+B^{+}_{1}E_{a_{1}}\right]\\
	\mathcal{H}_{+\frac{1}{2},0}(A)&=\frac{1}{M_{a_{1}}}\sqrt{2M(E^{\prime}+M^{\prime})}\left[(E^{\prime}-M^{\prime})(A^{+}_{1}-A^{+}_{2})+A^{+}_{1}E_{a_{1}}\right]
	\end{aligned}
\end{equation}
and other helicity amplitudes are obtained by parity transformation
\begin{equation}
	\begin{aligned}
	 \mathcal{H}_{\lambda_{1},\lambda_{K^{*}}}(V/A)=(+/-)\eta_{b}\eta_{\Lambda}\eta_{K^{*}}(-1)^{S_{b}-S_{\Lambda}-S_{K^{*}}}\mathcal{H}_{-\lambda_{1},-\lambda_{K^{*}}}=\mathcal{H}_{-\lambda_{1},-\lambda_{K^{*}}}
	\end{aligned}
\end{equation}
where $\eta_{b},\eta_{\Lambda},\eta_{K^{*}}$ are intrinsic parity of $\Lambda_{b},\Lambda,K^{*}$ respectively, and the amplitude $\mathcal{H}_{\lambda_{1},\lambda_{K^{*}}}$ finally is expressed 
\begin{equation}
	\begin{aligned}
	 \mathcal{H}_{\lambda_{1},\lambda_{K^{*}}}=\mathcal{H}_{\lambda_{1},\lambda_{K^{*}}}(V)-\mathcal{H}_{\lambda_{1},\lambda_{K^{*}}}(A)
	\end{aligned}
\end{equation}
In the experimental side, the branching fractions are measured as $Br(\Lambda_b^0\to \Lambda^0\pi^+\pi^-)=(4.7\pm1.9)\times10^{-6}$, $Br(\Lambda_b^0\to \Lambda^0 K^+\pi^-)=(5.7\pm1.3)\times10^{-6}$ and $Br(\Lambda_b^0\to \Lambda^0 K^+K^-)=(1.62\pm0.23)\times10^{-5}$ \cite{PDG}. The signal events of these three processes are $65\pm14$, $97\pm14$ and $185\pm15$, respectively, using the LHCb Run I data with the integrated luminosity of 3 fb$^{-1}$ \cite{LHCb:2016rja}. We hope these decay channels can be investigated in the future experiments.

The conventional definition of CP violation, such as direct CPV, involves the ratio of $\Gamma-\bar{\Gamma}$ to $\Gamma+\bar{\Gamma}$, which is constrained to the range of $-1$ to $1$. This is a well-defined concept. However, when considering CP violation induced by asymmetry parameters like $a^\alpha_{CP}$, $a^{\beta}_{CP}$, $a^{\gamma}_{CP}$, $a^{\mathcal{O}_{1}}_{CP}$, and $a^{\mathcal{O}_{2}}_{CP}$ as defined above, these quantities can have magnitudes larger than $1$, since there are no specific constraints or principles that limit them to a particular range. In particularly, the divergence issue arises in the case of $a^{\beta}_{CP}$ and $a^{\mathcal{O}_{2}}_{CP}$ when all the strong phases tend to zero. In other words, the possible enhancement due to a small denominator is not desirable although we indeed want a relative large CP violation in a certain process, since all of experimental measurements are ad-jointed with uncertainties that are also enlarged in it.

Furthermore, our proof of complementarity is based on the properties of interference terms. However, the actual experimental measurements involve the ratios of $\mathcal{O}_{1}/\Gamma$ and $\mathcal{O}_{2}/\Gamma$, rather than the pure quantities $\mathcal{O}_{1}$ and $\mathcal{O}_{2}$. Therefore, we propose a definition that avoids conflicts with our proof of complementarity and does not suffer from the shortcomings of small denominator enhancements. The suggested definition is as follows:
\begin{equation}
		\begin{aligned}
     \mathcal{O}^{\prime}_{1}=\frac{\Gamma(\mathcal{O}_{1}/\Gamma)}{\langle\Gamma\rangle}=\frac{\mathcal{O}_{1}}{\langle\Gamma\rangle},~ \mathcal{O}^{\prime}_{2}=\frac{\Gamma(\mathcal{O}_{2}/\Gamma)}{\langle\Gamma\rangle}=\frac{\mathcal{O}_{2}}{\langle\Gamma\rangle}
		\end{aligned}
\end{equation}
and corresponding charge conjugation
\begin{equation}
		\begin{aligned}
     \bar{\mathcal{O}}^{\prime}_{1}=\frac{\bar{\Gamma}(\bar{\mathcal{O}}_{1}/\bar{\Gamma})}{\langle\Gamma\rangle}=\frac{\bar{\mathcal{O}}_{1}}{\langle\Gamma\rangle},~ \bar{\mathcal{O}}^{\prime}_{2}=\frac{\bar{\Gamma}(\bar{\mathcal{O}}_{2}/\bar{\Gamma})}{\langle\Gamma\rangle}=\frac{\bar{\mathcal{O}}_{2}}{\langle\Gamma\rangle}
		\end{aligned}
\end{equation}
where $\langle\Gamma\rangle$ is the average decay width $(\Gamma+\bar{\Gamma})/2$. Then, the CPV is defined as
\begin{equation}\label{B24}
		\begin{aligned}
      a^{\mathcal{O}_{1}}_{CP}&=\frac{1}{2}\left(\mathcal{O}^{\prime}_{1}\pm\bar{\mathcal{O}}^{\prime}_{1}\right)\\     
      a^{\mathcal{O}_{2}}_{CP}&=\frac{1}{2}\left(\mathcal{O}^{\prime}_{2}\pm\bar{\mathcal{O}}^{\prime}_{2}\right)\\
		\end{aligned}
\end{equation}
The same definition can also be applied to the asymmetry parameters $\alpha, \beta, \gamma$. Firstly, it is important to note that the redefinition mentioned above provides a dimensionless measure of CP violation and does not alter the complementarity we have proven. Furthermore, the fact that the magnitudes of CP violation observations $a^{\mathcal{O}_{1}}_{CP}$ and $a^{\mathcal{O}_{2}}_{CP}$ are both smaller than 1. It can be established through the following analysis: The asymmetry parameters $\mathcal{O}_{1,2}/\Gamma$ represent the probabilities of certain events occurring in specific phase space regions, and as probabilities, they must be smaller than 1. Additionally, the summation or difference of $\mathcal{O}^{\prime}_{1}$ and $\bar{\mathcal{O}}^{\prime}_{1}$ is smaller than 2. Considering the direct CP violation, the coefficient $\Gamma/{\langle\Gamma\rangle}$ and $\bar{\Gamma}/{\langle\Gamma\rangle}$ just varies in the range of $\left[0,2\right]$, and has no singularities. Therefore, the observed CP violation in equation (\ref{B24}) falls within the range of -1 to 1. We agree that the suggested definition is an improvement over the previous one, and it offers advantages in both theoretical and experimental aspects, while the previous definition has been widely used and applied.  By considering the dimensionless CP violation and ensuring that the magnitudes of asymmetry parameters are within a reasonable range, the new definition offers a more robust and accurate representation of CP violation phenomena.

%% file: 4-angular-TPA.tex
\section{$\mathcal{T}$-odd and $\mathcal{T}$-even correlations}\label{sec.5}
One might naturally expect that CP violation could be revealed by examining whether time reversal symmetry is respected or not, assuming that CPT is a valid symmetry. However, we have demonstrated that the detection of time reversal violation is not straightforward by solely considering a $\mathcal{T}$-odd quantity due to the existence of the final state interactions pollution. Nevertheless, the cancellation of this background is anticipated by comparing a $\mathcal{T}$-odd quantity with its CP conjugation. This concept is referred to as $\mathcal{T}$-odd correlation-induced CP asymmetry. In the following section, we will initially focus on the fundamental properties of $\mathcal{T}$-odd correlation, and then give a general proposal to establish a pair of complementary $\mathcal{T}$-even and -odd CP asymmetries that exhibit distinct strong phase dependencies as sine and cosine functions respectively. The potential applications of these concepts in $b$-baryon decays, primarily focusing on the four-body modes, will be discussed.

\subsection{General properties}
The $\mathcal{T}$-odd correlations have been extensively studied as a potential probe of CP violation in the angular distributions of $B$ and $D$ meson four-body decays \cite{Valencia:1988it, Datta:2003mj}. Moreover, the exploration of new physics possibilities has been a focus of recent discussions \cite{Datta:2003mj, London:2020doj, Tiwari:2022pug, Dutta:2021del}. In essence, $\mathcal{T}$-odd correlations are quantities that exhibit odd behavior under time reversal. To understand their general properties, we can discuss them employing examples of triple products, which provide a framework that is applicable across various scenarios
\begin{equation}
	\begin{aligned}
     \psi=(\vec{v}_{1}\times\vec{v}_{2})\cdot\vec{v}_{3}
	\end{aligned}
\end{equation}
where $\vec{v}_{i}$ is the spin vector or momentum of $i-th$ particle. The decay $B\to V_{1}V_{2}$ serves as an excellent example for illustration purposes. This is particularly effective as both the polarizations and one final momentum in the $B$ rest frame are well-defined, hence
\begin{equation}
	\begin{aligned}
     \psi\left[B\to V_{1}V_{2}\right]=(\vec{\epsilon}_{1}\times\vec{\epsilon}_{2})\cdot\vec{p}
	\end{aligned}
\end{equation}
and $\vec{\epsilon}_{i}$ represents the polarization vector of $V_{i}$. Additionally, the observable $\beta$ can be viewed as a specific type of triple product. Utilizing Equation (\ref{eq: trace and square amplitude}), we can extract the term proportional to $\beta$
\begin{equation}
	\begin{aligned}
     Tr&\left\{SP^{*}(\vec{\xi}\cdot\vec{\sigma})(\vec{\eta}\cdot\vec{\sigma})(\vec{\sigma}\cdot\hat{p})+S^{*}P(\vec{\xi}\cdot\vec{\sigma})(\vec{\sigma}\cdot\hat{p})(\vec{\eta}\cdot\vec{\sigma})\right\}\\
     &=2i\left\{SP^{*}\epsilon_{ijk}\xi^{i}\eta^{j}p^{k}+S^{*}P\epsilon_{ijk}\xi^{i}p^{j}\eta^{k}\right\}\\
     &=2\mathcal{I}m(S^{*}P)(\vec{\xi}\times\vec{\eta})\cdot \vec{p}
	\end{aligned}
\end{equation}
Here, $\vec{\xi}$ and $\vec{\eta}$ represent the polarization vectors of the initial and final baryons, respectively. In experimental measurements, the determination of these quantities necessitates the derivation of the angular distribution of the relevant decay chain, as the spin or polarization cannot be directly reconstructed from current collider detectors. Consequently, all triple products involving spin and polarization vectors are ultimately transformed into pure momentum correlations to meet the requirements of experimental measurements.

The momentum triple-product correlations have been experimentally measured in $b$-baryon multi-body decays such as $\Lambda^{0}_{b}\to p3\pi, pK\pi\pi, p3K$, and $\Xi^{0}_{b}\to pK^{+}K^{-}\pi$ \cite{LHCb:2016yco, LHCb:2018fpt, LHCb:2019oke, LHCb:2019jyj}. To illustrate, let's consider a decay process $\mathcal{M}\to abcd$, where the final state particles $abcd$ are all light particles and can be detected directly in experiments. A $\mathcal{T}$-odd triple-momentum correlation can be defined as
\begin{equation}
	\begin{aligned}
     \psi=\vec{p}_{i}\cdot(\vec{p}_{j}\times\vec{p}_{k})
	\end{aligned}
\end{equation}
three momentum $\vec{p}_{i},\vec{p}_{j},\vec{p}_{k}$ must be independent to each other. One finds it is also Parity-odd, and lists the properties under the other transformations \cite{Bevan:2014nva}
\begin{equation}
	\begin{aligned}
     C\psi&=\bar{\psi}=\vec{p}_{\bar{i}}\cdot(\vec{p}_{\bar{j}}\times\vec{p}_{\bar{k}})\\
     CP\psi&=-\vec{p}_{\bar{i}}\cdot(\vec{p}_{\bar{j}}\times\vec{p}_{\bar{k}})=-\bar{\psi}\\
     \mathcal{T}\psi&=-\vec{p}_{i}\cdot(\vec{p}_{j}\times\vec{p}_{k})=-\psi
	\end{aligned}
\end{equation}
It is worth noting that the momentum triple products vanish in three-body and two-body decays due to the lack of sufficient independent momentum.

A triple product asymmetry $ A(\psi)$ is obtained by taking the difference between $\psi>0$ and $\psi<0$
\begin{equation}
	\begin{aligned}
     A(\psi)&=\frac{\Gamma(\psi>0)-\Gamma(\psi<0)}{\Gamma}\\
     \bar{A}(\bar{\psi})&=\frac{\bar{\Gamma}(\bar{\psi}>0)-\bar{\Gamma}(\bar{\psi}<0)}{\Gamma}\\
	\end{aligned}
\end{equation}
CP asymmetry $a_{CP}^{\psi}$ induced by $A(\psi)$, according to (\ref{2.26}), is defnined as
\begin{equation}
	\begin{aligned}
     a_{CP}^{\psi}&=\frac{1}{2}\left\{A_(\psi)-\hat{C}\hat{P}A(\psi)\right\}\\
     &=\frac{1}{2}\left\{A(\psi)+ \bar{A}(\bar{\psi})\right\}
	\end{aligned}
\end{equation}
Therefore, the CP-violating parameter $a_{CP}^{\psi}$ is also a local CP violation similar to $a_{CP}^{\alpha}$ and $a_{CP}^{\beta}$, indicating differences in the differential distribution between a decay and its CP-conjugate, rather than the overall width obtained by integrating over phase space as in the case of direct CP violation. It is worth noting that the systematic uncertainties arising from production and detection asymmetries are largely mitigated in the measurement of this type of CP violation. To further illustrate this point, let us explicitly express $a_{CP}^{\mathcal{T}}$
\begin{equation}
	\begin{aligned}
     a_{CP}^{\mathcal{T}}&=\frac{1}{2}\left\{A_{\mathcal{T}}(\psi)+ \bar{A}_{\mathcal{T}}(\bar{\psi})\right\}\\
     &=\frac{\Gamma(\psi>0)\bar{\Gamma}(\bar{\psi}>0)-\Gamma(\psi<0)\bar{\Gamma}(\bar{\psi}<0)}{\Gamma\bar{\Gamma}}
	\end{aligned}
\end{equation}
We observe that production and detection asymmetries are eliminated in the numerator. Hence, there is great potential for triple-product asymmetries in the experiment. In contrast to direct CPV, it is noteworthy that certain triple-product CP asymmetries are not subject to the suppression effect of the small strong phase difference $\Delta\delta$. Specifically, some of these asymmetries are dependent on the strong phases in a cosine function form, making them particularly sensitive to small $\Delta\delta$. In the following discussion, we will focus on this crucial property in more depth.

\subsection{Strong phase dependence}\label{sec5.2}
The discussion in Sec.2 tells us that the expectation value of a $\mathcal{T}$-odd operator $T$ is represented as an asymmetry parameter denoted as $\Delta_{t}$
\begin{equation}\label{def4.2}
	\begin{aligned}
	  \Delta_{t}=\bra{f}T\ket{f}=\Gamma(t)-\Gamma(-t)
	\end{aligned}
\end{equation}
Next step, the mathematical relation between a specific $\Delta_{t}$ and helicity amplitudes will be shown by taking into account the eigenstates of $\mathcal{T}$-odd operators\cite{Geng:2021sxe,Geng:2021lrc}.
Firstly, the eigenstates of a $\mathcal{T}$-odd operators always exist as a pair of objects with opposite eigenvalues. To be precise, assuming $T$ is such an operator with $\ket{\psi}$ as an eigenstate associated with the eigenvalue $\lambda_{t}$, we have
\begin{equation}
	\begin{aligned}
	 T\ket{\psi}=\lambda_{t}\ket{\psi}
	\end{aligned}
\end{equation}
Perform the time reversal operator $\mathcal{T}$ on $T$
\begin{equation}
	\begin{aligned}
	 \mathcal{T}T\mathcal{T}^{-1}\mathcal{T}\ket{\psi}=\lambda^{*}_{t}\mathcal{T}\ket{\psi}=\lambda_{t}\mathcal{T}\ket{\psi} \Rightarrow T\mathcal{T}\ket{\psi}=-\lambda_{t}\mathcal{T}\ket{\psi}
	\end{aligned}
\end{equation}
Hence, $\mathcal{T}\ket{\psi}$ is also an eigenstate of $T$ with the negative eigenvalue $-\lambda_{t}$, where $\lambda_{t}=\lambda^{*}_{t}$ is maintained since $T$ is Hermitian as an observable. The decomposition of the $T$ eigenstate into a superposition of helicity states is always valid due to the completeness of the helicity representation. A straightforward example is the parameter $\beta$, which is regarded as the expectation value of the operator $(\hat{S}_{i}\times\hat{S}_{f})\cdot \hat{p}$, where $\hat{p}$ aligns with $\vec{e}_{z}$. For simplicity, this operator can be written in ladder form
\begin{equation}
	\begin{aligned}
	 T_{\beta}= \frac{i}{2}\left\{S^{+}_{i}S^{-}_{f}-S^{-}_{i}S^{+}_{f}\right\}
	\end{aligned}
\end{equation}
It is easy to verify following equations
\begin{equation}
	\begin{aligned}
	 T_{\beta}\ket{\vec{p}=p\vec{e}_{z},\lambda_{i}=-\frac{1}{2},\lambda_{f}=+\frac{1}{2}}&=\frac{i}{2}\ket{\vec{p}=p\vec{e}_{z},\lambda_{i}=+\frac{1}{2},\lambda_{f}=-\frac{1}{2}}\\
	 T_{\beta}\ket{\vec{p}=p\vec{e}_{z},\lambda_{i}=+\frac{1}{2},\lambda_{f}=-\frac{1}{2}}&=-\frac{i}{2}\ket{\vec{p}=p\vec{e}_{z},\lambda_{i}=-\frac{1}{2},\lambda_{f}=+\frac{1}{2}}\\
	\end{aligned}
\end{equation}
Combining these two helicity states, we arrive the eigenequation of $T_{\beta}$
\begin{equation}
	\begin{aligned}
	 T_{\beta}&\left\{\ket{\lambda_{i}=+\frac{1}{2},\lambda_{f}=-\frac{1}{2}}\pm i\ket{\lambda_{i}=-\frac{1}{2},\lambda_{f}=+\frac{1}{2}}\right\}\\
	 &=\mp\frac{1}{2}\left\{\ket{\lambda_{i}=+\frac{1}{2},\lambda_{f}=-\frac{1}{2}}\pm i\ket{\lambda_{i}=-\frac{1}{2},\lambda_{f}=+\frac{1}{2}}\right\}
	\end{aligned}
\end{equation}
where the momentum symbol has been omitted in the ket. The asymmetry parameter $\beta$ is naturally determined based on the definition \eqref{def4.2}
\begin{equation}
	\begin{aligned}
	  \beta&=\left\{\left|\mathcal{H}_{+1/2,-1/2}-i\mathcal{H}_{-1/2,+1/2}\right|^{2}-\left|\mathcal{H}_{+1/2,-1/2}+i\mathcal{H}_{-1/2,+1/2}\right|^{2}\right\}/\left\{\left|\mathcal{H}_{+1/2,+1/2}\right|^{2}+\left|\mathcal{H}_{-1/2,-1/2}\right|^{2}\right\}\\
	  &=\frac{2\mathcal{I}m(\mathcal{H}_{+1/2}\mathcal{H}^{*}_{-1/2})}{\left|\mathcal{H}_{+1/2,+1/2}\right|^{2}+\left|\mathcal{H}_{-1/2,-1/2}\right|^{2}}
	\end{aligned}
\end{equation}
which is consistent with the expectation value of normal polarization in the discussion of hyperon decay. Let us now delve to the similar analysis for more complex case $B\to VV$. In the decay of $B\to V_{1}V_{2}$, $\psi=(\hat{S}_{1}\times\hat{S}_{2})\cdot\hat{p}$ is a natural T-odd triple product correlation due to the initial polarization being vanishing. We also re-express $\psi$ as
\begin{equation}
	\begin{aligned}
	 \psi= \frac{i}{2}\left\{S^{+}_{1}S^{-}_{2}-S^{-}_{1}S^{+}_{2}\right\}
	\end{aligned}
\end{equation}
The corresponding eigenstates are given as 
\begin{equation}
	\begin{aligned}
	 \psi\ket{\mathcal{H}_{\parallel}}&=0\ket{\mathcal{H}_{\parallel}}\\
	 \psi\left\{\mathcal{H}_{\perp}\mp i\mathcal{H}_{0}\right\}&=\pm\sqrt{2}\left\{\mathcal{H}_{\perp}\mp i\mathcal{H}_{0}\right\}
	\end{aligned}
\end{equation}
which has been studied in \cite{Geng:2021lrc}. Hence, the asymmetry parameter associated with $\psi$ is given by
\begin{equation}
	\begin{aligned}
	 \Delta_{\psi}=\frac{\left|\mathcal{H}_{\perp}- i\mathcal{H}_{0}\right|^{2}-\left|\mathcal{H}_{\perp}+ i\mathcal{H}_{0}\right|^{2}}{\sum_{j=0,\perp,\parallel}\left|\mathcal{H}_{j}\right|^{2}}=\frac{2\mathcal{I}m(\mathcal{H}_{\perp}\mathcal{H}^{*}_{0})}{\sum_{j=0,\perp,\parallel}\left|\mathcal{H}_{j}\right|^{2}}
 \end{aligned}
\end{equation}
Ultimately, this $\mathcal{T}$-odd correlation is anticipated to manifest in the final angular distribution of the cascade decay $B\to V_{1}V_{2}\to P_{1}P_{2}P_{3}P_{4}$ in the form of momentum correlation, as the spin vectors $\vec{S}_{i/f}$ are finally linked to the momentum distribution of the strong decays $V_{1}\to P_{1}P_{2}$ and $V_{2}\to P_{3}P_{4}$ \cite{Datta:2003mj}
	\begin{equation}
	\begin{aligned}
	\frac{d\Gamma}{d\cos\theta_{1}d\cos\theta_{2}d\varphi}&\propto\\&\left|\mathcal{H}_{0}\right|^{2}\cos^{2}\theta_{1}\cos^{2}\theta_{2}+\frac{\left|\mathcal{H}_{\perp}\right|^{2}}{2}\sin^{2}\theta_{1}\sin^{2}\theta_{2}\sin^{2}\varphi+\frac{\left|\mathcal{H}_{\parallel}\right|^{2}}{2}\sin^{2}\theta_{1}\sin^{2}\theta_{2}\cos^{2}\varphi\\&+\frac{Re(\mathcal{H}^{*}_{\parallel}\mathcal{H}_{0})} {2\sqrt{2}}\sin2\theta_{1}\sin2\theta_{2}\cos\varphi-\frac{Im(\mathcal{H}_{\perp}\mathcal{H}_{0}^{*})}{2\sqrt{2}}\sin2\theta_{1}\sin2\theta_{2}\sin\varphi\\
	 &+\frac{Re(\mathcal{H}^{*}_{\parallel}\mathcal{H}_{\perp})}{2\sqrt{2}}\sin^{2}\theta_{1}\sin^{2}\theta_{2}\cos2\varphi-\frac{Im(\mathcal{H}^{*}_{\parallel}\mathcal{H}_{\perp})}{2\sqrt{2}}\sin^{2}\theta_{1}\sin^{2}\theta_{2}\sin2\varphi
	\end{aligned}
    \end{equation}
where $\theta_{1}$ and $\theta_{2}$ represent the two polar angles in the strong decays $V_{1}\to P_{1}P_{2}$ and $V_{2}\to P_{3}P_{4}$ respectively, and $\varphi$ is the angle between the two decay planes, similar to Fig.\ref{fig} but with the final products replaced by $P_{1,2,3,4}$. In terms of final momentum correlations, $\sin\varphi$ is essentially equivalent to $\psi$, while another $\mathcal{T}$-odd correlation is associated with $\sin2\varphi$, as discussed in \cite{Geng:2021lrc}.

It is worth noting that both of the illustrated $\mathcal{T}$-odd correlations are closely connected to the imaginary part of interference terms of helicity amplitudes. We will demonstrate it as a general result based on the anti-unitary property of the time reversal operator. Furthermore, if a CP asymmetry observable is formulated using the definition in \eqref{2.26} and \eqref{2.27}, we will show that the quantity $a^{\mathcal{T}-odd}_{CP}$ is always proportional to the strong phase difference in the form of $\cos\Delta\delta$, subject to the fulfillment of following two conditions
\begin{itemize}
    \item In the Hilbert space of the final states of a physical process that we are interested, with a properly chosen basis \{$|\psi_n\rangle$, n =1,2,...\}, there exists a unitary transformation $\mathcal{U}$ that transforms $\mathcal{T}|\psi_n\rangle$ back to $|\psi_n \rangle$ up to a universal phase factor, {\it i.e.}, $\mathcal{U}\mathcal{T}|\psi_n\rangle  = e^{i\alpha}|\psi_n\rangle$; 
    \item $Q_-$ is conserved under this unitary transformation, {\it i.e.} $\mathcal{U}Q_-\mathcal{U}^\dagger = Q_-$.
\end{itemize}
The proof is as follows. Firstly, we decompose the final state in the helicity representation as
\begin{equation}
	\begin{aligned}
	 \ket{f}=\sum_{\lambda}h_{\lambda}\ket{\hat{p},J,M,\lambda}
	\end{aligned}
\end{equation}
where $\lambda$ represents all possible sets of final helicity combinations. For instance, $\lambda$ is $\left\{+\frac{1}{2},-\frac{1}{2}\right\}$ for the decay $\Lambda\to p\pi$, and $\left\{(+\frac{1}{2},+1),(-\frac{1}{2},-1),(+\frac{1}{2},0),(-\frac{1}{2},0)\right\}$ for $\Lambda_{b}\to pK^{*}$. The coefficients $h_{\lambda}$ are essentially helicity amplitudes that quantify the probability of final states being in the helicity state $\ket{\hat{p},J,M,\lambda}$. The expectation value $\langle \mathcal{O}_{-}\rangle$ is finally given by
\begin{equation}\label{eq5.22}
	\begin{aligned}
	 \bra{f}\mathcal{O}_{-}\ket{f}=\sum_{\lambda^{\prime},\lambda}h^{*}_{\lambda^{\prime}}h_{\lambda}\bra{\hat{p},J,M,\lambda^{\prime}}\mathcal{O}_{-}\ket{\hat{p},J,M,\lambda}
	\end{aligned}
\end{equation}
Perform the time reversal transformation
\begin{equation}
	\begin{aligned}
    \bra{\hat{p},J,M,\lambda^{\prime}}\mathcal{O}_{-}\ket{\hat{p},J,M,\lambda}&=\bra{\mathcal{T};\hat{p},J,M,\lambda^{\prime}}\mathcal{T}\mathcal{O}_{-}\mathcal{T}^{-1}\mathcal{T}\ket{\hat{p},J,M,\lambda}^{*}\\
    &=-\bra{-\hat{p},J,-M,\lambda^{\prime}}\mathcal{O}_{-}\ket{-\hat{p},J,-M,\lambda}^{*}\\
    &=-\bra{\hat{p},J,M,\lambda^{\prime}}\mathcal{O}_{-}\ket{\hat{p},J,M,\lambda}^{*}\\
	\end{aligned}
\end{equation}
Thus, it is purely imaginary. In the final step, we utilize rotational invariance, as $\mathcal{O}_{-}$ is a scalar under $SO(3)$. Combining this with hermiticity, we obtain
\begin{equation}
	\begin{aligned}
\bra{\hat{p},J,M,\lambda^{\prime}}\mathcal{O}_{-}\ket{\hat{p},J,M,\lambda}
&=-\bra{\hat{p},J,M,\lambda}\mathcal{O}_{-}\ket{\hat{p},J,M,\lambda^{\prime}}\\
	\end{aligned}
\end{equation}
Substituting above two equations into (\ref{eq5.22}), one arrives at
\begin{equation}
	\begin{aligned}
	 \bra{f}\mathcal{O}_{-}\ket{f}&=\sum_{\lambda^{\prime},\lambda}h_{\lambda^{\prime}}h^{*}_{\lambda}\bra{\hat{p},J,M,\lambda}\mathcal{O}_{-}\ket{\hat{p},J,M,\lambda^{\prime}}\\
	 &=-\sum_{\lambda^{\prime},\lambda}h^{*}_{\lambda^{\prime}}h_{\lambda}\bra{\hat{p},J,M,\lambda}\mathcal{O}_{-}\ket{\hat{p},J,M,\lambda^{\prime}}\\
	 &=\frac{1}{2}\sum_{\lambda^{\prime},\lambda}\left\{h_{\lambda^{\prime}}h^{*}_{\lambda}-h^{*}_{\lambda^{\prime}}h_{\lambda}\right\}\bra{\hat{p},J,M,\lambda}\mathcal{O}_{-}\ket{\hat{p},J,M,\lambda^{\prime}}\\
	 &\propto \sum_{\lambda^{\prime},\lambda}\mathcal{I}m(h_{\lambda^{\prime}}h^{*}_{\lambda})\propto \sum_{\omega^{\prime},\omega}\mathcal{I}m(h_{\omega^{\prime}}h^{*}_{\omega})
	\end{aligned}
\end{equation}
This is the first result we aimed to establish. Here, $\omega$ in the final line denotes parity, and $h_{\omega}$ is used to denote the partial wave amplitudes with parity $\omega$, given that the distinction between helicity and partial representations is a linear matrix. We consider two possible scenarios: (1) If $T$ is also a parity-odd, then its expectation value or asymmetry parameter $\Delta_{t}$ will solely involve interference terms from partial waves with opposite parity. Consequently, it is a generalization of the asymmetry parameter $\beta$. One can demonstrate that the associated CP-violating asymmetry defined in \eqref{2.26} is proportional to the cosine of the strong phases. (2) If $T$ is $\mathcal{T}$-odd but parity-even, then it comprises interference terms solely arising from waves with the same parity. Nevertheless, one can show that the CP-violating quantity defined in \eqref{2.27} is also proportional to the cosine of the strong phases, as the negative sign from Parity will be compensated by that from charge conjugation, as indicated by the partial wave amplitudes in Eq. \eqref{eq:S and P wave1}.

\subsection{Complementary observables}
In the previous subsection, we provided an exact proof that $\mathcal{T}$-odd asymmetry parameter must be dependent on the imaginary part of amplitude interference terms under two conditions. Here, we will derive their associated CP asymmetries and strong phase dependencies by revealing that they are all proportional to the cosine of the strong phases. A natural idea raises whether one can find a pair of complementary CPV observables that are dependent on strong phases as cosine and sine respectively, hence, it may enhance the probability to search for CPV in the decay involving baryons. To achieve this, a critical question should be clarified that how to determine whether two observations are complementary or not. Here we will give some criteria under the prescription of partial wave and helicity treatment respectively. It is easy to imagine that the different properties of complementarity appears in the different scheme since the definition of strong phase changes. Firstly, let us investigate the most simple case that only asymmetry parameters $\alpha,\beta,\gamma$ are involved, and then generalize it to the general cases.
\subsubsection{Complementary $a^{\alpha}_{CP},a^{\beta}_{CP},a^{\gamma}_{CP}$}
The asymmetry parameters $\alpha$, $\beta$, and $\gamma$ are defined and discussed in the partial wave and helicity schemes in Sec.\ref{sec.3}. Analogous to the parameterization of partial wave amplitudes in \eqref{eq:S and P wave1}, we can parameterize the helicity amplitudes in terms of tree and penguin contributions, through which the complementarity between $a^{\beta}_{CP}$ and $a^{\gamma}_{CP}$ can be established
\begin{equation}
	\begin{aligned}
		\mathcal{H}_{+}&=|\mathcal{H}_{+,t}|e^{i\delta_{+,t}}e^{i\phi_{t}}+|\mathcal{H}_{+,p}|e^{i\delta_{+,p}}e^{i\phi_{p}}\\
		\mathcal{H}_{-}&=|\mathcal{H}_{-,t}|e^{i\delta_{-,t}}e^{i\phi_{t}}+|\mathcal{H}_{-,p}|e^{i\delta_{-,p}}e^{i\phi_{p}}\\
		\bar{\mathcal{H}}_{+}&=|\mathcal{H}^{\prime}_{+,t}|e^{i\delta^{\prime}_{+,t}}e^{-i\phi_{t}}+|\mathcal{H}^{\prime}_{+,p}|e^{i\delta^{\prime}_{+,p}}e^{-i\phi_{p}}\\
		\bar{\mathcal{H}}_{-}&=|\mathcal{H}^{\prime}_{-,t}|e^{i\delta^{\prime}_{-,t}}e^{-i\phi_{t}}+|\mathcal{H}^{\prime}_{-,p}|e^{i\delta^{\prime}_{-,p}}e^{-i\phi_{p}}\\
	\end{aligned}
\end{equation}
where subscripts $t$ and $p$ denote the contributions from tree and penguin amplitudes, and $\delta$ and $\phi$ represent the strong and weak phases, respectively. An important distinction arises in that one cannot simply express the amplitude $\bar{\mathcal{H}}_{\pm}$ by reversing the weak phase of $\mathcal{H}$, as was done for $\bar{S}$ and $\bar{P}$. This is because the helicity amplitude $\mathcal{H}_{\pm}$ includes both vector and axial components simultaneously, and thus does not possess a definite transformation property under parity and charge conjugation. Consequently, the strong phase $\delta^{\prime}_{\pm,t}$ is not necessarily equal to $\delta_{\pm,t}$, nor are $\delta^{\prime}_{\pm,p}$ and $\delta_{\pm,p}$. However, this new set of strong phases, $\delta_{\pm,t}$, $\delta_{\pm,p}$, $\delta^{\prime}_{\pm,t}$, and $\delta^{\prime}_{\pm,p}$, are related to those of the partial waves $S$ and $P$ through the following equations
\begin{equation}\label{B.7}
	\begin{aligned}
	 \mathcal{H}_{+}&=|\mathcal{H}_{+,t}|e^{i\delta_{+,t}}e^{i\phi_{t}}+|\mathcal{H}_{+,p}|e^{i\delta_{+,p}}e^{i\phi_{p}}\\
	 &=\frac{1}{\sqrt{2}}\left\{\left[|S_{t}|e^{i\delta_{s,t}}+|P_{t}|e^{i\delta_{p,t}}\right]e^{i\phi_{t}}+\left[|S_{p}|e^{i\delta_{s,p}}+|P_{p}|e^{i\delta_{p,p}}\right]e^{i\phi_{p}}\right\}\\	 \bar{\mathcal{H}}_{+}&=|\mathcal{H}^{\prime}_{+,t}|e^{i\delta^{\prime}_{+,t}}e^{-i\phi_{t}}+|\mathcal{H}^{\prime}_{+,p}|e^{i\delta^{\prime}_{+,p}}e^{-i\phi_{p}}\\
	 &=\frac{1}{\sqrt{2}}\left\{\left[-|S_{t}|e^{i\delta_{s,t}}+|P_{t}|e^{i\delta_{p,t}}\right]e^{-i\phi_{t}}+\left[-|S_{p}|e^{i\delta_{s,p}}+|P_{p}|e^{i\delta_{p,p}}\right]e^{-i\phi_{p}}\right\}\\
	\end{aligned}
\end{equation}
The similar equations also apply for $\mathcal{H}_{-}$ and $\bar{\mathcal{H}}_{-}$. Consequently, one obtains the following constraints
\begin{equation}\label{446}
	\begin{aligned}
	  |\mathcal{H}_{+,t}|e^{i\delta_{+,t}}&=\frac{1}{\sqrt{2}}\left[|S_{t}|e^{i\delta_{s,t}}+|P_{t}|e^{i\delta_{p,t}}\right]\\
	  |\mathcal{H}_{+,p}|e^{i\delta_{+,p}}&=\frac{1}{\sqrt{2}}\left[|S_{p}|e^{i\delta_{s,p}}+|P_{p}|e^{i\delta_{p,p}}\right]\\	  
	  |\mathcal{H}^{\prime}_{+,t}|e^{i\delta^{\prime}_{+,t}}&=\frac{1}{\sqrt{2}}\left[-|S_{t}|e^{i\delta_{s,t}}+|P_{t}|e^{i\delta_{p,t}}\right]\\
	  |\mathcal{H}^{\prime}_{+,p}|e^{i\delta^{\prime}_{+,p}}&=\frac{1}{\sqrt{2}}\left[-|S_{p}|e^{i\delta_{s,p}}+|P_{p}|e^{i\delta_{p,p}}\right]\\
	\end{aligned}
\end{equation}
One can determine the explicit relationship between the two sets of strong phases defined under helicity and partial waves by solving the equations provided above. It is important to stress again
\begin{equation}
	\begin{aligned}
	 |\mathcal{H}_{+,t}|\neq |\mathcal{H}^{\prime}_{+,t}|,~~	 |\mathcal{H}_{+,p}|\neq |\mathcal{H}^{\prime}_{+,p}|,~~\delta_{s,t}\neq \delta^{\prime}_{s,t},~~\delta_{s,p}\neq \delta^{\prime}_{s,p}
	\end{aligned}
\end{equation}
In other words, it is not that there is no any symmetry implication from Charge Conjugation under the helicity scheme, rather hidden in the complex equations as \eqref{B.7}. This fact will be shown to be crucial for the self-consistency in the subsequent derivation. The definition of CP asymmetries $a^{\alpha,\beta,\gamma}_{CP}$ are modified as that of Eq.\eqref{B24} as discussed before. Specifically, the calculations within the helicity framework are conducted as follows
\begin{equation}
	\begin{aligned}
a^{\alpha}_{CP}&\propto\left[|\mathcal{H}_{+}|^{2}-|\bar{\mathcal{H}}_{+}|^{2}\right]-\left[|\mathcal{H}_{-}|^{2}-|\bar{\mathcal{H}}_{-}|^{2}\right]\\
&=\left[\left||\mathcal{H}_{+,t}|e^{i\delta_{+,t}}e^{i\phi_{t}}+|\mathcal{H}_{+,p}|e^{i\delta_{+,p}}e^{i\phi_{p}}\right|^{2}-\left||\mathcal{H}^{\prime}_{+,t}|e^{i\delta^{\prime}_{+,t}}e^{-i\phi_{t}}+|\mathcal{H}^{\prime}_{+,p}|e^{i\delta^{\prime}_{+,p}}e^{-i\phi_{p}}\right|^{2}\right]\\
&-\left[\left||\mathcal{H}_{-,t}|e^{i\delta_{-,t}}e^{i\phi_{t}}+|\mathcal{H}_{-,p}|e^{i\delta_{-,p}}e^{i\phi_{p}}\right|^{2}-\left||\mathcal{H}^{\prime}_{-,t}|e^{i\delta^{\prime}_{-,t}}e^{-i\phi_{t}}+|\mathcal{H}^{\prime}_{-,p}|e^{i\delta^{\prime}_{-,p}}e^{-i\phi_{p}}\right|^{2}\right]\\
&=\left[|\mathcal{H}_{+,t}|^{2}+|\mathcal{H}_{+,p}|^{2}+2|\mathcal{H}_{+,t}||\mathcal{H}_{+,p}|\mathcal{R}e(e^{i(\delta_{+,p}-\delta_{+,t})}e^{i\Delta\phi})\right]\\
&-\left[|\mathcal{H}^{\prime}_{+,t}|^{2}+|\mathcal{H}^{\prime}_{+,p}|^{2}+2|\mathcal{H}^{\prime}_{+,t}||\mathcal{H}^{\prime}_{+,p}|\mathcal{R}e(e^{i(\delta^{\prime}_{+,p}-\delta^{\prime}_{+,t})}e^{-i\Delta\phi})\right]\\
&-\left[|\mathcal{H}_{-,t}|^{2}+|\mathcal{H}_{-,p}|^{2}+2|\mathcal{H}_{-,t}||\mathcal{H}_{-,p}|\mathcal{R}e(e^{i(\delta_{-,p}-\delta_{-,t})}e^{i\Delta\phi})\right]\\
&+\left[|\mathcal{H}^{\prime}_{-,t}|^{2}+|\mathcal{H}^{\prime}_{-,p}|^{2}+2|\mathcal{H}^{\prime}_{-,t}||\mathcal{H}^{\prime}_{-,p}|\mathcal{R}e(e^{i(\delta^{\prime}_{-,p}-\delta^{\prime}_{-,t})}e^{-i\Delta\phi})\right]
	\end{aligned}
\end{equation}
With the help of equations (\ref{B.7}), or equivalently, the charge conjugation symmetry, one can reduce above fussy expression as
\begin{equation}
	\begin{aligned}
    a^{\alpha}_{CP}&\propto \left[ |\mathcal{H}_{+,t}||\mathcal{H}_{+,p}|\sin(\delta_{+,p}-\delta_{+,t})+|\mathcal{H}_{-,t}||\mathcal{H}_{-,p}|\sin(\delta_{-,p}-\delta_{-,t})\right]\sin\Delta\phi
	\end{aligned}
\end{equation}
This is what we want! It is also consistent with our expectations since only tree and penguin inside $\mathcal{H}_{+}$ and $\mathcal{H}_{-}$ interfere terms contribute to $a^{\alpha}_{CP}$. This structure of $a^{\alpha}_{CP}$ under helicity scheme is similar to $a^{\gamma}_{CP}$ in the partial waves. By repeating the analysis above, we obtain
\begin{equation}
	\begin{aligned}
a^{\beta}_{CP}&\propto \left[r_{+}\cos(\delta_{+,t}-\delta_{-,p})-r_{-}\cos(\delta_{+,p}-\delta_{-,t})\right]\sin\Delta\phi
	\end{aligned}
\end{equation}
and
\begin{equation}
	\begin{aligned}
a^{\gamma}_{CP}&\propto \left[r_{+}\sin(\delta_{+,t}-\delta_{-,p})-r_{-}\sin(\delta_{+,p}-\delta_{-,t})\right]\sin\Delta\phi
	\end{aligned}
\end{equation}
where $r_{+}=|\mathcal{H}_{+,t}|/|\mathcal{H}_{+,p}|$ and $r_{-}=|\mathcal{H}_{-,t}|/|\mathcal{H}_{-,p}|$. It is worth noting that the CP asymmetries induced by $\beta$ and $\gamma$ are complementary to each other with respect to the strong phases $\delta_{\pm,t}$ and $\delta_{\pm,p}$ defined by equations (\ref{B.7}). The dependence on $\delta^{\prime}_{\pm,t}$ and $\delta^{\prime}_{\pm,p}$ is not necessary, as they can be expressed as functions of the phases without primes due to the charge conjugation symmetry.

The final conclusion is that different complementary properties arise when different definitions of strong phases are considered. CP-violating observables $a^{\alpha,\beta}_{CP}$ are complementary when the partial wave scheme is utilized, whereas $a^{\beta,\gamma}_{CP}$ exhibit complementarity in the helicity convention. Another implication from the above discussion is that the two complementary observables are $\mathcal{T}$-odd and $\mathcal{T}$-even, respectively. This can be understood, as we will demonstrate that $\mathcal{T}$-odd CPV must depend on the cosine of the strong phase, and a similar argument can be made to show that $\mathcal{T}$-even CPV depends on the sine of the strong phase. A crucial question arises as to how one can determine whether the two $\mathcal{T}$-odd and $\mathcal{T}$-even observables depend on the same set of strong phases, as seen in $a^{\alpha,\beta,\gamma}_{CP}$, and ultimately establish the exact complementarity.

\subsubsection{General criteria and proof}
While we have proven the exact complementarity in CP asymmetries defined by asymmetry parameters, the generalization of it remains an open question. To directly address this question, a more effective approach is to provide the expectation value of these $\mathcal{T}-$odd and -even correlators in a specific Hilbert space, as discussed in the Sec.\eqref{sec5.2}. In this context, we outline the criteria for complementary observables within the partial wave and helicity treatments, respectively. It is important to note that the definition of strong phase varies between the two schemes, resulting in differences in the properties of complementarity.
\begin{itemize}
    \item Criterion 1: If two observables depend on the real and imaginary parts of the same interference term within the partial wave prescription, CP asymmetries associated with them are completely complementary.
\end{itemize}
Proof: The proof is straightforward. One can firstly confirm the Parity of these two observables must be identical, as parity wave amplitudes undergo specific parity transformations. This implies that two observables must depend on different interference terms if their parity is opposite, such as $\mathcal{R}e(S^{*}P)$ and $\mathcal{I}m(S^{*}D)$. Therefore, we only need to consider two cases: parity odd and even, respectively. (1) In the case of Parity odd, the situation is similar to $a^{\alpha}_{CP}$ and $a^{\beta}_{CP}$, which have been previously proven. (2) If two observable are parity even, for example, $\mathcal{R}e(S^{*}D)$ and $\mathcal{I}m(S^{*}D)$, the derivations in $a^{\alpha,\beta}_{CP}$ remain entirely valid since the minus sign from the definition of CPV in Eq.\eqref{2.27} is exactly compensated by the Charge Conjugation as the parameterization of $S,P$ waves in Eq.\eqref{eq:S and P wave1}. More complex scenarios can be expressed as linear combinations of interference terms with the same parity, and the above proof naturally holds true in such cases.
\begin{itemize}
    \item Criterion 2: If two observable depends on the real and imaginary parts of the same interference terms under the helicity amplitude prescription, CP asymmetries associated with them are completely complementary.
\end{itemize}
Proof: Now, let us consider two operators with expectation values given by the helicity amplitudes as
\begin{equation}
	\begin{aligned}    \langle\mathcal{O}_{1}\rangle&=\mathcal{R}e(\mathcal{H}_{\lambda_{i},\lambda_{j}}\mathcal{H}^{*}_{\lambda_{\mu},\lambda_{\nu}}+\mathcal{H}_{-\lambda_{i},-\lambda_{j}}\mathcal{H}^{*}_{-\lambda_{\mu},-\lambda_{\nu}})\\    \langle\mathcal{O}_{2}\rangle&=\mathcal{I}m(\mathcal{H}_{\lambda_{i},\lambda_{j}}\mathcal{H}^{*}_{\lambda_{\mu},\lambda_{\nu}}+\mathcal{H}_{-\lambda_{i},-\lambda_{j}}\mathcal{H}^{*}_{-\lambda_{\mu},-\lambda_{\nu}})\\
	\end{aligned}
\end{equation}
Under the parity transformation, it can be verified that both operators are even. The CP asymmetries induced by $\mathcal{O}_{1}$ and $\mathcal{O}_{2}$ are defined as
\begin{equation}
	\begin{aligned}
    a^{\mathcal{O}_{1}}_{CP}&=\frac{\langle\mathcal{O}_{1}\rangle-\langle\bar{\mathcal{O}}_{1}\rangle}{\langle\Gamma\rangle},~~ a^{\mathcal{O}_{2}}_{CP}&=\frac{\langle\mathcal{O}_{2}\rangle-\langle\bar{\mathcal{O}}_{2}\rangle}{\langle\Gamma\rangle}\\
	\end{aligned}
\end{equation}
Analogy to the Eq. \eqref{B.7}, the helicity amplitude $\mathcal{H}_{\lambda_{i},\lambda_{j}}$ can be decomposed as
\begin{equation}\label{B.14}
	\begin{aligned}
    \mathcal{H}_{\lambda_{i},\lambda_{j}}&=\left|\mathcal{H}^{t}_{i,j}\right|e^{i\phi_{t}}e^{i\delta^{t}_{i,j}}+\left|\mathcal{H}^{p}_{i,j}\right|e^{i\phi_{p}}e^{i\delta^{p}_{i,j}}\\
    &=\sum_{i=1}^{m}\left[S_{i,t}e^{i\delta^{s}_{i,t}}e^{i\phi_{t}}+S_{i,p}e^{i\delta^{s}_{i,p}}e^{i\phi_{p}}\right]+\sum_{j=1}^{n}\left[P_{j,t}e^{i\delta^{p}_{j,t}}e^{i\phi_{t}}+P_{j,p}e^{i\delta^{p}_{j,p}}e^{i\phi_{p}}\right]\\
    \bar{\mathcal{H}}_{\lambda_{i},\lambda_{j}}&=\left|\mathcal{H}^{\prime t}_{i,j}\right|e^{-i\phi_{t}}e^{i\delta^{\prime t}_{i,j}}+\left|\mathcal{H}^{\prime p}_{i,j}\right|e^{-i\phi_{p}}e^{i\delta^{\prime p}_{i,j}}\\
    &=-\sum_{i=1}^{m}\left[S_{i,t}e^{i\delta^{s}_{i,t}}e^{-i\phi_{t}}+S_{i,p}e^{i\delta^{s}_{i,p}}e^{-i\phi_{p}}\right]+\sum_{j=1}^{n}\left[P_{j,t}e^{i\delta^{p}_{j,t}}e^{-i\phi_{t}}+P_{j,p}e^{i\delta^{p}_{j,p}}e^{-i\phi_{p}}\right]\\
	\end{aligned}
\end{equation}
where we use the symbol $S_{i}$ to represent all the partial waves with the same Parity as the $S$ wave, such as $S_{0}=S, S_{1}=D$, and $P_{j}$ represents all partial waves with the same Parity as the $P$ wave for simplicity. $\lambda_{i}$ and $\lambda_{j}$ are helicity symbols. A similar decomposition exists for $\mathcal{H}_{-\lambda_{i},-\lambda_{j}}$. Now, let us derive the explicit phase dependence of $a^{\mathcal{O}_{1}}_{CP}$ and $a^{\mathcal{O}_{2}}_{CP}$, respectively.
\begin{equation}
	\begin{aligned}
    a^{\mathcal{O}_{1}}_{CP}&\propto \mathcal{R}e(\mathcal{H}_{\lambda_{i},\lambda_{j}}\mathcal{H}^{*}_{\lambda_{\mu},\lambda_{\nu}}+\mathcal{H}_{-\lambda_{i},-\lambda_{j}}\mathcal{H}^{*}_{-\lambda_{\mu},-\lambda_{\nu}})-\mathcal{R}e(\bar{\mathcal{H}}_{\lambda_{i},\lambda_{j}}\bar{\mathcal{H}}^{*}_{\lambda_{\mu},\lambda_{\nu}}+\bar{\mathcal{H}}_{-\lambda_{i},-\lambda_{j}}\bar{\mathcal{H}}^{*}_{-\lambda_{\mu},-\lambda_{\nu}})\\
    &=\mathcal{R}e\left[\left|\mathcal{H}^{t}_{i,j}\right|\left|\mathcal{H}^{t}_{\mu,\nu}\right|e^{i(\delta^{t}_{i,j}-\delta^{t}_{\mu,\nu})}+\left|\mathcal{H}^{p}_{i,j}\right|\left|\mathcal{H}^{p}_{\mu,\nu}\right|e^{i(\delta^{p}_{i,j}-\delta^{p}_{\mu,\nu})}\right.\\&\left.-\left|\mathcal{H}^{\prime t}_{i,j}\right|\left|\mathcal{H}^{\prime t}_{\mu,\nu}\right|e^{i(\delta^{\prime t}_{i,j}-\delta^{\prime t}_{\mu,\nu})}-\left|\mathcal{H}^{\prime p}_{i,j}\right|\left|\mathcal{H}^{\prime p}_{\mu,\nu}\right|e^{i(\delta^{\prime p}_{i,j}-\delta^{\prime p}_{\mu,\nu})}\right]\\
    &+\mathcal{R}e\left[\left|\mathcal{H}^{t}_{-i,-j}\right|\left|\mathcal{H}^{t}_{-\mu,-\nu}\right|e^{i(\delta^{t}_{-i,-j}-\delta^{t}_{-\mu,-\nu})}+\left|\mathcal{H}^{p}_{-i,-j}\right|\left|\mathcal{H}^{p}_{-\mu,-\nu}\right|e^{i(\delta^{p}_{-i,-j}-\delta^{p}_{-\mu,-\nu})}\right.\\&\left.-\left|\mathcal{H}^{\prime t}_{-i,-j}\right|\left|\mathcal{H}^{\prime t}_{-\mu,-\nu}\right|e^{i(\delta^{\prime t}_{-i,-j}-\delta^{\prime t}_{-\mu,-\nu})}-\left|\mathcal{H}^{\prime p}_{-i,-j}\right|\left|\mathcal{H}^{\prime p}_{-\mu,-\nu}\right|e^{i(\delta^{\prime p}_{-i,-j}-\delta^{\prime p}_{-\mu,-\nu})}\right]\\
    &+\mathcal{R}e\left[\left|\mathcal{H}^{t}_{i,j}\right|\left|\mathcal{H}^{p}_{\mu,\nu}\right|e^{i(\delta^{t}_{i,j}-\delta^{p}_{\mu,\nu})}e^{i\Delta\phi}+\left|\mathcal{H}^{p}_{i,j}\right|\left|\mathcal{H}^{t}_{\mu,\nu}\right|e^{i(\delta^{p}_{i,j}-\delta^{t}_{\mu,\nu})}e^{-i\Delta\phi}\right]\\
    &-\mathcal{R}e\left[\left|\mathcal{H}^{\prime t}_{i,j}\right|\left|\mathcal{H}^{\prime p}_{\mu,\nu}\right|e^{i(\delta^{\prime t}_{i,j}-\delta^{\prime p}_{\mu,\nu})}e^{-i\Delta\phi}+\left|\mathcal{H}^{\prime p}_{i,j}\right|\left|\mathcal{H}^{\prime t}_{\mu,\nu}\right|e^{i(\delta^{\prime p}_{i,j}-\delta^{\prime t}_{\mu,\nu})}e^{i\Delta\phi}\right]\\    &+\mathcal{R}e\left[\left|\mathcal{H}^{t}_{-i,-j}\right|\left|\mathcal{H}^{p}_{-\mu,-\nu}\right|e^{i(\delta^{t}_{-i,-j}-\delta^{p}_{-\mu,-\nu})}e^{i\Delta\phi}+\left|\mathcal{H}^{p}_{-i,-j}\right|\left|\mathcal{H}^{t}_{-\mu,-\nu}\right|e^{i(\delta^{p}_{-i,-j}-\delta^{t}_{-\mu,-\nu})}e^{-i\Delta\phi}\right]\\
    &-\mathcal{R}e\left[\left|\mathcal{H}^{\prime t}_{-i,-j}\right|\left|\mathcal{H}^{\prime p}_{-\mu,-\nu}\right|e^{i(\delta^{\prime t}_{-i,-j}-\delta^{\prime p}_{-\mu,-\nu})}e^{-i\Delta\phi}+\left|\mathcal{H}^{\prime p}_{-i,-j}\right|\left|\mathcal{H}^{\prime t}_{-\mu,-\nu}\right|e^{i(\delta^{\prime p}_{-i,-j}-\delta^{\prime t}_{-\mu,-\nu})}e^{i\Delta\phi}\right]\\
	\end{aligned}
\end{equation}
where the weak phase difference $\Delta\phi$ is defined as $\phi_{t}-\phi_{p}$. By utilizing equations similar to (\ref{B.14}), one can demonstrate that the sum of terms independent of the weak phase vanishes. This allows us to verify the consistency of our analysis, as an observable reflecting CP violation must vanish when the weak phase is zero. Finally, $a^{\mathcal{O}_{1}}_{CP}$ can be simplified as
\begin{equation}
	\begin{aligned}
    a^{\mathcal{O}_{1}}_{CP} \propto& \left\{\left|\mathcal{H}^{t}_{i,j}\right|\left|\mathcal{H}^{p}_{\mu,\nu}\right|\sin(\delta^{t}_{i,j}-\delta^{p}_{\mu,\nu})-\left|\mathcal{H}^{p}_{i,j}\right|\left|\mathcal{H}^{t}_{\mu,\nu}\right|\sin(\delta^{p}_{i,j}-\delta^{t}_{\mu,\nu})\right\}\sin\Delta\phi\\
    &\left\{\left|\mathcal{H}^{t}_{-i,-j}\right|\left|\mathcal{H}^{p}_{-\mu,-\nu}\right|\sin(\delta^{t}_{-i,-j}-\delta^{p}_{-\mu,-\nu})-\left|\mathcal{H}^{p}_{-i,-j}\right|\left|\mathcal{H}^{t}_{-\mu,-\nu}\right|\sin(\delta^{p}_{-i,-j}-\delta^{t}_{-\mu,-\nu})\right\}\sin\Delta\phi
	\end{aligned}
\end{equation}
This is what we want! The CP violating observable is ultimately expressed in terms of the strong phases $\delta_{\pm i,\pm j},\delta_{\pm\mu,\pm\nu}$ defined in the helicity framework. Using the same technique, another CPV observable is 
\begin{equation}
	\begin{aligned}
a^{\mathcal{O}_{2}}_{CP}&\propto\mathcal{I}m\left[\left|\mathcal{H}^{t}_{i,j}\right|\left|\mathcal{H}^{p}_{\mu,\nu}\right|e^{i(\delta^{t}_{i,j}-\delta^{p}_{\mu,\nu})}e^{i\Delta\phi}+\left|\mathcal{H}^{p}_{i,j}\right|\left|\mathcal{H}^{t}_{\mu,\nu}\right|e^{i(\delta^{p}_{i,j}-\delta^{t}_{\mu,\nu})}e^{-i\Delta\phi}\right]\\
    &-\mathcal{I}m\left[\left|\mathcal{H}^{\prime t}_{i,j}\right|\left|\mathcal{H}^{\prime p}_{\mu,\nu}\right|e^{i(\delta^{\prime t}_{i,j}-\delta^{\prime p}_{\mu,\nu})}e^{-i\Delta\phi}+\left|\mathcal{H}^{\prime p}_{i,j}\right|\left|\mathcal{H}^{\prime t}_{\mu,\nu}\right|e^{i(\delta^{\prime p}_{i,j}-\delta^{\prime t}_{\mu,\nu})}e^{i\Delta\phi}\right]\\    
    &+\mathcal{I}m\left[\left|\mathcal{H}^{t}_{-i,-j}\right|\left|\mathcal{H}^{p}_{-\mu,-\nu}\right|e^{i(\delta^{t}_{-i,-j}-\delta^{p}_{-\mu,-\nu})}e^{i\Delta\phi}+\left|\mathcal{H}^{p}_{-i,-j}\right|\left|\mathcal{H}^{t}_{-\mu,-\nu}\right|e^{i(\delta^{p}_{-i,-j}-\delta^{t}_{-\mu,-\nu})}e^{-i\Delta\phi}\right]\\
    &-\mathcal{I}m\left[\left|\mathcal{H}^{\prime t}_{-i,-j}\right|\left|\mathcal{H}^{\prime p}_{-\mu,-\nu}\right|e^{i(\delta^{\prime t}_{-i,-j}-\delta^{\prime p}_{-\mu,-\nu})}e^{-i\Delta\phi}+\left|\mathcal{H}^{\prime p}_{-i,-j}\right|\left|\mathcal{H}^{\prime t}_{-\mu,-\nu}\right|e^{i(\delta^{\prime p}_{-i,-j}-\delta^{\prime t}_{-\mu,-\nu})}e^{i\Delta\phi}\right]\\
	\end{aligned}
\end{equation}
Finally, one obtains a good results that
\begin{equation}
	\begin{aligned}
a^{\mathcal{O}_{2}}_{CP}&\propto \left\{\left|\mathcal{H}^{t}_{i,j}\right|\left|\mathcal{H}^{p}_{\mu,\nu}\right|\cos(\delta^{t}_{i,j}-\delta^{p}_{\mu,\nu})-\left|\mathcal{H}^{p}_{i,j}\right|\left|\mathcal{H}^{t}_{\mu,\nu}\right|\cos(\delta^{p}_{i,j}-\delta^{t}_{\mu,\nu})\right\}\sin\Delta\phi\\
    &\left\{\left|\mathcal{H}^{t}_{-i,-j}\right|\left|\mathcal{H}^{p}_{-\mu,-\nu}\right|\cos(\delta^{t}_{-i,-j}-\delta^{p}_{-\mu,-\nu})-\left|\mathcal{H}^{p}_{-i,-j}\right|\left|\mathcal{H}^{t}_{-\mu,-\nu}\right|\cos(\delta^{p}_{-i,-j}-\delta^{t}_{-\mu,-\nu})\right\}\sin\Delta\phi
	\end{aligned}
\end{equation}
Clearly, the definitions provided here yield the results that $a^{\mathcal{O}_{1}}_{CP}$ and $a^{\mathcal{O}_{2}}_{CP}$ are completely complementary. Furthermore, another pair of observables being Parity-odd are also completely complementary
  \begin{equation}
	\begin{aligned}    \langle\mathcal{O}_{1}\rangle&=\mathcal{R}e(\mathcal{H}_{\lambda_{i},\lambda_{j}}\mathcal{H}^{*}_{\lambda_{\mu},\lambda_{\nu}}-\mathcal{H}_{-\lambda_{i},-\lambda_{j}}\mathcal{H}^{*}_{-\lambda_{\mu},-\lambda_{\nu}})\\    \langle\mathcal{O}_{2}\rangle&=\mathcal{I}m(\mathcal{H}_{\lambda_{i},\lambda_{j}}\mathcal{H}^{*}_{\lambda_{\mu},\lambda_{\nu}}-\mathcal{H}_{-\lambda_{i},-\lambda_{j}}\mathcal{H}^{*}_{-\lambda_{\mu},-\lambda_{\nu}})\\
	\end{aligned}
\end{equation}

\subsection{Differential distribution}
The necessity of angular distribution has been emphasized based on the requirements of experimental measurement. Here, we will consider three types of angular analyses that may be instructive in searching for CP violation in $b$-baryon multibody decays. These include quasi-two-body modes $\Lambda^{0}_{b}\to N^{*}(\frac{3}{2}^{\pm},\frac{1}{2}^{\pm},\lambda_{1})M^{*}(1^{-},\lambda_{2})$ and quasi-three-body modes $\Lambda^{0}_{b}\to p(\lambda_{1})a^{-}_{1}(1260,\lambda_{2})\to p3\pi$, $\Lambda^{0}_{b}\to \Delta^{++}\pi^{-}\pi^{-}\to p3\pi$. In these decay modes, both $\mathcal{T}$-even and $\mathcal{T}$-odd observables are expected, especially when considering the initial b-baryon polarization.

\subsubsection{Quasi-two body modes}
Let's first examine the decay $\Lambda^{0}_{b}\to N^{*}(\frac{3}{2}^{\pm},\frac{1}{2}^{\pm})M^{*}(1^{-})$. Drawing an analogy to the decay $B\to V_{1}V_{2}$, it becomes straightforward to construct $\mathcal{T}$-odd correlation using the spin and momentum variables, as demonstrated in \cite{Geng:2021lrc}
\begin{equation}
	\begin{aligned}
     \mathcal{O}_{1}&=(\vec{S}_{1}\times\vec{S}_{2})\cdot\hat{p}\\
     \mathcal{O}_{2}&=(\vec{S}_{1}\cdot\vec{S}_{2})\mathcal{O}_{1}+\mathcal{O}_{1}(\vec{S}_{1}\cdot\vec{S}_{2})-(\vec{S}_{1}\cdot\hat{p})(\vec{S}_{2}\cdot\hat{p})\mathcal{O}_{1}-\mathcal{O}_{1}(\vec{S}_{1}\cdot\hat{p})(\vec{S}_{2}\cdot\hat{p})
	\end{aligned}
\end{equation}
Rewriting them in terms of ladder operators, we have
\begin{equation}
	\begin{aligned}
     \mathcal{O}_{1}&=\frac{i}{2}(S^{+}_{1}S^{-}_{2}-S^{-}_{1}S^{+}_{2})\\
     \mathcal{O}_{2}&=\frac{i}{2}(S^{+}_{1}S^{+}_{1}S^{-}_{2}S^{-}_{2}-S^{-}_{1}S^{-}_{1}S^{+}_{2}S^{+}_{2})
	\end{aligned}
\end{equation}
The eigenstates of $\mathcal{O}_{1}$ are solved in the helicity representation
\begin{equation}
	\begin{aligned}
      \ket{\lambda=-\frac{1}{2}, t_{1}=0}&=2\ket{\mathcal{H}_{+\frac{3}{2},+1}}+\sqrt{3}\ket{\mathcal{H}_{-\frac{1}{2},-1}}\\
      \ket{\lambda=-\frac{1}{2}, t_{1}=\pm\frac{7}{2}}&=-\sqrt{3}\ket{\mathcal{H}_{+\frac{3}{2},+1}}\pm\sqrt{2}i\ket{\mathcal{H}_{+\frac{1}{2},0}}+2\ket{\mathcal{H}_{-\frac{1}{2},-1}}
	\end{aligned}
\end{equation}
where $\lambda=\lambda_{1}-\lambda_{2}$, hence is equal to the helicity of the initial $\Lambda^{0}_{b}$, and $t_{1}$ represents the eigenvalues of $\mathcal{O}_{1}$. Therefore, the expectation value of $\mathcal{O}_{1}$ is directly proportional to the interference terms $\mathcal{I}m\left(\mathcal{H}_{+\frac{3}{2},+1}\mathcal{H}_{+\frac{1}{2},0}^{*}\right)$ and $\mathcal{I}m\left(\mathcal{H}_{-\frac{1}{2},-1}\mathcal{H}_{+\frac{1}{2},0}^{*}\right)$. The eigenstates with $\lambda=+\frac{1}{2}$ can be obtained through a parity transformation, as the $\mathcal{T}$-odd correlation $\mathcal{O}_{1}$ is also Parity-odd.
\begin{equation}
	\begin{aligned}
      \ket{\lambda=+\frac{1}{2}, t_{1}=0}&=2\ket{\mathcal{H}_{-\frac{3}{2},-1}}+\sqrt{3}\ket{\mathcal{H}_{+\frac{1}{2},+1}}\\
      \ket{\lambda=+\frac{1}{2}, t_{1}=\pm\frac{7}{2}}&=-\sqrt{3}\ket{\mathcal{H}_{-\frac{3}{2},-1}}\pm\sqrt{2}i\ket{\mathcal{H}_{-\frac{1}{2},0}}+2\ket{\mathcal{H}_{+\frac{1}{2},+1}}
	\end{aligned}
\end{equation}
The eigenstates of $\mathcal{O}_{2}$ can also be obtained
\begin{equation}
	\begin{aligned}
      \ket{\lambda=+\frac{1}{2}, t_{2}=\pm2\sqrt{3}}&=\ket{\mathcal{H}_{-\frac{3}{2},-1}}\pm i\ket{\mathcal{H}_{-\frac{1}{2},-1}}\\     
      \ket{\lambda=-\frac{1}{2}, t_{2}=\pm2\sqrt{3}}&=\ket{\mathcal{H}_{+\frac{3}{2},+1}}\pm i\ket{\mathcal{H}_{+\frac{1}{2},+1}}\\
	\end{aligned}
\end{equation}
Hence, the expectation value of $\mathcal{O}_{1}$ is proportional to the interference terms $\mathcal{I}m\left(\mathcal{H}_{+\frac{3}{2},+1}\mathcal{H}_{-\frac{1}{2},-1}^{*}\right)+\mathcal{I}m\left(\mathcal{H}_{+\frac{1}{2},+1}\mathcal{H}_{-\frac{3}{2},-1}^{*}\right)$. In order to completely measure these $\mathcal{T}$-odd correlations, one needs to examine the angular distribution of the decay chain $\Lambda_b^0\to B^*(\frac{3}{2}^{\pm})M^*(1^{-})$, $B^*\to p h$, $M^*\to P_1 P_2$ such as $\Lambda_b^0\to N^*(\frac{3}{2}^{\pm})\rho/K^{*}(1^{-})$, $N^*\to p \pi$, $\rho/K^{*}\to \pi/K,\pi$.

Next, we will discuss the angular distribution under the assumption of no initial $b$-baryon polarization. A detailed angular analysis considering the polarized initial $b$-baryon is provided in the \cite{Durieux:2016nqr}, so the focus of our discussion is not on deriving the angular distribution. Our aim is to investigate $\mathcal{T}$-odd and $\mathcal{T}$-even complementary observables. In the following formulation, the primary weak decay $\Lambda_b^0\to B^*(\frac{3}{2}^{\pm},\lambda_{1})M^*(1^{-},\lambda_{2})$ amplitudes are denoted by $\mathcal{H}_{\lambda_{1},\lambda_{2}}$, with the helicity symbols $\lambda_{1},\lambda_{2}$ representing $N^{*}(\frac{3}{2}^{\pm})$ and $\rho/K^{*}(1^{-})$ respectively. The initial helicity of $\Lambda^{0}_{b}$ is uniquely determined as $\lambda_{1}-\lambda_{2}$ through taking $N^{*}$ momentum as $\vec{z}$ direction. Consequently, we obtain
\begin{equation}\label{eq:angularNstarKstar}
	\begin{aligned}
		\frac{d\Gamma}{d\cos\theta_{R}d\cos\theta_{L}d\phi_{R}} &\propto\left[\sin^{2}\theta_{R}\sin^{2}\theta_L\left(\left|\mathcal{H}_{+1,+\frac{3}{2}}\right|^{2}+\left|\mathcal{H}_{-1,-\frac{3}{2}}\right|^{2}\right)\right.\\&\left.+\sin^{2}\theta_R(\frac{1}{3}+\cos^{2}\theta_L)\left(\left|\mathcal{H}_{+1,+\frac{1}{2}}\right|^{2}+\left|\mathcal{H}_{-1,-\frac{1}{2}}\right|^{2}\right)\right.\\&\left.+2\cos^{2}\theta_R(\frac{1}{3}+\cos^{2}\theta_L)\left(\left|\mathcal{H}_{0,-\frac{1}{2}}\right|^{2}+\left|\mathcal{H}_{0,+\frac{1}{2}}\right|^{2}\right)\right.\\&\left.-\frac{1}{\sqrt{3}}\sin^{2}\theta_R\sin^{2}\theta_L\left\{\mathcal{I}m\left(\mathcal{H}_{+\frac{3}{2},+1}\mathcal{H}_{-\frac{1}{2},-1}^{*}\right)+
		\mathcal{I}m\left(\mathcal{H}_{+\frac{1}{2},+1}\mathcal{H}_{-\frac{3}{2},-1}^{*}\right)\right\}\sin2\varphi\right.\\&\left.
		+\frac{1}{\sqrt{3}}\sin^{2}\theta_R\sin^{2}\theta_L\left\{\mathcal{R}e\left(\mathcal{H}_{+\frac{3}{2},+1}\mathcal{H}_{-\frac{1}{2},-1}^{*}\right)+\mathcal{R}e\left(\mathcal{H}_{+\frac{1}{2},+1}\mathcal{H}_{-\frac{3}{2},-1}^{*}\right)\right\}\cos2\varphi\right.\\&\left.
		-\frac{1}{\sqrt{6}}\sin2\theta_R\sin2\theta_L\left\{\mathcal{I}m\left(\mathcal{H}_{+\frac{3}{2},+1}\mathcal{H}_{+\frac{1}{2},0}^{*}\right)+\mathcal{I}m\left(\mathcal{H}_{-\frac{1}{2},0}\mathcal{H}_{-\frac{3}{2},-1}^{*}\right)\right\}\sin\varphi\right.\\&\left.
		+\frac{1}{\sqrt{6}}\sin2\theta_R\sin2\theta_L\left\{\mathcal{R}e\left(\mathcal{H}_{+\frac{3}{2},+1}\mathcal{H}_{+\frac{1}{2},0}^{*}\right)+\mathcal{R}e\left(\mathcal{H}_{-\frac{1}{2},0}\mathcal{H}_{-\frac{3}{2},-1}^{*}\right)\right\}\cos\varphi\right]
	\end{aligned}
\end{equation}
The kinematics is analogous to that of $B^{0}\to \Lambda^{+}_{c}\bar{\Xi}^{-}_{c}$ as illustrated in Fig. \ref{fig}. $\theta_{R}$ and $\theta_{L}$ represent the polar angles of $K^{*}/\rho$ and $N^{*}(\frac{3}{2}^{\pm})$ decays in their respective frames. The difference in angular distribution between $N^*(3/2^+)\rho$ and $N^*(3/2^-)\rho$ is solely due to an overall factor arising from the subsequent strong decay amplitudes of $N^{*}(3/2^{+})$ and $N^{*}(3/2^{-})$. It will be distinctive provided angular distribution is formulated as rotational functions
\begin{equation}\label{eq:angular1}
	\begin{aligned}
		\frac{d\Gamma}{d\cos\theta_{R}d\cos\theta_{L} d\phi_{R}}\propto \sum_{\lambda_{K^{*}},\lambda^{\prime}_{K^{*}},\lambda_{f},\lambda_{1},\lambda_{2}}&\delta_{\lambda_{1}-\lambda_{K^{*}},\lambda_{2}-\lambda^{\prime}_{K^{*}}}\mathcal{H}_{\lambda_{1},\lambda_{K^{*}}}\mathcal{H}^{*}_{\lambda_{2},\lambda^{\prime}_{K^{*}}}D^{*1}_{\lambda_{K^{*}},0}(\phi_{R},\theta_{R},0)\\
		&D^{1}_{\lambda^{\prime}_{K^{*}},0}(\phi_{R},\theta_{R},0)\left|h^{\prime}\right|^{2}d^{\frac{3}{2}}_{\lambda_{1},\lambda_{f}}(\theta_{L})d^{\frac{3}{2}}_{\lambda_{2},\lambda_{f}}(\theta_{L})\left|h_{\lambda_{f}}\right|^{2}
	\end{aligned}
\end{equation}
where $h_{\lambda_{f}}$ and $h^{\prime}$ are the dynamical amplitudes of the strong decays $N^{*}\to p\pi$ and $K^{*}\to K\pi$ respectively. It follows that $\left|h_{\lambda_{f},0}\right|^{2}=\left|h_{-\lambda_{f},0}\right|^{2}$ due to the conservation of parity in strong decays. The Kronecker delta function $\delta_{\lambda_{1}-\lambda_{K^{*}},\lambda_{2}-\lambda^{\prime}_{K^{*}}}$ expresses the condition of helicity conservation. 

It is crucial to conduct an angular analysis on $\Lambda_{b}\to B^{*}(1/2^{\pm})V $ with $B^{*}(1/2^{\pm})\to ph$ and $V\to P_{1}P_{2}$, as this process may contribute a significant background to the observables in $\Lambda_{b}\to B^{*}(3/2^{\pm})V $ and vice versa. Given the baryon spectrum, there exist numerous excited baryons like $N^{*}(1440),N^{*}(1520),N^{*}(1535)$ that are very close, possibly overlapping due to width effects. Consequently, there is a high likelihood of misidentifying them in experiments. However, they can be clearly distinguished based on the angular distribution, as the cascade decay $\Lambda_{b}\to B^{*}(1/2)V\to p\pi P_{1}P_{2}$ lacks azimuthal $\sin\varphi_{R},\sin2\varphi_{R}$ dependencies. Analogous to Eq. (\ref{eq:angular1}), one obtains
\begin{equation}\label{eq:angular11}
	\begin{aligned}
		\frac{d\Gamma}{d\cos\theta_{R}d\cos\theta_{L} d\phi_{R}}\propto \sum_{\lambda_{V},\lambda^{\prime}_{V},\lambda_{f},\lambda_{1},\lambda_{2}}&\delta_{\lambda_{1}-\lambda_{V},\lambda_{2}-\lambda^{\prime}_{V}}\mathcal{H}_{\lambda_{1},\lambda_{V}}\mathcal{H}^{*}_{\lambda_{2},\lambda^{\prime}_{V}}D^{*1}_{\lambda_{V},0}(\phi_{R},\theta_{R},0)\\
		&D^{1}_{\lambda^{\prime}_{V},0}(\phi_{R},\theta_{R},0)\left|h^{\prime}\right|^{2}d^{\frac{1}{2}}_{\lambda_{1},\lambda_{f}}(\theta_{L})d^{\frac{1}{2}}_{\lambda_{2},\lambda_{f}}(\theta_{L})\left|h_{\lambda_{f}}\right|^{2}
	\end{aligned}
\end{equation}
Explicitly, we obtain
\begin{equation}\label{eq:angular2}
	\begin{aligned}
	\frac{d\Gamma}{d\cos\theta_{R}}\propto \left[(|\mathcal{H}_{+\frac{1}{2},0}|^{2}+|\mathcal{H}_{-\frac{1}{2},0}|^{2})\cos^{2}\theta_{R}+(|\mathcal{H}_{+\frac{1}{2},+1}|^{2}+|\mathcal{H}_{-\frac{1}{2},-1}|^{2})\frac{\sin^{2}\theta_{R}}{2}\right]
	\end{aligned}
\end{equation}
with the same kinematic conventions as those for $N^{*}(3/2^{\pm})M^{*}$. This scenario offers a promising opportunity to extract CPV observables in the decay $\Lambda_b^0\to B^*(\frac{3}{2}^{-})M^*(1^{-})$ with minimal background from $\Lambda_b^0\to B^*(\frac{1}{2}^{\pm})M^*(1^{-})$ thanks to their distinguishable angular distributions.

It is worth mentioning that the distribution given in Eq. (\ref{eq:angular2}) is entirely distinct from Eq. (\ref{eq:angularLb2LzKstar}) and Eq. (\ref{eq:angularNstarKstar}) in terms of azimuthal dependence. In fact, by utilizing the orthonormality of Wigner rotational functions, it can be shown that the discrepancy between the $\Lambda K^{*}$ and $N^{*}(1/2^{\pm})K^{*}$ modes arises from the parity non-conservation in the weak decay $\Lambda\to p\pi$. One can sum over the index $\lambda_{f}$ in Eq. (\ref{eq:angular11}) separately
\begin{equation}\label{eq: summation}
	\begin{aligned}
    \sum_{\lambda_{f}}d^{\frac{1}{2}}_{\lambda_{1},\lambda_{f}}(\theta_{L})d^{*\frac{1}{2}}_{\lambda_{2},\lambda_{f}}(\theta_{L})\left|h_{\lambda_{f}}\right|^{2}=\delta_{\lambda_{1},\lambda_{2}}\left|h_{\lambda_{f}}\right|^{2}
	\end{aligned}
\end{equation}
It is clear that when parity is conserved, the magnitude amplitude $\left|h_{\lambda_{f}}\right|^{2}$ can be factored out as an overall factor, leading to the disappearance of the $\theta_{L}$ angle dependence in the angular distribution after summing over $\lambda_{f}$. Moreover, the azimuth angle $\varphi$ dependence also vanishes for the $N^*(1/2^\pm)K^*$ decay due to the condition $\delta_{\lambda_{1},\lambda_{2}}$ in Eq. (\ref{eq: summation}) and the helicity conservation in Eq. (\ref{eq:angular11}). However, differences arise in cases where parity is violated, leading to $\left|h_{\lambda_{f}}\right|^{2}\neq \left|h_{-\lambda_{f}}\right|^{2}$. In such scenarios, the discussion above is no longer valid, and the term $ \sum_{\lambda_{f}}d^{\frac{1}{2}}_{\lambda_{1},\lambda_{f}}(\theta_{\Lambda})d^{*\frac{1}{2}}_{\lambda_{2},\lambda_{f}}(\theta_{\Lambda})\left|h_{\lambda_{f}}\right|^{2}$ will involve the asymmetry parameter $\alpha$ from the decay $\Lambda\to p\pi$ as in Eq. (\ref{eq:angularLb2LzKstar}). Naturally, the azimuth $\varphi$ dependence does not vanish in such cases. This conclusion is also supported by Eq. (\ref{eq:angularLb2LzKstar}), where $\alpha$ being zero indicates parity conservation, and thus $\cos\varphi_{R}$ and $\sin\varphi_{R}$ dependencies vanish. Additionally, in the decay $\Lambda_{b}\to N^{*}(3/2^{-})\rho(770)\to p\pi\pi\pi$, another source of uncertainty, $\Lambda_{b}\to N^{*}(3/2^{-})f_{0}(500)\to p\pi\pi\pi$, is also suppressed through angular analysis
\begin{equation}
	\begin{aligned}
	\frac{d\Gamma}{d\cos\theta_{R}}\propto (|\mathcal{H}_{+\frac{1}{2}}|^{2}+|\mathcal{H}_{-\frac{1}{2}}|^{2})(9\cos^{2}\theta_{R}-3\sin2\theta_{R}+1)
	\end{aligned}
\end{equation}
Here, $\theta_{R}$ represents the polar angle between the proton momentum in the $N^{*}$ frame and that of the $N^{*}$ in the $b$-baryon frame in the strong decay $N^{*}(3/2^{-})\to p\pi$. It is noteworthy that distinguishing between $N^{*}(3/2^{-})\rho(770)$ and $N^{*}(3/2^{-})f_{0}(500)$ based on their azimuthal dependencies is also available.

As previous discussion, two asymmetry parameters, $\mathcal{A}_{T}(\sin2\varphi_{R})$ and $\mathcal{A}_{T}(\sin\varphi_{R})$, are introduced for the decay process $\Lambda_{b}^{0}\to N^{*}(3/2^{\pm})M^{*}(1^{-})$
\begin{equation}
	\begin{aligned}
	 \mathcal{A}_{T}(\sin2\varphi_{R})&=\mathcal{I}m\left(\mathcal{H}_{+\frac{3}{2},+1}\mathcal{H}_{-\frac{1}{2},-1}^{*}\right)-\mathcal{I}m\left(\mathcal{H}^{*}_{+\frac{1}{2},+1}\mathcal{H}_{-\frac{3}{2},-1}\right)\\
	 \bar{\mathcal{A}}_{T}(\sin2\varphi_{R})&=\mathcal{I}m\left(\bar{\mathcal{H}}_{+\frac{3}{2},+1}\bar{\mathcal{H}}_{-\frac{1}{2},-1}^{*}\right)-\mathcal{I}m\left(\bar{\mathcal{H}}^{*}_{+\frac{1}{2},+1}\bar{\mathcal{H}}_{-\frac{3}{2},-1}\right)\\
	\end{aligned}
\end{equation}
\begin{equation}
	\begin{aligned}
	 \mathcal{A}_{T}(\sin\varphi_{R})&=\mathcal{I}m\left(\mathcal{H}_{+\frac{3}{2},+1}\mathcal{H}_{+\frac{1}{2},0}^{*}\right)-\mathcal{I}m\left(\mathcal{H}^{*}_{-\frac{1}{2},0}\mathcal{H}_{-\frac{3}{2},-1}\right)\\
    \bar{\mathcal{A}}_{T}(\sin\varphi_{R})&= \mathcal{I}m\left(\bar{\mathcal{H}}_{+\frac{3}{2},+1}\bar{\mathcal{H}}_{+\frac{1}{2},0}^{*}\right)-\mathcal{I}m\left(\bar{\mathcal{H}}^{*}_{-\frac{1}{2},0}\bar{\mathcal{H}}_{-\frac{3}{2},-1}\right)\\
	\end{aligned}
\end{equation}
Clearly, both of these parameters are $\mathcal{T}$-odd and Parity-odd quantities as $\beta$. Following the previous discussion, we define
\begin{equation}
	\begin{aligned}
	a_{CP}(\sin\varphi_{R})&={\langle \mathcal{A}_{T}(\sin\varphi_{R})\rangle -\langle({CP}) \mathcal{A}_{T}(\sin\varphi_{R})({CP})^\dag\rangle\over 2\langle \Gamma\rangle}=\frac{\mathcal{A}_{T}(\sin\varphi_{R})+\bar{\mathcal{A}}_{T}(\sin\varphi_{R})}{2\langle\Gamma\rangle}
		\end{aligned}
\end{equation}
\begin{equation}
	\begin{aligned}
	a_{CP}(\sin2\varphi_{R})&={\langle \mathcal{A}_{T}(\sin2\varphi_{R})\rangle -\langle({CP}) \mathcal{A}_{T}(\sin2\varphi_{R})({CP})^\dag\rangle\over 2\langle \Gamma\rangle}=\frac{\mathcal{A}_{T}(\sin2\varphi_{R})+\bar{\mathcal{A}}_{T}(\sin2\varphi_{R})}{2\langle \Gamma\rangle}\\
	\end{aligned}
\end{equation}
These two CP asymmetries should naturally be $\beta$-like. In fact, the helicity amplitude can be expressed as linear combinations of partial waves. Consequently, the asymmetry parameters $\mathcal{A}_{T}(\sin\varphi_{R})$ and $\mathcal{A}_{T}(\sin2\varphi_{R})$ are simplified to the imaginary parts of interference terms involving different partial waves with opposite parity.

In addition to the $\mathcal{T}$-odd observables $\sin2\varphi_{R}$ and $\sin\varphi_{R}$, the $\cos2\varphi_{R}$ and $\cos\varphi_{R}$ terms are also valuable in constructing $\mathcal{T}$-even CP-violating observables. Analogous to $\mathcal{A}_{T}(\sin2\varphi_{R})$ and $\mathcal{A}_{T}(\sin\varphi_{R})$, we can define
\begin{equation}
	\begin{aligned}
	 \mathcal{A}_{T}(\cos2\varphi_{R})&=\mathcal{R}e\left(\mathcal{H}_{+\frac{3}{2},+1}\mathcal{H}_{-\frac{1}{2},-1}^{*}\right)+\mathcal{R}e\left(\mathcal{H}_{+\frac{1}{2},+1}\mathcal{H}_{-\frac{3}{2},-1}^{*}\right)\\
	 \bar{\mathcal{A}}_{T}(\cos2\varphi_{R})&=\mathcal{R}e\left(\bar{\mathcal{H}}_{+\frac{3}{2},+1}\bar{\mathcal{H}}_{-\frac{1}{2},-1}^{*}\right)+\mathcal{R}e\left(\bar{\mathcal{H}}_{+\frac{1}{2},+1}\bar{\mathcal{H}}_{-\frac{3}{2},-1}^{*}\right)\\
	\end{aligned}
\end{equation}
\begin{equation}
	\begin{aligned}
	 \mathcal{A}_{T}(\cos\varphi_{R})&=\mathcal{R}e\left(\mathcal{H}_{+\frac{3}{2},+1}\mathcal{H}_{+\frac{1}{2},0}^{*}\right)+\mathcal{R}e\left(\mathcal{H}_{-\frac{1}{2},0}\mathcal{H}_{-\frac{3}{2},-1}^{*}\right)\\
    \bar{\mathcal{A}}_{T}(\cos\varphi_{R})&=\mathcal{R}e\left(\bar{\mathcal{H}}_{+\frac{3}{2},+1}\bar{\mathcal{H}}_{+\frac{1}{2},0}^{*}\right)+\mathcal{R}e\left(\bar{\mathcal{H}}_{-\frac{1}{2},0}\bar{\mathcal{H}}_{-\frac{3}{2},-1}^{*}\right)\\
	\end{aligned}
\end{equation}
The other two CPV observables are introduced as
\begin{equation}
	\begin{aligned}
	a_{CP}(\cos\varphi_{R})&={\langle \mathcal{A}_{T}(\cos\varphi_{R})\rangle -\langle({CP}) \mathcal{A}_{T}(\cos\varphi_{R})({CP})^\dag\rangle\over 2\langle \Gamma\rangle}=\frac{\mathcal{A}_{T}(\cos\varphi_{R})-\bar{\mathcal{A}}_{T}(\cos\varphi_{R})}{2\langle\Gamma\rangle}
		\end{aligned}
\end{equation}
\begin{equation}
	\begin{aligned}
	a_{CP}(\cos2\varphi_{R})&={\langle \mathcal{A}_{T}(\cos2\varphi_{R})\rangle -\langle({CP}) \mathcal{A}_{T}(\cos2\varphi_{R})({CP})^\dag\rangle\over 2\langle \Gamma\rangle}=\frac{\mathcal{A}_{T}(\cos2\varphi_{R})-\bar{\mathcal{A}}_{T}(\cos2\varphi_{R})}{2\langle\Gamma\rangle}\\
	\end{aligned}
\end{equation}
According to the criterion given in the previous section, CP asymmetries $a_{CP}(\cos\varphi_{R}), a_{CP}(\cos2\varphi_{R})$ are complementary with $a_{CP}(\sin\varphi_{R})$ and $a_{CP}(\sin2\varphi_{R})$. Based on Eq. (\ref{eq:amplitudes}), the helicity amplitudes $\mathcal{H}_{\lambda_{1},\lambda_{2}}$ that appear in Eq. (\ref{eq:angularNstarKstar}) are always expressed as $C_{i}$ and $D_{i}$, which is convenient from a theoretical standpoint. Initially, we separate the helicity amplitude $\mathcal{H}_{\lambda_{1},\lambda_{2}}=\mathcal{H}_{\lambda_{1},\lambda_{2}}(V)+\mathcal{H}_{\lambda_{1},\lambda_{2}}(A)$ into two parts, comprising the vector and axial vector components \cite{Cheng:1996cs}. Under a parity transformation, we have
\begin{equation}\label{eq: parity transformation 1}
	\begin{aligned}
    \mathcal{H}_{\lambda_{1},\lambda_{2}}(V)&=(+)\eta_{b}\eta_{1}\eta_{2}(-1)^{S_{b}-S_{1}-S_{2}}\mathcal{H}_{-\lambda_{1},-\lambda_{2}}(V)=-\eta_{1}\mathcal{H}_{-\lambda_{1},-\lambda_{2}}(V)\\    
    \mathcal{H}_{\lambda_{1},\lambda_{2}}(A)&=(-)\eta_{b}\eta_{1}\eta_{2}(-1)^{S_{b}-S_{1}-S_{2}}\mathcal{H}_{-\lambda_{1},-\lambda_{2}}(A)=+\eta_{1}\mathcal{H}_{-\lambda_{1},-\lambda_{2}}(A)\\
	\end{aligned}
\end{equation}
where $\eta_{b},\eta_{1},\eta_{2}$ are intrinsic parity of $\Lambda^{0}_{b},N^{*}(\frac{3}{2}^{\pm})$ and $V(1^{-})$ involved in weak decays, $S_{b},S_{1},S_{2}$ are spin of them, respectively. Hence, we obtain
\begin{equation}
	\begin{aligned}
     \mathcal{H}_{\lambda_{1},\lambda_{2}}(V)&=\frac{1}{2}\left[\mathcal{H}_{\lambda_{1},\lambda_{2}}\pm\mathcal{H}_{-\lambda_{1},-\lambda_{2}}\right]\\     \mathcal{H}_{\lambda_{1},\lambda_{2}}(A)&=\frac{1}{2}\left[\mathcal{H}_{\lambda_{1},\lambda_{2}}\mp\mathcal{H}_{-\lambda_{1},-\lambda_{2}}\right]\\
	\end{aligned}
\end{equation}
for $\eta_{1}=\mp$. Explicitly, $V$ and $A$ amplitude are derived as
\begin{equation}
	\begin{aligned}
    \mathcal{H}_{+\frac{3}{2},+1}(V/A)=2\left\{\begin{array}{c}
    		                          -\sqrt{Q_{+}}C_{1} \\\\
    		                           \sqrt{Q_{-}}D_{1}
                                              \end{array}
                                        \right\}
	\end{aligned}
\end{equation}
\begin{equation}
	\begin{aligned}
    \mathcal{H}_{-\frac{1}{2},-1}(V/A)=\frac{2}{\sqrt{3}}\left\{\begin{array}{c}
    		   -\sqrt{Q_{+}}\left[C_{1}-\frac{2Q_{-}}{M_{i}M_{f}}C_{2}\right] \\\\
    		   \sqrt{Q_{-}}\left[D_{1}-\frac{2Q_{+}}{M_{i}M_{f}}D_{2}\right]
                                              \end{array}
                                        \right\}
	\end{aligned}
\end{equation}
\begin{equation}
	\begin{aligned}
    \mathcal{H}_{+\frac{1}{2},0}(V/A)=\frac{2\sqrt{2}}{\sqrt{3}M_{f}M_{V}}\left\{\begin{array}{c}
    		   -\sqrt{Q_{+}}\left[\frac{1}{2}(M^{2}_{i}-M^{2}_{f}-M^{2}_{V})C_{1}-\frac{Q_{-}(M_{i}+M_{f})}{M_{i}}C_{2}+p^{2}_{c}C_{3}\right] \\\\
    		   \sqrt{Q_{-}}\left[\frac{1}{2}(M^{2}_{i}-M^{2}_{f}-M^{2}_{V})D_{1}-\frac{Q_{+}(M_{i}-M_{f})}{M_{i}}D_{2}+p^{2}_{c}D_{3}\right]
                                              \end{array}
                                        \right\}
	\end{aligned}
\end{equation}
where the upper (lower) entry is for $V$ ($A$) helicity amplitude, $p_{c}$ is the c.m. momentum. Above results are general for decay $\frac{1}{2}^{+}[M_{i}]\to \frac{3}{2}^{\pm}[M_{f}]V[M_{V}]$, for simplicity, here we suppress the superscript of $C^{\pm}_{i},D^{\pm}_{i}$, and $Q_{\pm}$ is given by
\begin{equation}
	\begin{aligned}
      Q_{\pm}=(M_{i}\pm M_{f})^{2}-M^{2}_{V}=2M_{i}(E_{f}\pm M_{f})
	\end{aligned}
\end{equation}
The helicity amplitudes $\mathcal{H}_{\lambda_{1},\lambda_{2}}$ appearing in asymmetry parameters are 
\begin{equation}
	\begin{aligned}
     \mathcal{H}_{\lambda_{1},\lambda_{2}}&=\frac{1}{2}\left[\mathcal{H}_{\lambda_{1},\lambda_{2}}(V)+\mathcal{H}_{\lambda_{1},\lambda_{2}}(A)\right]\\    
     \mathcal{H}_{-\lambda_{1},-\lambda_{2}}&=-\frac{1}{2}\eta_{1}\left[\mathcal{H}_{\lambda_{1},\lambda_{2}}(V)-\mathcal{H}_{\lambda_{1},\lambda_{2}}(A)\right]\\
	\end{aligned}
\end{equation}
As an example, for $N^{*}(1520)$
\begin{equation}
	\begin{aligned}
     \mathcal{H}_{+\frac{3}{2},+1}=-\sqrt{Q_{+}}C_{1}+\sqrt{Q_{-}}D_{1},~~\mathcal{H}_{-\frac{3}{2},-1}=-\sqrt{Q_{+}}C_{1}-\sqrt{Q_{-}}D_{1}
	\end{aligned}
\end{equation}
For convenience, let's define some dimensionless quantities
\begin{equation}
	\begin{aligned}
     \mathcal{Y}_{1}&=\frac{2Q_{-}}{M_{i}M_{f}},~~~~~~~~~~~~\mathcal{Y}_{2}=\frac{2Q_{+}}{M_{i}M_{f}},~~~~~~~~~~~~~\mathcal{Y}_{3}=\frac{M^{2}_{i}-M^{2}_{f}-M^{2}_{V}}{2M_{V}M_{f}}\\
     \mathcal{Y}_{4}&=\frac{Q_{-}(M_{i}+M_{f})}{M_{i}M_{V}M_{f}},~~\mathcal{Y}_{5}=\frac{Q_{+}(M_{i}-M_{f})}{M_{i}M_{V}M_{f}},~~\mathcal{Y}_{6}=\frac{p^{2}_{c}}{M_{V}M_{f}}\\
	\end{aligned}
\end{equation}
By leveraging these simplifications, we can determine the dependence of $a_{CP}(\sin\varphi_{R})$ and $a_{CP}(\sin2\varphi_{R})$ on $C_{i}$ and $D_{i}$
\begin{equation}
	\begin{aligned}
     a_{CP}(\sin2\varphi_{R})\propto\frac{2}{\sqrt{3}}\sqrt{Q_{+}Q_{-}}\left[\mathcal{Y}_{2}\mathcal{I}m\left(\bar{C}_{1}\bar{D}^{*}_{2}+C_{1}D^{*}_{2}\right)+\mathcal{Y}_{1}\mathcal{I}m\left(\bar{C}_{2}^{*}\bar{D}_{1}+C_{2}^{*}D_{1}\right)\right]
	\end{aligned}
\end{equation}
 \begin{equation}
	\begin{aligned}
		a_{CP}(\sin\varphi_{R})\propto&\frac{2\sqrt{2}}{\sqrt{3}}\sqrt{Q_{+}Q_{-}}\left[\mathcal{Y}_{4}\mathcal{I}m\left(C^{*}_{2}D_{1}+\bar{C}^{*}_{2}\bar{D}_{1}\right)+\mathcal{Y}_{5} \mathcal{I}m\left(C_{1}D^{*}_{2}+\bar{C}_{1}\bar{D}^{*}_{2}\right)\right.\\&\left.
		-\mathcal{Y}_{6}\mathcal{I}m\left(C^{*}_{3}D_{1}+\bar{C}^{*}_{3}\bar{D}_{1}\right)-\mathcal{Y}_{6} \mathcal{I}m\left(C_{1}D^{*}_{3}+\bar{C}_{1}\bar{D}^{*}_{3}\right)\right]
	\end{aligned}
\end{equation}
When compared with the Lee-Yang asymmetry parameter $\beta$, $C_{i}$ and $D_{i}$ exhibit similar properties to $S$ and $P$ under parity transformations. It may be more natural, perhaps, to decompose $C_{i}$ and $D_{i}$ into tree and penguin contributions, particularly for those interested in performing specific QCD calculations on them. Drawing an analogy to the partial wave discussion earlier (\ref{eq:S and P wave1}), we have
\begin{equation}\label{eq: C and D}
	\begin{aligned}
	 C_{j}&=C_{j,t}e^{i\delta^{C}_{j,t}}e^{i\phi_{t}}+C_{j,p}e^{i\delta^{C}_{j,p}}e^{i\phi_{p}}\\	 
	 D_{j}&=D_{j,t}e^{i\delta^{D}_{j,t}}e^{i\phi_{t}}+D_{j,p}e^{i\delta^{D}_{j,p}}e^{i\phi_{p}}\\
	 \bar{C}_{j}&=-\left[C_{j,t}e^{i\delta^{C}_{j,t}}e^{-i\phi_{t}}+C_{j,p}e^{i\delta^{C}_{j,p}}e^{-i\phi_{p}}\right]\\	 
	 \bar{D}_{j}&=D_{j,t}e^{i\delta^{D}_{j,t}}e^{-i\phi_{t}}+D_{j,p}e^{i\delta^{D}_{j,p}}e^{-i\phi_{p}}\\
	\end{aligned}
\end{equation}

\begin{equation}
	\begin{aligned}
		a_{CP}(\sin 2\varphi_{R})
		&=\frac{4}{\sqrt{3}}\sqrt{Q_{+}Q_{-}} \sin\Delta\phi\left[\mathcal{Y}_{2}C_{1,t}D_{2,p}\cos(\delta^{C}_{1,t}-\delta^{D}_{2,p})-\mathcal{Y}_{2}C_{1,p}D_{2,t}\cos(\delta^{C}_{1,p}-\delta^{D}_{2,t})\right.\\&\left.+\mathcal{Y}_{1}C_{2,p}D_{1,t}\cos(\delta^{D}_{1,t}-\delta^{C}_{2,p})-\mathcal{Y}_{1}C_{2,t}D_{1,p}\cos(\delta^{D}_{1,p}-\delta^{C}_{2,t})\right]
	\end{aligned}
\end{equation}
\begin{equation}
	\begin{aligned}
		a_{CP}(\sin \varphi_{R})&=\frac{4\sqrt{2}}{\sqrt{3}}\sqrt{Q_{+}Q_{-}}\sin\Delta\phi\left[\mathcal{Y}_{4}C_{2,p}D_{1,t}\cos(\delta^{D}_{1,t}-\delta^{C}_{2,p})-\mathcal{Y}_{4}C_{2,t}D_{1,p}\cos(\delta^{D}_{1,p}-\delta^{C}_{2,t})\right.\\&\left.+\mathcal{Y}_{5}C_{1,t}D_{2,p}\cos(\delta^{C}_{1,t}-\delta^{D}_{2,p})-\mathcal{Y}_{5}C_{1,p}D_{2,t}\cos(\delta^{C}_{1,p}-\delta^{D}_{2,t})\right.\\&\left.-\mathcal{Y}_{6}C_{3,p}D_{1,t}\cos(\delta^{D}_{1,t}-\delta^{C}_{3,p})+\mathcal{Y}_{6}C_{3,t}D_{1,p}\cos(\delta^{D}_{1,p}-\delta^{C}_{3,t})\right.\\&\left.-\mathcal{Y}_{6}C_{1,t}D_{3,p}\cos(\delta^{C}_{1,t}-\delta^{D}_{3,p})+\mathcal{Y}_{6}C_{1,p}D_{3,t}\cos(\delta^{C}_{1,p}-\delta^{D}_{3,t})\right]
	\end{aligned}
\end{equation}
Here, $\Delta\phi$ represents the weak phase difference between the tree and penguin amplitudes, given by $\phi_{t}-\phi_{p}$. The strong phase difference is often unknown. $C_{i,t(p)}$ and $D_{i,t(p)}$ denote the magnitudes of the tree (penguin) contributions to $C$ and $D$. Similarly, the expression for $a_{CP}(\cos2\varphi),a_{CP}(\cos\varphi)$ are obtained
\begin{equation}
	\begin{aligned}
	 a_{CP}(\cos2\varphi_{R})
	 &\propto \frac{2}{\sqrt{3}}\mathcal{R}e\left[Q_{+}\left(|C_{1}|^{2}-\mathcal{Y}_{1}C_{1}C^{*}_{2}\right)+Q_{-}\left(|D_{1}|^{2}-\mathcal{Y}_{2}D_{1}D^{*}_{2}\right)\right]\\
	 &-\frac{2}{\sqrt{3}}\mathcal{R}e\left[Q_{+}\left(|\bar{C}_{1}|^{2}-\mathcal{Y}_{1}\bar{C}_{1}\bar{C}^{*}_{2}\right)+Q_{-}\left(|\bar{D}_{1}|^{2}-\mathcal{Y}_{2}\bar{D}_{1}\bar{D}^{*}_{2}\right)\right]
	\end{aligned}
\end{equation}
\begin{equation}
	\begin{aligned}
	 a_{CP}(cos\varphi_{R})
	 &\propto \sqrt{\frac{2}{3}}\mathcal{R}e\left[Q_{+}(\mathcal{Y}_{3}|C_{1}|^{2}-\mathcal{Y}_{4}C_{1}C^{*}_{2}+\mathcal{Y}_{6}C_{1}C^{*}_{3})\right]\\&+\sqrt{\frac{2}{3}}\mathcal{R}e\left[Q_{+}(\mathcal{Y}_{3}|D_{1}|^{2}-\mathcal{Y}_{5}D_{1}D^{*}_{2}+\mathcal{Y}_{6}D_{1}D^{*}_{3})\right]\\
	 &-\sqrt{\frac{2}{3}}\mathcal{R}e\left[Q_{+}(\mathcal{Y}_{3}|\bar{C}_{1}|^{2}-\mathcal{Y}_{4}\bar{C}_{1}\bar{C}^{*}_{2}+\mathcal{Y}_{6}\bar{C}_{1}\bar{C}^{*}_{3})\right]\\&-\sqrt{\frac{2}{3}}\mathcal{R}e\left[Q_{+}(\mathcal{Y}_{3}|\bar{D}_{1}|^{2}-\mathcal{Y}_{5}\bar{D}_{1}\bar{D}^{*}_{2}+\mathcal{Y}_{6}\bar{D}_{1}\bar{D}^{*}_{3})\right]
	\end{aligned}
\end{equation}
The presence of $|C_{1}|^{2}-|\bar{C}_{1}|^{2}$ in $a_{CP}(\cos\varphi_{R})$ and $a_{CP}(\cos2\varphi_{R})$ implies direct CP violation, where the strong phase dependence takes the form of a sine function. Specifically, we obtain
\begin{equation}
	\begin{aligned}
	 a_{CP}(\cos2\varphi_{R})&\propto \frac{2}{\sqrt{3}}[\Delta(C_{1},D_{1})-\mathcal{Y}_{1} \mathcal{R}e(C_{1},C_{2})-\mathcal{Y}_{2}\mathcal{R}e(D_{1},D_{2})]\sin\Delta\phi
	\end{aligned}
\end{equation}
\begin{equation}
	\begin{aligned}
	 a_{CP}(\cos\varphi_{R})\propto \frac{2}{\sqrt{3}}&\left[\mathcal{Y}_{3}\Delta(C_{1},D_{1})-\mathcal{Y}_{4}\mathcal{R}e(C_{1},C_{2})-\mathcal{Y}_{5}\mathcal{R}e(D_{1},D_{2})+\right.\\&\left.\mathcal{Y}_{6}\mathcal{R}e(C_{1},C_{3})+\mathcal{Y}_{6}\mathcal{R}e(D_{1},D_{3})\right]\sin\Delta\phi
	\end{aligned}
\end{equation}
For simplicity, we define 
\begin{equation}
	\begin{aligned}
	\Delta(C_{1},D_{1})&=4Q_{+}C_{1,t}C_{1,p}\sin(\delta^{C}_{1,t}-\delta^{C}_{1,p})+4Q_{-}D_{1,t}D_{1,p}\sin(\delta^{D}_{1,t}-\delta^{D}_{1,p})
	\end{aligned}
\end{equation}
\begin{equation}
	\begin{aligned}
	\mathcal{R}e(C_{i},C_{j})&=-2\left[C_{i,t}C_{j,p}\sin(\delta^{C}_{i,t}-\delta^{C}_{j,p})-C_{i,p}C_{j,t}\sin(\delta^{C}_{i,p}-\delta^{C}_{j,t})\right]\\
	\mathcal{R}e(D_{i},D_{j})&=-2\left[D_{i,t}D_{j,p}\sin(\delta^{D}_{i,t}-\delta^{D}_{j,p})-D_{i,p}D_{j,t}\sin(\delta^{D}_{i,p}-\delta^{D}_{j,t})\right]\\
	\end{aligned}
\end{equation}

\begin{figure}[tbph]
\centering
\includegraphics[width=5cm]{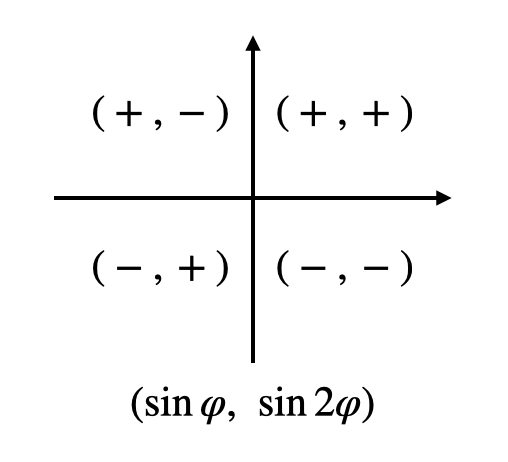}
\caption{The quadrant diagram of $(\sin\phi_{R}, \sin2\phi_{R})$.}\label{fig-quadrant}
\end{figure}

As discussed before, all spin and polarization vectors in the $\mathcal{T}-$odd correlations are ultimately transformed as momentum correlations for the requirement of experimental measurements. In a quasi-two-body decay chain of $A\to ab$, $a\to 1 2$,  $b\to 3 4$, $\varphi_{R}$ is the angle between the planes of $12$ and $34$, as depicted in following Figure. One could view $a,b$ as $\Lambda^{+}_{c}$ and $\bar{\Xi}^{-}_{c}$ for $B\to\Lambda^{+}_{c}\bar{\Xi}^{-}_{c} $ decay, and $N^{*}V$ for the decay $\Lambda^{0}_{b}\to N^{*}V$ with $N^{*}(\frac{3}{2}^{\pm})$, $V=K^{*},\rho...$. Define $\vec n_a=\vec p_1\times \vec p_2/|\vec p_1\times \vec p_2|$, $\vec n_b=\vec p_3\times \vec p_4/|\vec p_3\times \vec p_4|$, $\hat p_i=\vec p_i/|\vec p_i|$.
\begin{figure}[h!]
\centering
\includegraphics[width=0.65\textwidth]{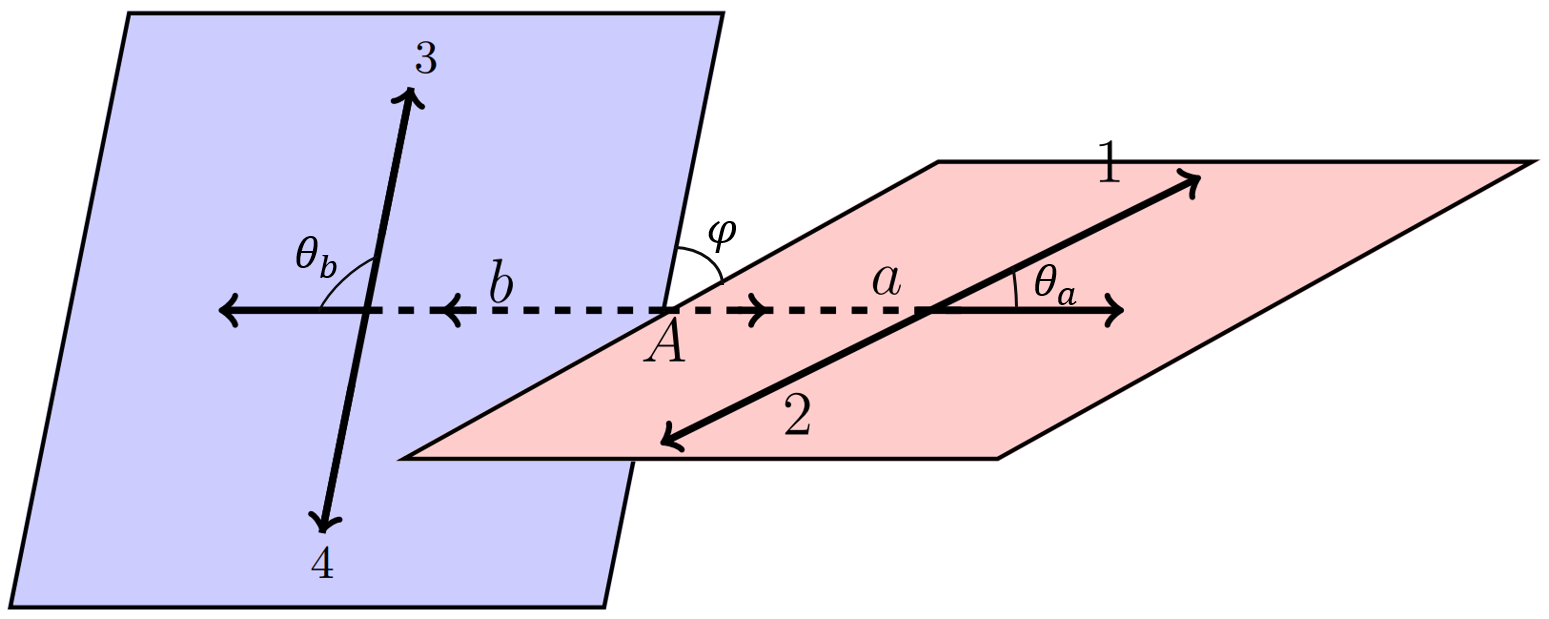}
\caption{Decay planes in the rest frame of mother particle.}\label{fig-planes}
\end{figure}
In the rest frame of the mother particle, $\vec p_1+\vec p_2+\vec p_3+\vec p_4=0$.
\begin{align}\label{eq:sinphi-TPA}
\sin\varphi_{R}&=(\vec n_a\times \vec n_b)\cdot \hat p_b=\vec n_a\cdot (\vec n_b\times \hat p_b)
={1\over |\vec p_1\times \vec p_2||\vec p_3\times\vec p_4||\vec p_b|}(\vec p_1\times \vec p_2)\cdot [(\vec p_3\times \vec p_4)\times\vec p_b]\nonumber\\
&={1\over |\vec p_1\times \vec p_2||\vec p_3\times\vec p_4||\vec p_b|}(\vec p_1\times \vec p_2)\cdot [(\vec p_3\cdot \vec p_b)\vec p_4-(\vec p_4\cdot \vec p_b)\vec p_3]\nonumber\\
&={|\vec p_b|\over |\vec p_1\times \vec p_2||\vec p_3\times\vec p_4|}(\vec p_1\times \vec p_2)\cdot \vec p_4
\end{align}
since ${|\vec p_b|/ |\vec p_1\times \vec p_2||\vec p_3\times\vec p_4|}$ is a positive quantity, 
$\sin\varphi_{R}$ is therefore equivalent to the momentum triple-product $(\vec p_1\times \vec p_2)\cdot \vec p_4$.
However, $\sin2\varphi_{R}$ is different from the momentum triple-products. 
\begin{align}
\sin2\varphi_{R}&=2\cos\varphi_{R}\sin\varphi_{R}=2(\vec n_a\cdot \vec n_b)(\vec n_a\times \vec n_b)\cdot \hat p_b
\nonumber\\
&={2|\vec p_b|\over |\vec p_1\times \vec p_2|^2|\vec p_3\times\vec p_4|^2}[(\vec p_1\times \vec p_2)\cdot (\vec p_3\times \vec p_4)][(\vec p_1\times \vec p_2)\cdot \vec p_4]
\end{align}
From the quadrant diagram of $(\sin\varphi_{R}, \sin2\varphi_{R})$ in Fig. \ref{fig-quadrant}, it is clear that $\sin2\varphi_{R}$ is different from $\sin\varphi_{R}$, thereby different from the momentum triple product.

\begin{figure}[tbph]
\centering
\includegraphics[width=4cm]{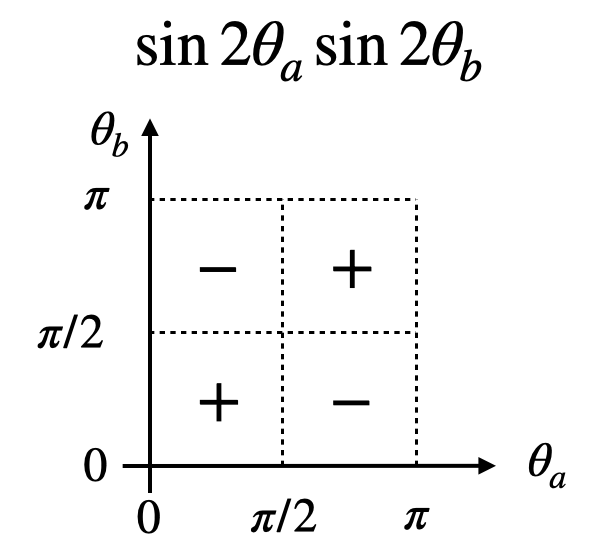}
\caption{The signs of $\sin2\theta_L\sin2\theta_R$.}\label{fig-sin2theta}
\end{figure}

Asymmetry parameters $\mathcal{A}(\sin\varphi_{R})$ and $\mathcal{A}(\sin2\varphi_{R})$ are therefore extracted by following two observation
\begin{align}
\mathcal{A}(\sin2\varphi_{R})={\Gamma(\sin2\varphi_{R}>0)-\Gamma(\sin2\varphi_{R}<0)\over \Gamma(\sin2\varphi_{R}>0)+\Gamma(\sin2\varphi_{R}<0)}
\end{align}
\begin{align}\label{eq-ATsinphi}
\mathcal{A}(\sin\varphi_{R})={\Gamma[{\rm Sign}(\sin2\theta_L\sin2\theta_R)\sin\varphi_{R}>0]-\Gamma[{\rm Sign}(\sin2\theta_L\sin2\theta_R)\sin\varphi_{R}<0]\over \Gamma[{\rm Sign}(\sin2\theta_L\sin2\theta_R)\sin\varphi_{R}>0]+\Gamma[{\rm Sign}(\sin2\theta_L\sin2\theta_R)\sin\varphi_{R}<0]}
\end{align}
The signs of $\sin2\theta_a\sin2\theta_b$ are shown in Fig.\ref{fig-sin2theta}. 
Besides, in the rest frame of the mother particle, all the momentum triple products are equivalent with each other. 
\begin{align}
&(\vec p_1\times \vec p_2)\cdot \vec p_1=(\vec p_1\times \vec p_2)\cdot \vec p_2=0
\end{align}
\begin{align}
&(\vec p_1\times \vec p_2)\cdot \vec p_3=(\vec p_2\times \vec p_3)\cdot \vec p_1=(\vec p_3\times \vec p_1)\cdot \vec p_2\nonumber\\
&=-(\vec p_2\times \vec p_1)\cdot \vec p_3=-(\vec p_3\times \vec p_2)\cdot \vec p_1=-(\vec p_1\times \vec p_3)\cdot \vec p_2
\end{align}
\begin{align}
&(\vec p_1\times \vec p_2)\cdot \vec p_4=-(\vec p_1\times \vec p_2)\cdot \vec p_3
\end{align}
\begin{align}
&(\vec p_3\times \vec p_4)\cdot \vec p_1=-[\vec p_3\times (\vec p_1+\vec p_2+\vec p_3)]\cdot \vec p_1=-(\vec p_3\times \vec p_2)\cdot \vec p_1=(\vec p_1\times \vec p_2)\cdot \vec p_3
\end{align}
\begin{align}
&(\vec p_3\times \vec p_4)\cdot \vec p_2=-(\vec p_3\times \vec p_4)\cdot \vec p_1
\end{align}
From the above equations, it can be seen that the triple-product asymmetries are all equivalent with each other except for a minus sign for some asymmetries. 

From the angular distributions of  $\Lambda_b\to N^*\rho$ or $N^* K^*$ in Eq.(\ref{eq:angularNstarKstar}), it can be seen that the term of $\sin\varphi_{R}$ vanishes with the integration of $\theta_L$ and $\theta_R$.
That might be a reason why the current measurements of triple-product asymmetries in $\Lambda_b$ decays do not find CPV \cite{Aaij:2016cla,LHCb:2019oke,LHCb:2018fpt}. Although the current measurements use the binned method, but the bins are not related to the angles of $\theta_L$ and $\theta_R$. In other words, one should take some of weight functions in the binned methods. The observable of Eq.(\ref{eq-ATsinphi}) with signs of $\sin2\theta_L\sin2\theta_R$ can be used to measure the triple-product asymmetries and thereby measure the CPV. The above analysis will also be applied to the case of polarized $b$-baryons, and the corresponding quasi-two-body decay angular distribution are derived in \cite{Durieux:2016nqr}.

\subsubsection{Quasi-three-body decays}
In the following sections, we shall discuss the quasi-three-body decays of b-baryons which is lack of explorations up to now. For example, some observables are also promising in $\Lambda^{0}_{b}\to pa_{1}(1260)\to p 3\pi$ and $\Lambda^{0}_{b}\to \Delta^{++}\pi^{-}\pi^{-} \to p3\pi$. Before performing the angular analysis on it, however, we shall think about associated $\mathcal{T}$-odd correlations firstly. If there is no initial $\Lambda^{0}_{b}$ polarization, only one $\mathcal{T}$-odd triple correlation $(\hat{S}_{a_{1}}\times\hat{S}_{p})\cdot\hat{p}$ makes sense for $\Lambda^{0}_{b}\to pa_{1}$, however, it vanishes in the final angular distribution since one can not reveal the spin of proton by final momentum distributions. Nevertheless, someone argues that triple momentum correlation $(\hat{p}_{1}\times\hat{p}_{2})\cdot\hat{p}_{3}$ can be constructed for $pa_{1}$ with $ a_{1}\to 3\pi$ mode. However, one must note that the expectation value of this object is not necessary to be the imaginary part of weak decay amplitudes like discussion before. The reason is that the $\mathcal{T}$-odd triple product $(\vec{p}_{1}\times\vec{p}_2)\cdot\vec{p}_3$ concist of three momentum in four-body decays can not satisfy the conditions {\bf(i)} and {\bf(ii)} in Sec.\ref{sec5.2} simultaneously. The $\mathcal{T}$ transformation flips all the particle momenta, so the condition {\bf(i)} requires that $\mathcal{U}$ flips the momenta back. Then, we must have $\mathcal{U}\; (\vec{p}_{1}\times\vec{p}_2)\cdot\vec{p}_3 \;\mathcal{U}^\dagger = - (\vec{p}_{1}\times\vec{p}_2)\cdot\vec{p}_3$ and thus the condition {\bf(ii)} is not satisfied. Therefore, the corresponding CPV is not necessarily proportional to $\cos\delta_s$~\cite{Bevan:2014nva}. The presence of $\Lb$ polarization brings us hopes to find non-trivial $\mathcal{T}$-odd correlations in $\Lb\to p 3\pi$. Therefore, let's investigate the angular distribution with  initial polarization.

To illustrate this distribution clearly, two steps are required. Firstly, one can get the polarization density matrix of $a_{1}$, and secondly, distribution is obtained by addressing the three body decay of $a_{1}$ using the formulas discussed in Sec.\ref{sec: three body}. For simplicity, we take the assumption that $\Lambda^{0}_{b}$ polarization is completely normal to it's production plane indicting that induced only through the $pp$ collisions. A decay plane is well-defined using initial polarization direction $\vec{P}$ and proton momentum $\vec{p}$, in which the $a_{1}$ rest frame is determined using $-\vec{p}/|\vec{p}|\equiv\vec{e}_{z}$ and $\vec{P}\times\vec{p}/|\vec{P}\times\vec{p}|\equiv\vec{e}_{y}$, $\vec{e}_{x}=\vec{e}_{y}\times\vec{e}_{z}$ with right hand convention. Correspondingly, the angular distribution normal to $a_{1}\to 3\pi$ plane is formulated using (\ref{eq: compact formula}), which is explicitly expressed as
\begin{equation}\label{eq: distribution for a1}
	\begin{aligned}
	\left(\frac{d\Gamma}{d\Omega}\right)&=R^{+}_{1}\left\{\left[\mathcal{R}e\rho_{-1,-1}+\mathcal{R}e\rho_{+1,+1}\right]\frac{1+\cos^{2}\beta}{2}+\mathcal{R}e\rho_{0,0}\sin^{2}\beta
	\right.\\&\left.+\sqrt{2}\cos\beta\sin\beta\left\{\left(\mathcal{R}e\rho_{+1,0}-\mathcal{R}e\rho_{-1,0}\right)\cos\alpha-\left(\mathcal{I}m\rho_{+1,0}+\mathcal{I}m\rho_{-1,0}\right)\sin\alpha\right\}
	\right.\\&\left.+\sin^{2}\beta\left(\mathcal{R}e\rho_{1,-1}\cos2\alpha-\mathcal{I}m\rho_{1,-1}\sin2\alpha\right)\right\}\\
	&+R^{-}_{1}\left\{\left(\mathcal{R}e\rho_{+1,+1}-\mathcal{R}e\rho_{-1,-1}\right)\cos\beta\right.\\&\left.+\sqrt{2}\sin\beta\left[\left(\mathcal{R}e\rho_{+1,0}+\mathcal{R}e\rho_{-1,0}\right)\cos\alpha-\left(\mathcal{I}m\rho_{+1,0}-\mathcal{I}m\rho_{-1,0}\right)\sin\alpha\right]\right\}
	\end{aligned}
\end{equation}
The elements of the density matrix for $a_{1}$ are given by
\begin{equation}
	\begin{aligned}
	\rho_{+1,+1}&=\frac{1-P_{b}\cos\theta_{P}}{2}\left|\mathcal{H}_{+1/2,+1}\right|^{2}\\
 \rho_{-1,-1}&=\frac{1+P_{b}\cos\theta_{P}}{2}\left|\mathcal{H}_{-1/2,-1}\right|^{2}\\
	\rho_{+1,0}&=\rho^{*}_{0,+1}=P_{b}\sin\theta_{P}\mathcal{H}_{+1/2,+1}\mathcal{H}^{*}_{+1/2,0}\\
 \rho_{0,-1}&=\rho^{*}_{-1,0}=P_{b}\sin\theta_{P}\mathcal{H}_{-1/2,0}\mathcal{H}^{*}_{-1/2,-1}\\	
	\rho_{0,0}&=\frac{1-P_{b}\cos\theta_{P}}{2}\left|\mathcal{H}_{-1/2,0}\right|^{2}+\frac{1+P_{b}\cos\theta_{P}}{2}\left|\mathcal{H}_{+1/2,0}\right|^{2}\\
	\end{aligned}
\end{equation}
Here, $\mathcal{H}_{\lambda_{1},\lambda_{2}}$ represents the weak decay amplitudes for $\Lambda^{0}_{b}\to pa_{1}(1260)$, where $\lambda_{1}$ and $\lambda_{2}$ correspond to the helicities of the proton and vector meson $a_{1}$, respectively. Some simplifications can be made in $d\Gamma/d\Omega$ if we set the polarization of $\Lambda^{0}_{b}$ to zero
\begin{equation}\label{eq: No initial polarization}
	\begin{aligned}
		\left(\frac{dN}{d\Omega}\right)&\propto R^{+}_{1}\left\{\left[\left|\mathcal{H}_{-1/2,-1}\right|^{2}+\left|\mathcal{H}_{+1/2,+1}\right|^{2}\right]\frac{1+\cos^{2}\beta}{2}+\left[\left|\mathcal{H}_{-1/2,0}\right|^{2}+\left|\mathcal{H}_{+1/2,0}\right|^{2}\right]\sin^{2}\beta\right\}\\
		&+R^{-}_{1}\left(\left|\mathcal{H}_{+1/2,+1}\right|^{2}-\left|\mathcal{H}_{-1/2,-1}\right|^{2}\right)\cos\beta
	\end{aligned}
\end{equation}
where $R^{\pm}_{1}$ are  
\begin{equation}
	\begin{aligned}
	 R^{+}_{1}&=\frac{1}{2}\left[|R_{1}|^{2}+|R_{-1}|^{2}\right]\\
	 R^{-}_{1}&=\frac{1}{2}\left[|R_{1}|^{2}-|R_{-1}|^{2}\right]\\
	\end{aligned}
\end{equation}
It is important to note that $R^{-}_{M}$ vanishes when identical particles are present in the final state, due to an ambiguity in the normal direction definition of $a_{1}\to 3\pi$. Consequently, a degeneracy arises between $M>0$ and $M<0$, which is unavoidable when dealing with identical particles. To address this and preserve $R^{-}_{M}$ in the angular distribution, we can divide the Dalitz space into two regions with $\omega_{1}>\omega_{2}$ and $\omega_{1}<\omega_{2}$, respectively. By introducing a sign function to distinguish between them, similar to the discussion of $\mathcal{A}_{T}(\sin\varphi_{R})$ in the previous section. It follows from Eq. \eqref{eq: No initial polarization} that in the absence of $\Lambda^{0}_{b}$ polarization, an $\alpha$-like quantity appears
\begin{equation}
	\begin{aligned}
	 \mathcal{A}(\cos\beta)=R^{-}_{1}\left(\left|\mathcal{H}_{+1/2,+1}\right|^{2}-\left|\mathcal{H}_{-1/2,-1}\right|^{2}\right)
	\end{aligned}
\end{equation}
where $\cos\beta$ is expected to be proportional to the three-momentum correlation $(\vec{p}_{1}\times\vec{p}_{2})\cdot\vec{p}$. This relationship is intuitive to understand, as $\vec{p}_{1}\times\vec{p}_{2}$ defines the normal vector of the $a_{1}(1260)$ decay plane, and hence $(\vec{p}_{1}\times\vec{p}_{2})\cdot\vec{p}$ is proportional to $\cos\beta$. Naturally, this implies a CP-violating observable
\begin{equation}
	\begin{aligned}
		a_{CP}(\cos\beta)=\frac{\mathcal{A}(\cos\beta)+\mathcal{\bar{A}}(\cos\beta)}{2\langle\Gamma\rangle}
	\end{aligned}
\end{equation}
which does not require the initial $b-$baryon polarization. Let's now focus on the derivation of $a_{CP}(\cos\beta)$ in terms of $A_{i}$ and $B_{i}$. Explicitly, the helicity amplitudes $\mathcal{H}_{\lambda_{1},\lambda_{2}}$, which are linear combinations of the vector part $\mathcal{H}_{\lambda_{1},\lambda_{2}}(V)$ and the axial part $\mathcal{H}_{\lambda_{1},\lambda_{2}}(A)$, are expressed in the $\Lambda_{b}$ rest frame
\begin{equation}
	\begin{aligned}
	\mathcal{H}_{+\frac{1}{2},+1}(V/A)=2\left\{\begin{array}{c}
    		                          -\sqrt{Q_{-}}B_{1} \\\\
    		                          -\sqrt{Q_{+}}A_{1}
                                              \end{array}
                                        \right\}
	\end{aligned}
\end{equation}
\begin{equation}
	\begin{aligned}
	\mathcal{H}_{+\frac{1}{2},0}(V/A)=\frac{2}{M_{V}}\left\{\begin{array}{c}
    		          \sqrt{Q_{-}}(M_{i}+M_{f})B_{1}+\sqrt{Q}_{+}p_{c}B_{2} \\\\
    		          \sqrt{Q_{+}}(M_{i}-M_{f})A_{1}-\sqrt{Q}_{-}p_{c}A_{2}
                                              \end{array}
                                        \right\}
	\end{aligned}
\end{equation}
Under Parity transformation, one obtains
\begin{equation}
	\begin{aligned}
    \mathcal{H}_{\lambda_{1},\lambda_{2}}(V/A)=(+/-)\eta_{b}\eta_{p}\eta_{a}(-1)^{S_{b}-S_{a}-S_{p}}\mathcal{H}_{-\lambda_{1},-\lambda_{2}}(V/A)=(-/+)\mathcal{H}_{-\lambda_{1},-\lambda_{2}}(V/A)
	\end{aligned}
\end{equation}
Hence, the total weak amplitudes are given by:
\begin{equation}
	\begin{aligned}
    \mathcal{H}_{\lambda_{1},\lambda_{2}}&=\mathcal{H}_{\lambda_{1},\lambda_{2}}(A)+\mathcal{H}_{\lambda_{1},\lambda_{2}}(V)\\    \mathcal{H}_{-\lambda_{1},-\lambda_{2}}&=\mathcal{H}_{\lambda_{1},\lambda_{2}}(A)-\mathcal{H}_{\lambda_{1},\lambda_{2}}(V)\\
	\end{aligned}
\end{equation}
For simplicity, we also define dimensionless parameters
\begin{equation}
	\begin{aligned}
     \mathcal{W}_{1}=\frac{M_{i}+M_{f}}{M_{V}},~~\mathcal{W}_{2}=\frac{M_{i}-M_{f}}{M_{V}},~~\mathcal{W}_{3}=\frac{p_{c}}{M_{V}}
	\end{aligned}
\end{equation}
The dependence of $a_{CP}(\cos\beta)$ on $A_{i}$ and $B_{i}$ can be determined as follows
\begin{equation}
	\begin{aligned}
     \mathcal{A}(\cos\beta)\propto -4\sqrt{Q_{+}Q_{-}}\mathcal{R}e(A_{1}B^{*}_{1}),~~     \mathcal{\bar{A}}(\cos\beta)\propto -4\sqrt{Q_{+}Q_{-}}\mathcal{R}e(\bar{A}_{1}\bar{B}^{*}_{1})
	\end{aligned}
\end{equation}
and
\begin{equation}
	\begin{aligned}
    a_{CP}(\cos\beta)&\propto \mathcal{A}(\cos\beta)+\mathcal{\bar{A}}(\cos\beta)\\
    &=-4\sqrt{Q_{+}Q_{-}}\mathcal{R}e(A_{1}B^{*}_{1}+\bar{A}_{1}\bar{B}^{*}_{1})\\
    &=-4\sqrt{Q_{+}Q_{-}}\mathcal{R}e(A_{1},B_{1})\sin\Delta\phi
	\end{aligned}
\end{equation}
which is exactly a $\alpha$-like observable. To derive the $\mathcal{T}$-odd correlations and their corresponding CP asymmetries, it is essential to utilize the initial polarized $\Lambda^{0}_{b}$. From the distribution given in Eq. (\ref{eq: distribution for a1}), two observables related to the interference of the imaginary parts of amplitudes can be extracted
\begin{equation}
	\begin{aligned}
     \mathcal{A}_{2}&=\mathcal{I}m(\mathcal{H}_{+1/2,+1}\mathcal{H}^{*}_{+1/2,0}-\mathcal{H}_{-1/2,-1}\mathcal{H}^{*}_{-1/2,0})\\     \mathcal{A}^{P}_{2}&=\mathcal{I}m(\mathcal{H}_{+1/2,+1}\mathcal{H}^{*}_{+1/2,0}+\mathcal{H}_{-1/2,-1}\mathcal{H}^{*}_{-1/2,0})
	\end{aligned}
\end{equation}
In fact, these two quantities are induced by a $\mathcal{T}$-odd operator referred to as $T_{2}$ in the work \cite{Geng:2021lrc}
\begin{equation}
	\begin{aligned}
     T_{2}=(\hat{S}_{a}\times\hat{p})\cdot\hat{S}_{i},~~T^{P}_{2}=T_{2}\cdot(\hat{S}_{p}\cdot\hat{p})
	\end{aligned}
\end{equation}
Here, $\hat{p}$ represents the unit proton momentum, while $\hat{S}_{i}$, $\hat{S}_{a}$, and $\hat{S}_{p}$ denote the polarizations of $\Lambda^{0}_{b}$, $a_{1}$, and the proton, respectively. Similar to the discussion of $\mathcal{A}_{T}(\sin\varphi_{R})$ and $\mathcal{A}_{T}(\sin2\varphi_{R})$ in $\Lambda^{0}_{b}\to N^{*}(3/2^{\pm})M^{*}$ decays, a similar analysis can also be carried out for the $pa_{1}$ modes. Explicitly, we have
\begin{equation}
	\begin{aligned}
     \mathcal{A}_{2}=2\left\{\sqrt{Q_{+}Q_{-}}\left[\mathcal{W}_{1}\mathcal{I}m(A_{1}B^{*}_{1})+\mathcal{W}_{2}\mathcal{I}m(A^{*}_{1}B_{1})\right]-\mathcal{W}_{3}\left[Q_{+}\mathcal{I}m(A_{1}B^{*}_{2})+Q_{-}\mathcal{I}m(A^{*}_{2}B_{1})\right]\right\}
	\end{aligned}
\end{equation}
\begin{equation}
	\begin{aligned}
     \mathcal{A}^{P}_{2}=2\sqrt{Q_{+}Q_{-}}\mathcal{W}_{3}\left[\mathcal{I}m(A_{1}A^{*}_{2})+\mathcal{I}m(B_{1}B^{*}_{2})\right]
	\end{aligned}
\end{equation}
Two essential CP-violating observables are defined based on their Parity properties
\begin{equation}
	\begin{aligned}
     a^{T_{2}}_{CP}=\frac{\mathcal{A}_{2}+\bar{\mathcal{A}}_{2}}{2\langle\Gamma\rangle},~~     a^{T^{P}_{2}}_{CP}=\frac{\mathcal{A}^{P}_{2}-\bar{\mathcal{A}}^{P}_{2}}{2\langle\Gamma\rangle}
	\end{aligned}
\end{equation}
$A_{i}$ and $B_{i}$ can be decomposed into contributions from tree and penguin diagrams with strong phases and weak phases, as done for $\Lambda^{0}_{b}\to N^{*}(3/2^{\pm})M^{*}$ in Eq. \eqref{eq: C and D}. Ultimately, one obtains
\begin{equation}
	\begin{aligned}
     a^{T_{2}}_{CP}&\propto \sqrt{Q_{+}Q_{-}}(\mathcal{W}_{1}-\mathcal{W}_{2})\left[A_{1,t}B_{1,p}\cos(\delta^{A}_{1,t}-\delta^{B}_{1,p})-A_{1,p}B_{1,t}\cos(\delta^{A}_{1,p}-\delta^{B}_{1,t})\right]\sin\Delta\phi\\
     &-Q_{+}\mathcal{W}_{3}\left[A_{1,t}B_{2,p}\cos(\delta^{A}_{1,t}-\delta^{B}_{2,p})-A_{1,p}B_{2,t}\cos(\delta^{A}_{1,p}-\delta^{B}_{2,t})\right]\sin\Delta\phi\\
     &+Q_{-}\mathcal{W}_{3}\left[A_{2,t}B_{1,p}\cos(\delta^{A}_{2,t}-\delta^{B}_{1,p})-A_{2,p}B_{1,t}\cos(\delta^{A}_{2,p}-\delta^{B}_{1,t})\right]\sin\Delta\phi
	\end{aligned}
\end{equation}
\begin{equation}
	\begin{aligned}
     a^{T^{P}_{2}}_{CP}&\propto \sqrt{Q_{+}Q_{-}}\mathcal{W}_{3}\left[A_{1,t}A_{2,p}\cos(\delta^{A}_{1,t}-\delta^{A}_{2,p})-A_{1,p}A_{2,t}\cos(\delta^{A}_{1,p}-\delta^{A}_{2,t})\right]\sin\Delta\phi\\
     &+\sqrt{Q_{+}Q_{-}}\mathcal{W}_{3}\left[B_{1,t}B_{2,p}\cos(\delta^{B}_{1,t}-\delta^{B}_{2,p})-B_{1,p}B_{2,t}\cos(\delta^{B}_{1,p}-\delta^{B}_{2,t})\right]\sin\Delta\phi
	\end{aligned}
\end{equation}
The corresponding $\mathcal{T}$-even correlations induce the asymmetry parameters
\begin{equation}
	\begin{aligned}
     \mathcal{B}_{2}&=\mathcal{R}e(\mathcal{H}_{+1/2,+1}\mathcal{H}^{*}_{+1/2,0}-\mathcal{H}_{-1/2,-1}\mathcal{H}^{*}_{-1/2,0})\\     \mathcal{B}^{P}_{2}&=\mathcal{R}e(\mathcal{H}_{+1/2,+1}\mathcal{H}^{*}_{+1/2,0}+\mathcal{H}_{-1/2,-1}\mathcal{H}^{*}_{-1/2,0})
	\end{aligned}
\end{equation}
The CP asymmetries induced by them are
\begin{equation}
	\begin{aligned}
     a^{B_{2}}_{CP}=\frac{\mathcal{B}_{2}+\bar{\mathcal{B}}_{2}}{2\langle\Gamma\rangle},~~     a^{B^{P}_{2}}_{CP}=\frac{\mathcal{B}^{P}_{2}-\bar{\mathcal{B}}^{P}_{2}}{2\langle\Gamma\rangle}
	\end{aligned}
\end{equation}
These two CP asymmetries are expected to be complementary to those of $ a^{T_{2}}_{CP},a^{T^{P}_{2}}_{CP}$.

Another quasi-three-body decay that we will focus on is $\Lambda^{0}_{b}\to \Delta^{++}(\to p\pi^{+})\pim\pim$ with a polarized $\Lambda^{0}_{b}$. Naively, the $\beta$-like quantity in $\Lambda^{0}_{b}\to \Delta^{++}\pip\pim$ could be measured analogously to $\Lambda^{0}_{b}\to pa_{1}$. Therefore, it is promising to observe the CP violation of this decay mode. The three-body decay $\Lambda_{b}\to \Delta^{++}\pi^{-}\pi^{-}\to p\pi^{+}\pi^{-}\pi^{-}$ is meaningful to investigate since it has relatively large statistics in the four-body decay of $\Lambda^{0}_{b}$ \cite{Aaij:2019rkf}. It was explored by employing the $N\pi \to \Delta^{++}\pi^{-}$ scattering data within the framework of final state re-scatterings. Referring to Fig. \ref{fig-T Polarization1}, the coordinate system $\left\{\hat{e}_{x},\hat{e}_{y},\hat{e}_{z}\right\}$ in which $\Lambda^{0}_{b}$ is located in is determined with the help of the proton beam momentum. Correspondingly, the angular distribution is labeled with five variables $\alpha,\beta,\gamma,\theta,\varphi$, which are the respective three Euler angles to describe $\Lambda^{0}_{b}\to \Delta^{++}\pim\pim$, and the proton momentum direction $(\theta,\varphi)$ with respect to the normal vector of the weak decay plane. As discussed earlier, the initial $\Lambda^{0}_{b}$ is assumed to be polarized along the $\hat{e}_{y}$ axis and is described by a density matrix.
\begin{equation}
	\begin{aligned}
		\rho_{m,m^{\prime}}=\frac{1}{2}\left[1+P_{b}\hat{\sigma}_{y}\right]
	\end{aligned}
\end{equation}
For simplicity, we only deal with the polarized part of angular distribution. Here the polarized part means only $P_{b}\hat{\sigma}_{y}$ in $\rho_{m,m^{\prime}}$ is considered.
\begin{equation}\label{eq:Delta}
	\begin{aligned}
 	 \frac{d\Gamma}{d\Omega}=&\sum_{M,m,m^{\prime}}\sum_{\lambda_{3},\lambda^{\prime}_{3}}(2\pi)P_{b}\sigma_{y;m,m^{\prime}}D^{*j=1/2}_{m,M}(\alpha,\beta,0)D^{j=1/2}_{m^{\prime},M}(\alpha,\beta,0)\\&
 	 \sum_{\lambda_{p}}F_{M}(\lambda_{3})F^{*}_{M}(\lambda^{\prime}_{3})D^{*j=3/2}_{\lambda_{3},\lambda_{p}}(\varphi,\theta,0)D^{j=3/2}_{\lambda^{\prime}_{3},\lambda_{p}}(\varphi,\theta,0)\left|h_{\lambda_{p}}\right|^{2}
	\end{aligned}
\end{equation}
where $h_{\lambda_{p}}$ is the helicity amplitudes of $\Delta^{++}\to p\pi$ with proton helicity $\lambda_{p}$, and the spin density matrix of $\Delta^{++}$ is defined with $\lambda_{3}$ as the helicity of $\Delta^{++}$
\begin{equation}\label{eq:Delta polarization} 
	\begin{aligned}
	 \rho_{\lambda_{3},\lambda^{\prime}_{3}}=(2\pi)P_{b}\sin\beta\sin\alpha \left\{F_{-1/2}(\lambda_{3})F^{*}_{-1/2}(\lambda^{\prime}_{3})-F_{+1/2}(\lambda_{3})F^{*}_{+1/2}(\lambda^{\prime}_{3})\right\} 
	\end{aligned}
\end{equation}
Therefore, the angular distribution is expressed as
\begin{equation}\label{eq: angular distribution of Delat}
	\begin{aligned}
		\frac{d\Gamma}{d\Omega}&=\sum_{\lambda_{3},\lambda^{\prime}_{3},\lambda_{p}}\rho_{\lambda_{3},\lambda^{\prime}_{3}}D^{*j=3/2}_{\lambda_{3},\lambda_{p}}(\varphi,\theta,0)D^{j=3/2}_{\lambda^{\prime}_{3},\lambda_{p}}(\varphi,\theta,0)\left|h_{\lambda_{p}}\right|^{2}\\
	\end{aligned}
\end{equation}
Obviously, the density matrix $\rho_{\lambda_{3},\lambda^{\prime}_{3}}$ satisfies the Hermitian condition $\rho_{\lambda_{3},\lambda^{\prime}_{3}}=\rho^{*}_{\lambda^{\prime}_{3},\lambda_{3}}$.

To obtain a non-trivial angular distribution, one needs to divide the phase space into two parts with the condition $\omega_{12}>\omega_{23}$ and $\omega_{12}<\omega_{23}$ in experimental measurements. Explicitly, the distribution given in Eq. (\ref{eq: angular distribution of Delat}) is formulated as
\begin{equation}
	\begin{aligned}
		\frac{d\Gamma}{d\Omega}
		&\supset\frac{3}{4}\left(\rho_{3/2,3/2}+\rho_{-3/2,-3/2}\right)\sin^{2}\theta+\frac{1}{4}\left(\rho_{1/2,1/2}+\rho_{-1/2,-1/2}\right)\left(1+3\cos^{2}\theta\right)\\
		&-\sqrt{3}\sin\theta\cos\theta\mathcal{R}e\left[\rho_{3/2,1/2}e^{i\varphi}\right]-\frac{\sqrt{3}}{4}(1+3\cos2\theta)\sin\theta\mathcal{R}e\left[\rho_{-1/2,-3/2}e^{i\varphi}\right]\\
		&+\frac{\sqrt{3}}{2}\sin^2\theta\left\{3\cos\theta\mathcal{R}e\left[\rho_{+1/2,-3/2}e^{i2\varphi}\right]-\mathcal{R}e\left[\rho_{3/2,-1/2}e^{i2\varphi}\right]\right\}
		-\frac{3}{2}\sin^{3}\theta\mathcal{R}e\left[\rho_{3/2,-3/2}e^{i3\varphi}\right]
	\end{aligned}
\end{equation}
The true dynamical aspect of the three-body decay is highly non-trivial. Nevertheless, it is also significant to measure its angular distribution in experiments since each term in the above formulas could be utilized to define some non-trivial CP asymmetries. Another important point is that our proof of complementarity between $\mathcal{T}$-even and $\mathcal{T}$-odd in the helicity scheme relies on the linear relation between the helicity amplitudes and partial waves, which might be invalid in the three-body case due to the presence of complex intermediate states. However, the angular distribution contains many interference terms and could be used to reveal CP violation in the three-body decay of $\Lambda^{0}_{b}\to \Delta^{++}\pi^{-}\pi^{-}$. Therefore, it is interesting to investigate this in experiments. If the initial polarization is very small and can be ignored, then one can simplify the distribution function
\begin{equation}\label{eq:Delta1}
	\begin{aligned}
 	 \frac{d\Gamma}{d\Omega}=&\sum_{M,m,m^{\prime}}\sum_{\lambda_{3},\lambda^{\prime}_{3}}(2\pi)\delta_{m,m^{\prime}}D^{*j=1/2}_{m,M}(\alpha,\beta,0)D^{j=1/2}_{m^{\prime},M}(\alpha,\beta,0)\\&
 	 \sum_{\lambda_{p}}F_{M}(\lambda_{3})F^{*}_{M}(\lambda^{\prime}_{3})D^{*j=3/2}_{\lambda_{3},\lambda_{p}}(\varphi,\theta,0)D^{j=3/2}_{\lambda^{\prime}_{3},\lambda_{p}}(\varphi,\theta,0)\left|h_{\lambda_{p}}\right|^{2}
	\end{aligned}
\end{equation}
and explicitly
\begin{equation}
	\begin{aligned}
		\frac{d\Gamma}{d\Omega}
		&=\frac{3}{4}\sin^{2}\theta\left[\left|F_{+1/2}(+3/2)\right|^{2}+\left|F_{-1/2}(+3/2)\right|^{2}+\left|F_{+1/2}(-3/2)\right|^{2}+\left|F_{-1/2}(-3/2)\right|^{2}\right]\\
		&+\frac{1}{4}\left(1+3\cos^{2}\theta\right)\left[\left|F_{+1/2}(+1/2)\right|^{2}+\left|F_{-1/2}(+1/2)\right|^{2}+\left|F_{+1/2}(-1/2)\right|^{2}+\left|F_{-1/2}(-1/2)\right|^{2}\right]
	\end{aligned}
\end{equation}
Above formula is further reduced to
\begin{equation}
	\begin{aligned}
		\frac{d\Gamma}{d\Omega}\propto 1+\kappa\cos^{2}\theta
	\end{aligned}
\end{equation}
with
\begin{equation}
	\begin{aligned}
		\kappa=\frac{3\rho_{1/2,1/2}-\rho_{3/2,3/2}+3\rho_{-1/2,-1/2}-\rho_{-3/2,-3/2}}{\rho_{3/2,3/2}+\rho_{-3/2,-3/2}+\rho_{1/2,1/2}+\rho_{-1/2,-1/2}}
	\end{aligned}
\end{equation}
Another important three-body decay chain used to reveal CP violation in the baryon sector is of the type $\Lambda^{0}_{b}\to \Lambda\pi^{+}\pi^{-},\Lambda K\pi,\Lambda K K$ with $\Lambda\to p\pi^{-}$, which defines a kinematic plane where a coordinate system $\vec{e}_{xyz}$ exists
\begin{equation}
		\begin{aligned}
         \vec{e}_{z}&=\frac{\vec{p}_{\Lambda}}{|\vec{p}_{\Lambda}|}\\
         \vec{e}_{y}&=\vec{e}_{z}\times \frac{\vec{p}_{+}}{|\vec{p}_{+}|}\\
         \vec{e}_{x}&=\vec{e}_{y}\times\vec{e}_{z}
		\end{aligned}
\end{equation}
where $\hat{p}_{+}$ is the momentum of the $\pi^{+}$ or $K^{+}$ meson used to define the normal direction of the decay plane. In this scenario, the distribution function is given by
	\begin{equation}
		\begin{aligned}
	  \frac{d\Gamma}{d\cos\theta}\propto 1+\langle\vec{P}_{z}\rangle \alpha_{\Lambda}\cos\theta+\langle\vec{P}_{x}\rangle \alpha_{\Lambda}\sin\theta\cos\phi+\langle\vec{P}_{y}\rangle \alpha_{\Lambda}\sin\theta\sin\phi
		\end{aligned}
\end{equation}
The value of $\alpha_{\Lambda}$ serves as an input to determine the polarization components $\langle\vec{P}_{xyz}\rangle$, and the associated CP asymmetries are defined as
\begin{equation}
		\begin{aligned}
a^{z}_{CP}&=\frac{\langle\vec{P}_{z}\rangle+\langle\bar{\vec{P}}_{z}\rangle}{2}\\
a^{y}_{CP}&=\frac{\langle\vec{P}_{y}\rangle+\langle\bar{\vec{P}}_{y}\rangle}{2}\\
a^{x}_{CP}&=\frac{\langle\vec{P}_{x}\rangle-\langle\bar{\vec{P}}_{x}\rangle}{2}\\
		\end{aligned}
\end{equation}
The above distribution can also be used to discuss the case of $B$ meson decay, such as $B^{+}\to p\bar{\Lambda}\pi^{0}$ with $\Lambda\to p\pi$. The decay channel $B^{+}\to p\bar{\Lambda}\rho^{0}\to p\bar{\Lambda}\pi^{+}\pi^{-}$ with $\Lambda\to p\pi$ could also be considered under similar observable.

%% file: 5-discussions.tex
\section{Summary}\label{sec.6}
Various observables for numerous decay modes of $b$-baryons, including two-body and three-, four-body decays, have been discussed in the sections above in absence of specific dynamical calculations. As emphasized, the complex dynamical predictions for the observables we focus on are well beyond the scope of the current work. Our primary goal in this study is to offer some phenomenological insights into baryon multibody decays since they are extensively existed in the real physical world.

In this work, we provide a complete and comprehensive review for the asymmetry parameters, polarization in strong and weak productions and their associated CP asymmetries. Basically, there are three essential parts in our work that are polarization, CP asymmetries and angular distributions respectively. Spin and polarization, as a probe, are sensitive to the underline dynamics of a reaction, hence always show some implications for Parity, Charge conjugation, and induce some CP violations located in the regional phase space. However, the experimental realization of these quantities are not straightforward in the modern colliders since the spin and polarization can not be measured directly. Therefore, the angular distribution, as a powerful bridge to connect these observables and experiment, are emphasised in this paper.

Specifically, in this work, we firstly study the properties of polarization components of baryons produced in the decays and collisions. They are characterised by asymmetry parameters $\alpha,\beta,\gamma$ and normal polarization. The CP asymmetries defined by these parameters are analysed in detail, including their strong phase dependence under partial wave and helicity representation, complementarity, and potential applications in the multi-body decays. We also propose a novel method to explore the produced normal polarization of $\Lambda^{0}_{b}$ baryon in experiment. Our key point is $\mathcal{T}$-odd and -even correlations and their associated CP asymmetries hat are expected to be promising for the detection of baryon CPV in the future. We prove the CP asymmetries induced by $\mathcal{T}$-odd correlation always depends on the strong phase $\Delta\delta$ as a cosine function which means it is no longer suppressed when the small $\Delta\delta$ occurs. Therefore, it provide more opportunities for baryon CPV provided the direct CPV is not observed. More importantly, a pair of CP asymmetries defined by $\mathcal{T}$-odd and -even correlations are proved to be complementary with each other in the strong phase dependence. We give the exact criterion in the Sec.\eqref{sec.5} by taking different definitions of strong phases. In order to show it's potential applications in the $b$-baryon decays, we explore three kinds of b-baryons decaying modes $\Lambda^{0}_{b}\to B^{*}(\frac{3}{2}^{\pm})V, \Lambda^{0}_{b}\to pa_{1}, \Lambda^{0}_{b}\to \Delta^{++}\pi^{-}\pi^{-}$ by incorporating their $\mathcal{T}$-odd correlations and CP asymmetries with comprehensive derivations. Finally, we arrange some discussions and suggestions for the experimental investigations in the future.

%% file: 6-summary.tex
\section*{ACKNOWLEDGMENTS}
We are very grateful to Jin-Lin Fu, Xiao-Rui Lyu, Wen-Bin Qian and Yang-Heng Zheng for the useful discussions on the experimental measurements, to Ke-Sheng Huang and Sheng-Qi Zhang for their contributions at the very early stage of this project. This work is supported by the National Natural Science Foundation of China under Grant Nos. 12335003, 12375086 and by the Fundamental Research Funds for the Central Universities under No. lzujbky-2024-oy02 and lzujbky-2023-it12.

%% file: 7-appendix.tex
\appendix
\section{Approaches to calculate differential distributions}\label{App.}
The angular distribution of both decays and scatterings can be derived using the helicity amplitude approach, which conveniently includes the spin and polarization of particles compared to the conventional partial wave method. In this appendix, we will provide a comprehensive review of the Jacob-Wick helicity formalism \cite{Jacob:1959at}, emphasizing its importance in our work. We will introduce the construction of helicity states and amplitudes for scatterings and decays, and also explore the symmetry properties of helicity amplitudes and three-body states.

The helicity amplitude approach has been widely adopted in scattering and decay analyses due to its numerous advantages over outdated angular analysis methods.
\begin{itemize}
\item {The polarization is taken into consideration conveniently in the helicity formalism, while it is tedious in the conventional angular momentum coupling scheme where many $C$-$G$ symbols are involved in the derivations.}

\item {The helicity is invariant under rotation such that corresponding eigenstates can be obtained with a definite total angular momentum $J$.}

\item {The helicity formalism is also well-defined for massless particle like photon.}
\end{itemize}

The helicity of one particle, denoted as $\lambda$, is defined as the projection of the total angular momentum along its moving direction. It is evident that $\lambda$ can be expressed as the dot product of the particle's spin ($\vec{s}$) and the unit momentum vector ($\hat{p}$). For a massive free particle, we can construct a complete basis represented by the state $\ket{\vec{p},\lambda}$, where  $\lambda$ takes values from $-s$ to $s$ inclusively. In general, the helicity state $\ket{p,\theta,\phi;\lambda}$, which corresponds to an arbitrary direction in space, can be obtained by applying a rotation to the state $\ket{p,0,0;\lambda}$.
\begin{equation}
	\begin{aligned}
	\ket{p,\theta,\phi,\lambda}=\mathcal{R}(\phi,\theta,0)\ket{p,0,0,\lambda}
	\end{aligned}
\end{equation}
where $\ket{p,0,0,\lambda}$ is the state orienting to $z$ axis with the helicity $\lambda$. Meanwhile, which can also be decomposed as the superposition of angular momentum states since the rotational matrix is definite under angular representation
\begin{equation}
	\begin{aligned}
	\ket{p,0,0,\lambda}&=\sum_{j,m}C_{j,m}\ket{p,j,m;\lambda}\delta_{\lambda,m}=\sum_{j}C_{j}\ket{p,j,\lambda;\lambda}
	\end{aligned}
\end{equation}
Applying a rotation to the equation mentioned above, we have:
\begin{equation}\label{eq: expansion of plane wave}
	\begin{aligned}
	\ket{p,\theta,\phi,\lambda}=\sum_{j}C_{j}\mathcal{R}(\phi,\theta,0)\ket{p,j,\lambda;\lambda}=\sum_{j,m}C_{j}D^{j}_{m,\lambda}(\phi,\theta,0)\ket{p,j,m;\lambda}
	\end{aligned}
\end{equation}
Coefficient $C_{j}$ can be extracted using the normalization condition 
 \begin{equation}\label{Normalization condition 1}
	\begin{aligned}
	\bra{p^{\prime},\theta^{\prime},\phi^{\prime},\lambda^{\prime}}\ket{p,\theta,\phi,\lambda}&=\delta_{p,p^{\prime}}\delta(\cos\theta-\cos\theta^{\prime})\delta(\phi-\phi^{\prime})\delta_{\lambda,\lambda^{\prime}}
	\end{aligned}
\end{equation}
 \begin{equation}\label{Normalization condition 2}
	\begin{aligned}
    \bra{p^{\prime},j^{\prime},m^{\prime},\lambda^{\prime}}\ket{p,j,m;\lambda}&=\delta_{p,p^{\prime}}\delta_{j,j^{\prime}}\delta_{m,m^{\prime}}\delta_{\lambda,\lambda^{\prime}}
 	\end{aligned}
 \end{equation}
$C_{j}$ is determined as
 \begin{equation}
	\begin{aligned}
	 C_{j}=\sqrt{\frac{2j+1}{4\pi}}
	\end{aligned}
\end{equation}
Therefore, one obtains the transformation function between two different representation
\begin{equation}
	\begin{aligned}
	\ket{p,\theta,\phi,\lambda}=\sum_{j,m}\sqrt{\frac{2j+1}{4\pi}}D^{j}_{m,\lambda}(\phi,\theta,0)\ket{p,j,m;\lambda}
	\end{aligned}
\end{equation}
and
\begin{equation}
	\begin{aligned}
		\ket{p,j,m;\lambda}=\int d\Omega\sqrt{\frac{2j+1}{4\pi}}D^{*j}_{m,\lambda}(\phi,\theta,0)\ket{p,\theta,\phi,\lambda}
	\end{aligned}
\end{equation}
The self-consistency of normalization conditions (\ref{Normalization condition 1}) and (\ref{Normalization condition 2}) could be verified using above transformation relation.

The helicity state of a system composed of two particles can be constructed as the direct product of the helicity states of the individual particles
\begin{equation}
	\begin{aligned}
	 \ket{\vec{p_{1}},\lambda_{1};\vec{p_{2}},\lambda_{2}}= \ket{\vec{p_{1}},\lambda_{1}}\otimes\ket{\vec{p_{2}},\lambda_{2}}
	\end{aligned}
\end{equation}
For convenience, we consider the center of mass frame
\begin{equation}
	\begin{aligned}
		\ket{\vec{p},\lambda_{1};-\vec{p},\lambda_{2}}=\psi_{\vec{p},\lambda_{1}}\chi_{\vec{p},\lambda_{2}}
	\end{aligned}
\end{equation}
where $\chi_{\vec{p},\lambda_{2}}$ is the state directing opposite $z$ with the helicity $\lambda_{2}$. The transformation of two particle state between two different representations is analogous as single particle case.

\begin{figure}[h!]
\centering
\includegraphics[width=0.9\textwidth]{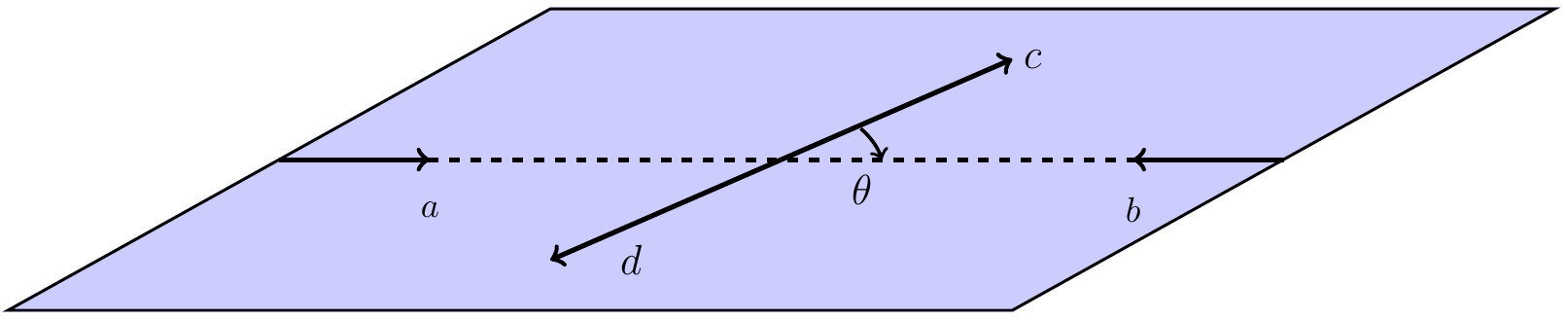}
\caption{The kinematic diagram for the scattering process $ab\to cd$ is depicted, where $\theta$ is defined as the dot product between $\hat{p}_{a}$ and $\hat{p}_{c}$.}
\label{fig-two body scattering}
\end{figure}

Now, let's derive the helicity formalism for  $ab\to cd$ scattering amplitude. The differential cross section is defined as
\begin{equation}
	\begin{aligned}
	 \frac{d\sigma}{d\Omega}=\left(\frac{2\pi}{p}\right)^{2}\left|\bra{q,\theta,\phi,\lambda_{c},\lambda_{d}}T(W)\ket{p,0,0,\lambda_{a},\lambda_{b}}\right|^{2}
	\end{aligned}
\end{equation}
where $\lambda_{a,b}$ and $\lambda_{c,d}$ are helicity symbols, $\theta$ is the polar angle of final particles, seeing Fig(\ref{fig-two body scattering}).
The scattering matrix element $\bra{q,\theta,\phi,\lambda_{c},\lambda_{d}}T(W)\ket{p,0,0,\lambda_{a},\lambda_{b}}$ can be calculated by a representation transformation
\begin{equation}\label{A12}
	\begin{aligned}
		\bra{q,\theta,\phi,\lambda_{c},\lambda_{d}}&T(W)\ket{p,0,0,\lambda_{a},\lambda_{b}}\\
		&=\sum_{j,m}\frac{2j+1}{4\pi}D^{*j}_{m,\lambda_{c}-\lambda_{d}}(\phi,\theta,0)\bra{\lambda_{c},\lambda_{d}}T^{j}(W)\ket{\lambda_{a},\lambda_{b}}D^{j}_{m,\lambda_{a}-\lambda_{b}}(0,0,0)\\
		&=\sum_{j,m}\frac{2j+1}{4\pi}D^{*j}_{m,\lambda_{c}-\lambda_{d}}(\phi,\theta,0)\bra{\lambda_{c},\lambda_{d}}T^{j}(W)\ket{\lambda_{a},\lambda_{b}}\delta_{m,\lambda_{a}-\lambda_{b}}\\
		&=\sum_{j}\frac{2j+1}{4\pi}D^{*j}_{\lambda_{i},\lambda_{f}}(\phi,\theta,0)\bra{\lambda_{c},\lambda_{d}}T^{j}(W)\ket{\lambda_{a},\lambda_{b}}\\
	\end{aligned}
\end{equation}
where we denote initial and final helicity as $\lambda_{i}$ and $\lambda_{f}$
\begin{equation}
	\begin{aligned}
	 \lambda_{i}=\lambda_{a}-\lambda_{b},~~ \lambda_{f}=\lambda_{c}-\lambda_{d}
	\end{aligned}
\end{equation}
Analogy to non-relativistic scattering, we define scattering amplitude $f_{\lambda_{a},\lambda_{b},\lambda_{c},\lambda_{d}}$ 
\begin{equation}\label{eq: scattering amplitudes 1}
	\begin{aligned}
		\frac{d\sigma}{d\Omega}=\left|f_{\lambda_{a},\lambda_{b},\lambda_{c},\lambda_{d}}(\theta,\phi)\right|^{2}
	\end{aligned}
\end{equation}
and 
\begin{equation}\label{eq: scattering amplitudes}
	\begin{aligned}
    f_{\lambda_{a},\lambda_{b},\lambda_{c},\lambda_{d}}(\theta,\phi)=\sum_{j}(2j+1)D^{*j}_{\lambda_{i},\lambda_{f}}(\phi,\theta,0)f^{j}_{\lambda_{i},\lambda_{f}}
	\end{aligned}
\end{equation}
where $f^{j}_{\lambda_{i},\lambda_{f}}$ represents the magnitude of scattering amplitude
\begin{equation}
	\begin{aligned}
	f^{j}_{\lambda_{i},\lambda_{f}}=\frac{1}{2p}\bra{\lambda_{c},\lambda_{d}}T^{j}(W)\ket{\lambda_{a},\lambda_{b}}
	\end{aligned}
\end{equation}
Let's consider the simplest scenario, where all the helicities are set to zero ($\lambda_{a}=\lambda_{b}=\lambda_{c}=\lambda_{d}=0$), and the total angular momentum $j$ is equal to the orbital angular momentum $L$.
\begin{equation}\label{eq: limit}
	\begin{aligned}
		f_{\lambda_{a},\lambda_{b},\lambda_{c},\lambda_{d}}(\theta,\phi)=\sum_{l}(2l+1)D^{*l}_{0,0}(\phi,\theta,0)f_{l}=\sum_{l}(2l+1)P_{l}(\cos\theta)f_{l}
	\end{aligned}
\end{equation}
The above formula is just non-relativistic scattering partial wave amplitude. Therefore, the helicity formalism derived here is a general description for scattering with spin degree of freedom. It should be noted that the factorization of the angular dependence from the total amplitudes in equation (\ref{A12}) is solely based on the rotational symmetry of the scattering operator $T^{j}(W)$ or the interaction Hamiltonian. This can be viewed as a specific instance of the Wigner-Eckart theorem.

The decaying amplitude can also be investigated in the framework of helicity formalism. The corresponding matrix element is 
\begin{equation}\label{eq: decaying matrix element}
	\begin{aligned}
	 \bra{p,\theta,\phi;\lambda_{1},\lambda_{2}}\mathcal{H}\ket{jm}&=\sum_{j^{\prime},m^{\prime}}\sqrt{\frac{2j^{\prime}+1}{4\pi}}D^{j^{\prime}*}_{m^{\prime},\lambda_{1}-\lambda_{2}}(\phi,\theta,0)\bra{p,j^{\prime},m^{\prime},\lambda_{1},\lambda_{2}}\mathcal{H}\ket{jm}\\
	 &=\sum_{j^{\prime},m^{\prime}}\sqrt{\frac{2j^{\prime}+1}{4\pi}}D^{j^{\prime}*}_{m^{\prime},\lambda_{1}-\lambda_{2}}(\phi,\theta,0)\delta_{j,j^{\prime}}\delta_{m,m^{\prime}}\mathcal{H}_{\lambda_{1},\lambda_{2}}\\
	 &=\sqrt{\frac{2j+1}{4\pi}}D^{j*}_{m,\lambda_{1}-\lambda_{2}}(\phi,\theta,0)\mathcal{H}_{\lambda_{1},\lambda_{2}}
	\end{aligned}
\end{equation}
Differential decaying width is defined as
\begin{equation}\label{eq: decaying width}
	\begin{aligned}
	 \frac{d\Gamma}{d\Omega}=\frac{(2\pi)^{4}}{2M}\left|\bra{p,\theta,\phi;\lambda_{1},\lambda_{2}}\mathcal{H}\ket{jm}\right|^{2}
	\end{aligned}
\end{equation}
Substituting (\ref{eq: decaying matrix element}) into (\ref{eq: decaying width}), we arrive at
\begin{equation}
	\begin{aligned}
	\left|\bra{p,\theta,\phi;\lambda_{1},\lambda_{2}}\mathcal{H}\ket{jm}\right|^{2}=\frac{2j+1}{4\pi}\sum_{m,m^{\prime}}D^{j*}_{m,\lambda_{1}-\lambda_{2}}(\phi,\theta,0)D^{j}_{m^{\prime},\lambda_{1}-\lambda_{2}}(\phi,\theta,0)\mathcal{H}_{\lambda_{1},\lambda_{2}}\mathcal{H}^{*}_{\lambda_{1},\lambda_{2}}
	\end{aligned}
\end{equation}
The decaying width is obtained by integrating over phase space
\begin{equation}
	\begin{aligned}
	 \Gamma=\frac{(2\pi)^{4}}{2M}\sum_{\lambda_{1},\lambda_{2}}\left|\mathcal{H}_{\lambda_{1},\lambda_{2}}\right|^{2}
	\end{aligned}
\end{equation}
Similar to the partial wave amplitude, orthogonality arises when integrating out the phase space.

The number of independent helicity amplitude are always reduced further in the practical calculation due to the consideration of symmetry like parity and identical principle, even time reversal for some of special scatterings. Therefore, we must discuss the symmetry property of helicity state and amplitude. For single particle state, one can prove 
\begin{equation}
	\begin{aligned}
\mathcal{P}\psi_{\vec{p},\lambda}=e^{i\pi J_{y}}e^{-i\pi J_{y}}\mathcal{P}\ket{p,0,0,\lambda}=\eta(-1)^{j-\lambda}e^{i\pi J_{y}}\psi_{\vec{p},-\lambda}
	\end{aligned}
\end{equation}
\begin{equation}
	\begin{aligned}
	\mathcal{P}\chi_{\vec{p},\lambda}=e^{i\pi J_{y}}e^{-i\pi J_{y}}\mathcal{P}\chi_{\vec{p},\lambda}=\eta(-1)^{j+\lambda}e^{i\pi J_{y}}\chi_{\vec{p},-\lambda}
	\end{aligned}
\end{equation}
where $\eta$ denote the intrinsic parity. Hence, we obtain the transformation relation of state $\ket{p,0,0,\lambda_{1},\lambda_{2}}$ or $\psi_{\vec{p},\lambda}\chi_{\vec{p},\lambda}$ under parity 
\begin{equation}
	\begin{aligned}
		\mathcal{P}\ket{p,0,0,\lambda_{1},\lambda_{2}}=\eta_{1}\eta_{2}(-1)^{j_{1}-\lambda_{1}+j_{2}+\lambda_{2}}e^{i\pi J_{y}}\ket{p,0,0,-\lambda_{1},-\lambda_{2}}
	\end{aligned}
\end{equation}
For angular momentum state
\begin{equation}
	\begin{aligned}
		\mathcal{P}\ket{p,j,m;\lambda_{1},\lambda_{2}}&=\sqrt{\frac{2j+1}{4\pi}}\int d\Omega D^{*j}_{m,\lambda}(\phi,\theta,0)\mathcal{P}\mathcal{R}(\phi,\theta,0)\ket{\vec{p},0,0,\lambda_{1},\lambda_{2}}
	\end{aligned}
\end{equation}
Using the property of rotation function, it can be calculated as
\begin{equation}
	\begin{aligned}
		\mathcal{P}\ket{p,j,m;\lambda_{1},\lambda_{2}}=\eta_{1}\eta_{2}(-1)^{j-j_{1}-j_{2}}\ket{p,j,m;-\lambda_{1},-\lambda_{2}}
	\end{aligned}
\end{equation}
 If the scattering and decaying interactions respect parity, the transition operator or Hamiltonian remain invariant under the transformation $\mathcal{P}^{\dagger}\left[T(W)/\mathcal{H}\right]\mathcal{P}$.
\begin{equation}
	\begin{aligned}
	 f^{j}_{\lambda_{i},\lambda_{f}}&=\frac{1}{2p}\bra{\lambda_{c},\lambda_{d}}T^{j}(W)\ket{\lambda_{a},\lambda_{b}}=\frac{1}{2p}\bra{\lambda_{c},\lambda_{d}}\mathcal{P}^{\dagger}T^{j}(W)\mathcal{P}\ket{\lambda_{a},\lambda_{b}}\\
	 &=\frac{1}{2p}\eta_{a}\eta_{b}\eta_{c}\eta_{d}(-1)^{2j-j_{a}-j_{b}-j_{c}-j_{d}}\bra{-\lambda_{c},-\lambda_{d}}T^{j}(W)\ket{-\lambda_{a},-\lambda_{b}}\\
	 &=\eta_{a}\eta_{b}\eta_{c}\eta_{d}(-1)^{j_{a}+j_{b}-j_{c}-j_{d}}f^{j}_{-\lambda_{i},-\lambda_{f}}
	\end{aligned}
\end{equation} 
and
\begin{equation}
	\begin{aligned}
	\bra{p,\theta,\phi;\lambda_{1},\lambda_{2}}\mathcal{H}\ket{jm}&=\bra{p,\theta,\phi;\lambda_{1},\lambda_{2}}\mathcal{P}^{\dagger}\mathcal{H}\mathcal{P}\ket{jm}\\
	&=\eta\eta_{1}\eta_{2}(-1)^{j-j_{1}-j_{2}}\bra{p,\theta,\phi;-\lambda_{1},-\lambda_{2}}\mathcal{H}\ket{j,m}
	\end{aligned}
\end{equation}
where $j_{i}$ is the spin of $ith$ particle. For weak decay processes, the aforementioned discussion becomes invalid due to parity violation. However, we can still separate the decaying matrix element into vector and axial-vector parts.
\begin{equation}
	\begin{aligned}
	 \mathcal{H}_{-\lambda_{1},-\lambda_{2}}(V)&=\eta\eta_{1}\eta_{2}(-1)^{j-j_{1}-j_{2}}\mathcal{H}_{\lambda_{1},\lambda_{2}}(V)\\
	 \mathcal{H}_{-\lambda_{1},-\lambda_{2}}(A)&=-\eta\eta_{1}\eta_{2}(-1)^{j-j_{1}-j_{2}}\mathcal{H}_{\lambda_{1},\lambda_{2}}(A)\\
	\end{aligned}
\end{equation}
which we have used for $\Lambda_{b}$ multi-body decay.

The identical principle is also useful in the helicity formalism, especially, for three body decay like $\Lambda_{b}\to \Delta^{++}\pi^{-}\pi^{-}$. For two particle state, we define operator $\mathcal{P}_{12}$ which interchanges the particle $1$ and $2$. Therefore
\begin{equation}
	\begin{aligned}
	\mathcal{P}_{12}\psi_{p,\lambda_{1},\lambda_{2}}&=\psi_{p,\lambda_{1}}(2)\chi_{p,\lambda_{2}}(1)=(-1)^{2s-\lambda_{1}+\lambda_{2}}e^{-i\pi J_{y}}\psi_{p,\lambda_{2},\lambda_{1}}
	\end{aligned}
\end{equation}
For angular momentum state, we obtain
\begin{equation}
	\begin{aligned}
		\mathcal{P}_{12}\ket{JM;\lambda_{1},\lambda_{2}}
		&=(-1)^{2s-\lambda_{1}+\lambda_{2}}\int d\Omega D^{*J}_{M,\lambda}(\alpha,\beta,\gamma)\mathcal{R}(\alpha,\beta,\gamma)e^{-i\pi J_{y}}\psi_{p,\lambda_{2},\lambda_{1}}\\
		&=(-1)^{J+2s}\ket{JM;\lambda_{2},\lambda_{1}}
	\end{aligned}
\end{equation}
The symmetrical or anti-symmetrical state is constructed as
\begin{equation}
	\begin{aligned}
	 \left(1+(-1)^{2s}\mathcal{P}_{12}\right)\ket{JM;\lambda_{1},\lambda_{2}}=\ket{JM;\lambda_{1},\lambda_{2}}+(-1)^{J}\ket{JM;\lambda_{2},\lambda_{1}}
	\end{aligned}
\end{equation}
From above equation, one can easily confirm Landau-Yang theorem.

In order to account for polarization within the helicity framework, we utilize the spin density matrix approach. In this appendix, we will give the formula of the polarization of the final state in both scattering and decay processes. The density matrix of the final states can be obtained as following assuming the initial particles are unpolarized
\begin{equation}
	\begin{aligned}
		\rho_{\lambda_{f},\lambda_{f}^{\prime}}=\frac{1}{2s_{b}+1}\frac{1}{2s_{a}+1}\sum_{\lambda_{i}}f_{\lambda_{f},\lambda_{i}}f^{*}_{\lambda_{i},\lambda_{f}^{\prime}}
	\end{aligned}
\end{equation}
Polarization is defined as the expectation value of spin operator
\begin{equation}
	\begin{aligned}
	Tr(S\rho)=\sum_{\lambda_{f},\lambda_{f}^{\prime}}\rho_{\lambda_{f},\lambda_{f}^{\prime}}
	S_{\lambda_{f}^{\prime},\lambda_{f}}
	\end{aligned}
\end{equation}
Referring to Fig (\ref{fig-T Polarization1}), polarization components $S_{x},S_{z}$ vanish if we impose the constrain of parity conservation.

The polarization in the decay process can also be expressed using the density matrix formalism
\begin{equation}
	\begin{aligned}
		\rho_{\lambda_{f},\lambda^{\prime}_{f}}=\frac{2j+1}{4\pi}\sum_{m,m^{\prime}}\rho_{m,m^{\prime}}D^{j*}_{m,\lambda_{1}-\lambda_{2}}(\phi,\theta,0)D^{j}_{m^{\prime},\lambda_{1}-\lambda_{2}}(\phi,\theta,0)\mathcal{H}_{\lambda_{f}}\mathcal{H}^{*}_{\lambda_{f}^{\prime}}
	\end{aligned}
\end{equation}
Indeed, the interference between different helicity amplitudes, which depends on the phase space, reflects the polarization of the final particle in the decay process. Similarly, if the initial particle is polarized, the formula for the angular distribution will be more complex. This can be observed, for example, in the decay of a spin-1 particle into two spin-0 particles. 
\begin{equation}
	\begin{aligned}
		\mathcal{I}(\theta,\phi)&=\frac{3}{4\pi}\sum_{m,m^{\prime}}\rho_{m,m^{\prime}}D^{j*}_{m,\lambda_{1}-\lambda_{2}}(\phi,\theta,0)D^{j}_{m^{\prime},\lambda_{1}-\lambda_{2}}(\phi,\theta,0)\left|\mathcal{H}_{\lambda_{1},\lambda_{2}}\right|^{2}\\
		&=\frac{3}{4\pi}\left\{\rho_{1,1}\frac{1}{2}\sin^{2}\theta+\rho_{-1,-1}\frac{1}{2}\sin^{2}\theta+\rho_{0,0}\cos^{2}\theta-\frac{1}{\sqrt{2}}\sin2\theta\mathcal{R}e(\rho_{1,0}e^{i\phi})\right.\\&\left.~~~~~~~+\frac{1}{\sqrt{2}}\sin2\theta\mathcal{R}e(\rho_{-1,0}e^{i\phi})-\sin^{2}\theta\mathcal{R}e(\rho_{1,-1})e^{2i\phi}\right\}\left|\mathcal{H}_{0,0}\right|^{2}
	\end{aligned}
\end{equation}
It is noted that this angular distribution is trivial if the initial density matrix $\rho_{m,m^{\prime}}$ is diagonal. Of course, it is consistent with the case of general two body decays.

Analogous to the two-body problem, the helicity approach for three-body decays allows us to extract the angular dependence without explicitly considering the specific dynamics. Instead, we utilize rotational invariance and other symmetries \cite{Berman:1965gi}. In comparison to the two-body case, an additional quantum number called $M$ is introduced to describe the quantization of the normal direction to the decaying plane in the center-of-mass reference frame for the three-body state. As a result, we construct a three-body state as follows:
\begin{equation}
	\begin{aligned}
	\ket{\omega_{1},\lambda_{1};\omega_{2},\lambda_{2};\omega_{3},\lambda_{3}; \alpha,\beta,\gamma}
	\end{aligned}
\end{equation}
where $\omega_{i},\lambda_{i}$ are respective energy and helicity of $ith$ particle, and $\alpha,\beta,\gamma$ are three Euler angles referring to Fig (\ref{fig-T Polarization3body}). Above state describing a specific configuration is connected to angular state $	\ket{\omega_{1},\lambda_{1};\omega_{2},\lambda_{2};\omega_{3},\lambda_{3}; JmM}$ by a transformation function similar to the two body
\begin{equation}
	\begin{aligned}
	 \ket{\omega_{1},\lambda_{1};\omega_{2},\lambda_{2};\omega_{3},\lambda_{3}; jmM}=\int &D^{*j}_{m,M}(\alpha,\beta,\gamma)\ket{\omega_{1},\lambda_{1};\omega_{2},\lambda_{2};\omega_{3},\lambda_{3};\alpha,\beta,\gamma}\\
	 &d\alpha sin\beta d\beta d\gamma
	\end{aligned}
\end{equation}
When studying the three-body decay, it is customary to derive the angular distribution relative to the normal direction. In simpler terms, we determine the distribution of the normal vector of the decaying plane. For a particular final helicity configuration, we obtain the distribution of the normal vector as
\begin{equation}
	\begin{aligned}
	 \left(\frac{d\Gamma}{d\Omega}\right)_{\lambda_{1},\lambda_{2},\lambda_{2}}=\sum_{M,M^{\prime}}\sum_{m,m^{\prime}}\int \rho_{m,m^{\prime}}D^{j*}_{m,M}(\alpha,\beta,\gamma)D^{j}_{m^{\prime},M^{\prime}}(\alpha,\beta,\gamma)d\gamma \mathcal{F}_{M,M^{\prime}}
	\end{aligned}
\end{equation}
where $\rho_{m,m^{\prime}}$ denotes initial density matrix and $\mathcal{F}_{M,M^{\prime}}$ is 
\begin{equation}
	\begin{aligned}
	 \mathcal{F}_{M,M^{\prime}}=\int d\omega_{1}d\omega_{2}F_{M}(\omega_{1},\lambda_{1};\omega_{2},\lambda_{2};\omega_{3},\lambda_{3})F^{*}_{M^{\prime}}(\omega_{1},\lambda_{1};\omega_{2},\lambda_{2};\omega_{3},\lambda_{3})
	\end{aligned}
\end{equation}
The rotation angle of decaying plane about it's normal vector is always integrated out
\begin{equation}\label{eq:decay width}
	\begin{aligned}
		\left(\frac{d\Gamma}{d\Omega}\right)=\sum_{M}\sum_{m,m^{\prime}} \rho_{m,m^{\prime}}D^{j*}_{m,M}(\alpha,\beta,\gamma)D^{j}_{m^{\prime},M}(\alpha,\beta,\gamma)\left|\mathcal{R}_{M}\right|^{2}
	\end{aligned}
\end{equation}
where 
\begin{equation}
	\begin{aligned}
	 \left|\mathcal{R}_{M}\right|^{2}=2\pi\sum_{\lambda_{1},\lambda_{2},\lambda_{2}}\int d\omega_{1}d\omega_{2}\left|F_{M}(\omega_{1},\lambda_{1};\omega_{2},\lambda_{2};\omega_{3},\lambda_{3})\right|^{2}
	\end{aligned}
\end{equation}
For convenience, the formula (\ref{eq:decay width}) is expressed by $M\geq0$
\begin{equation}\label{eq:Mfeifushu}
	\begin{aligned}
		\left(\frac{d\Gamma}{d\Omega}\right)&=\sum_{M\geq0}\sum_{m,m^{\prime}}\left\{\left(\mathcal{R}e\rho_{m,m^{\prime}}\cos(m-m^{\prime})\alpha-\mathcal{I}m\rho_{m,m^{\prime}}sin(m-m^{\prime})\alpha\right)\right.\\&\left.~~~~~~~~\left[R^{+}_{M}Z_{m,m^{\prime}}^{jM+}(\beta)+R^{-}_{M}Z_{m,m^{\prime}}^{jM-}(\beta)\right]\right\}
	\end{aligned}
\end{equation}
and 
\begin{equation}
	\begin{aligned}
	 R_{M}^{\pm}&=\frac{1}{2}\left[\left|R_{M}\right|^{2}\pm\left|R_{-M}\right|^{2}\right]\\
	 Z_{m,m^{\prime}}^{jM\pm}(\beta)&=d^{j}_{m,M}(\beta)d^{j}_{m^{\prime},M}(\beta)\pm d^{j}_{m,-M}(\beta)d^{j}_{m^{\prime},-M}(\beta)
	\end{aligned}
\end{equation}
Therefore, if identical particles exist, $R_{M}^{-}$ will vanish due to the degeneracy between $M>0$ and $M<0$. To preserve the information of $R_{M}^{-}$, we can partition the phase space based on $\omega_{1}>\omega_{2}$ and $\omega_{1}<\omega_{2}$ and establish a universal direction for the normal vector.

The identical and parity symmetry is also useful for three body case, especially when we consider the strong decay like $\Lambda_{b}^{0*}\to \Lambda^{0}_{b}\pi^{+}\pi^{-}$. Under the parity transformation, the configuration $(\alpha,\beta)$ changes to $(\alpha+\pi,\pi-\beta)$ which can also be viewed as another operation consisting of a reflection $\mathcal{Y}$ about $XY$ plane and a rotation $e^{iJ_{z}\pi}$. Therefore, we have
\begin{equation}
	\begin{aligned}
	 \mathcal{P}\ket{\omega_{1},\lambda_{1};\omega_{2},\lambda_{2};\omega_{3},\lambda_{3}; jmM}=&\int D^{j*}_{m,M}(\alpha,\beta,\gamma)R(\alpha,\beta,\gamma)\mathcal{P}\\
	 &\ket{\omega_{1},\lambda_{1};\omega_{2},\lambda_{2};\omega_{3},\lambda_{3};0,0,0}d\alpha sin\beta d\beta d\gamma
	\end{aligned}
\end{equation}
where 
\begin{equation}
	\begin{aligned}
      \mathcal{P}\ket{\omega_{1},\lambda_{1};\omega_{2},\lambda_{2};\omega_{3},\lambda_{3};0,0,0}&=e^{iJ_{z}\pi}\mathcal{Y}\ket{\omega_{1},\lambda_{1};\omega_{2},\lambda_{2};\omega_{3},\lambda_{3};0,0,0}
	\end{aligned}
\end{equation}
It can be proved that angular state $\ket{\omega_{1},\lambda_{1};\omega_{2},\lambda_{2};\omega_{3},\lambda_{3}; jmM}$ transforms as
\begin{equation}
	\begin{aligned}
		\mathcal{P}\ket{\omega_{1},\lambda_{1};\omega_{2},\lambda_{2};\omega_{3},\lambda_{3}; jmM}=&\eta_{1}\eta_{2}\eta_{3}(-1)^{s_{1}-\lambda_{1}+s_{2}-\lambda_{2}+s_{3}-\lambda_{3}}e^{i\pi M}\\
		&\ket{\omega_{1},-\lambda_{1};\omega_{2},-\lambda_{2};\omega_{3},-\lambda_{3}; jmM}
	\end{aligned}
\end{equation}
Obviously, the factor $e^{i\pi M}$ is equal to $(-1)^{M}$. Like the $3\pi$ system, it satisfies 
\begin{equation}
	\begin{aligned}
		\mathcal{P}\ket{\omega_{1};\omega_{2};\omega_{3}; jmM}&=(-1)^{M+1}\ket{\omega_{1};\omega_{2};\omega_{3}; jmM}
	\end{aligned}
\end{equation}
Hence, if parity is conserved and determined by the initial parity, only even or odd values of $M$ contribute to the angular dependence.

 As previously mentioned, the normalized transverse polarization $\langle S_{y}\rangle$ of $\Lambda_{b}$ in the decay $\Lambda^{0*}_{b}\to \Lambda^{0}_{b}\pi^{+}\pi^{-}$ is obtained as the trace of $\rho\sigma_{y}$. The density matrix of final $\Lambda^{0}_{b}$ is formulated as
\begin{equation}
	\begin{aligned}
		\rho_{\lambda,\lambda^{\prime}}=2\pi\sum_{M}\sum_{m,m^{\prime}} \rho_{m,m^{\prime}}D^{j*}_{m,M}(\alpha,\beta,\gamma)D^{j}_{m^{\prime},M}(\alpha,\beta,\gamma)\mathcal{F}_{M,M}(\lambda,\lambda^{\prime})
	\end{aligned}
\end{equation}
where 
\begin{equation}
	\begin{aligned}
	 \mathcal{F}_{M,M}(\lambda,\lambda^{\prime})=\int d\omega_{1}d\omega_{2}F_{M}(\omega_{1},0;\omega_{2},0;\omega_{3},\lambda)F^{*}_{M}(\omega_{1},0;\omega_{2},0;\omega_{3},\lambda^{\prime})
	\end{aligned}
\end{equation}
The normalized transverse polarization $\langle S_{y}\rangle$ is expressed as
\begin{equation}
	\begin{aligned}
    Tr(\rho\sigma_{y})\left(\frac{d\Gamma}{d\Omega}\right)=&2\pi\sum_{M}\sum_{m,m^{\prime}} \rho_{m,m^{\prime}}D^{j*}_{m,M}(\alpha,\beta,\gamma)D^{j}_{m^{\prime},M}(\alpha,\beta,\gamma)\\
    &\left[\mathcal{F}_{M,M}(\lambda=+1/2,\lambda^{\prime}=-1/2)(-i)+i\mathcal{F}_{M,M}(\lambda=-1/2,\lambda^{\prime}=+1/2)\right]
	\end{aligned}
\end{equation}
We can transform the helicity amplitude $F_{M}(\omega_{1},0;\omega_{2},0;\omega_{3},\lambda)$ to $F_{M}(\omega_{1},0;\omega_{2},0;\omega_{3},-\lambda)$ by multiplying a constant due to the parity conservation and we arrive at
\begin{equation}
	\begin{aligned}
    Tr(\rho\sigma_{y})\left(\frac{d\Gamma}{d\Omega}\right)&=2\pi\sum_{M}\sum_{m,m^{\prime}} \rho_{m,m^{\prime}}D^{j*}_{m,M}(\alpha,\beta,\gamma)D^{j}_{m^{\prime},M}(\alpha,\beta,\gamma)\\
    &\left[\mathcal{F}_{M,M}(\lambda=+1/2,\lambda^{\prime}=-1/2)(-i)+i\mathcal{F}_{M,M}(\lambda=-1/2,\lambda^{\prime}=+1/2)\right]\\
    &=\sum_{M}\sum_{m}\epsilon(-1)^{M+\frac{1}{2}}d^{j}_{m,M}(\beta)d^{j}_{m,M}(\beta)\left|\mathcal{R}_{M}\right|^{2}
	\end{aligned}
\end{equation}
which is equivalent to the equation (\ref{polarization 1}) and (\ref{polarization 2}).

The angular dependence for the three-body decay $B\to K_{1}\gamma\to \gamma K\pi\pi$ can also be investigated using the aforementioned approach, rather than referring to \cite{Gronau:2017kyq}. 
In this case, we consider a similar weak decay scenario, namely $\Lambda_{b}\to pa_{1}^{-}(1260)$. We assume the density matrix of $a_{1}^{-}(1260)$ is given by $\rho_{m,m^{\prime}}$, which is determined by the weak interaction in the $\Lambda_{b}\to pa_{1}^{-}(1260)$. Furthermore, we are interested in studying the angular distribution of the normal vector of the decaying plane in the decay $a_{1}^{-}(1260)\to \pi\pi\pi$
 \begin{equation}
 	\begin{aligned}
 		\left(\frac{d\Gamma}{d\Omega}\right)&=R^{+}_{1}\left[\frac{1+\cos^{2}\beta}{2}\left(\mathcal{R}e\rho_{+1,+1}+\mathcal{R}e\rho_{-1,-1}\right)+\mathcal{R}e\rho_{0,0}\sin^{2}\beta\right]\\
 		&+R^{-}_{1}\cos\beta\left(\mathcal{R}e\rho_{+1,+1}-\mathcal{R}e\rho_{-1,-1}\right)
 	\end{aligned}
\end{equation}
the Euler variable $\beta$ is the polar angle of the momenta of proton with respect to the normal vector of decaying plane. It is self-consistent with above convention. Comparing with \cite{Gronau:2017kyq}, we find two result are similar to each other, and the factor $\left(\mathcal{R}e\rho_{+1,+1}-\mathcal{R}e\rho_{-1,-1}\right)$ denotes the longitudinal polarization of $a_{1}^{-}(1260)$.

By the way, the angular dependence in helicity formalism can also be derived by utilizing angular-dependent spinors and polarization vectors, instead of rotational matrices.\cite{Auvil:1966eao}
 \begin{equation}       
 u(\vec{p},1)= N_{p}\left(                 
 \begin{array}{ccc}  
 \phi^{1} \\  
 \frac{\lvert \vec{p} \rvert}{E+m}\phi^{1}\\  
 \end{array}
 \right),~~       
 u(\vec{p},2)= N_{p}\left(                 
 \begin{array}{ccc}  
 \phi^{2} \\  
 -\frac{\lvert \vec{p} \rvert}{E+m}\phi^{2}\\  
 \end{array}
 \right)          
 \end{equation}
 where $ N_{p}=\sqrt{E+m}$ and
 \begin{equation}
 \phi^{1}=\binom{\cos(\frac{\theta}{2})}{\sin(\frac{\theta}{2})e^{i\phi}}, \phi^{2}=\binom{-\sin(\frac{\theta}{2})e^{-i\phi}}{\cos(\frac{\theta}{2})}
 \end{equation}
 \begin{equation}
 	\begin{aligned}
 	 \epsilon_{\mu}(+)&=\frac{1}{\sqrt{2}}e^{i\varphi}(0; \cos\theta\cos\varphi-i\sin\varphi,\cos\theta\sin\varphi+i\cos\varphi,-\sin\theta)\\
 	 \epsilon_{\mu}(-)&=\frac{1}{\sqrt{2}}e^{-i\varphi}(0; -\cos\theta\cos\varphi-i\sin\varphi,-\cos\theta\sin\varphi+i\cos\varphi,\sin\theta)\\
 	 \epsilon_{\mu}(0)&=\frac{1}{\sqrt{2}}(\frac{|\vec{p}|}{m}; -\frac{E}{m}\sin\theta\cos\varphi,-\frac{E}{m}\sin\theta\sin\varphi,-\frac{E}{m}\cos\theta)\\
 	\end{aligned}
\end{equation} 
\section{Some useful angular distributions for bottom baryons}
As a supplement, the asymmetry parameters in many b-baryons decays are not yet explored in the experiment, therefore, the angular distribution of chain decay $\Lb\to \Xi^-h^+$ and $\Xibz\to \Xi^-h^+$ with $\Xi^{-}\to \Lambda\pi^{-}$ are meaningful to discuss. However, the statistics of these decay channel is challenge to experimentalist. Nevertheless, we will give the angular distribution of decay chain $\Lambda^{0}_{b}\to \Xi^{-}h^{+}$ with $\Xi^{-}\to \Lambda^{0}\pi^{-}\to p\pi^{-}\pi^{-}$ that might be useful in the further. 
	\begin{table}[ht] 
		\centering
		\caption{\label{tab:test} Convention of symbols} 
		\begin{tabular}{llll} 
			\toprule\toprule
			~~~~~~~~~~decay   &  helicity angle &  helicity amplitudes & Lee-Yang parameters   \\ 
			\midrule 
			$ \Lambda^{0}_{b}/\Xi_{b}\to \Xi^{-}h $ & ~~$(\theta,\phi)$ &   ~~~~~~~~~~$\mathcal{H}_{\lambda},\mathcal{H}_{\lambda^{\prime}}$ & ~~~~$\alpha,\beta,\gamma$ \\ 
            $ \Xi^{-}\to \Lambda^{0}(K^{-}/\pi^{-}) $ & ~~$(\theta_{1},\phi_{1})$ &   ~~~~~~~~~~$h_{\lambda_{1}},h_{\lambda^{\prime}_{1}}$ & ~~~~$\alpha_{1},\beta_{1},\gamma_{1}$ \\ 
            $ \Lambda^{0}\to p\pi^{-} $ & ~~$(\theta_{2},\phi_{2})$ &   ~~~~~~~~~~$f_{\lambda_{2}},f_{\lambda^{\prime}_{2}}$ & ~~~~$\alpha_{2},\beta_{2},\gamma_{2}$ \\
			\bottomrule\bottomrule 
		\end{tabular} 
	\end{table}
	
Only three angle $\theta_{1},\theta_{2},\phi_{2}$ contribute to angular distribution if the initial polarization is ignored. Under which, one obtains
\begin{equation}
	\begin{aligned}
		\frac{d\Gamma}{d\Omega}\propto\sum_{\left[h\right]}&\mathcal{H}_{\lambda}\mathcal{H}^{*}_{\lambda^{\prime}} h_{\lambda_{1}}h^{*}_{\lambda^{\prime}_{1}}d^{j=1/2}_{\lambda,\lambda_{1}}(\theta_{1})d^{j=1/2}_{\lambda^{\prime},\lambda^{\prime}_{1}}(\theta_{1})
	f_{\lambda_{p}}f^{*}_{\lambda^{\prime}_{p}}D^{*j=1/2}_{\lambda_{1},\lambda_{p}}(\phi_{2},\theta_{2},0)D^{j=1/2}_{\lambda^{\prime}_{1},\lambda_{p}}(\phi_{2},\theta_{2},0)  
	\end{aligned}
\end{equation}
where the summation $\left[h\right]$ means all possible combinations of $\lambda,\lambda^{\prime},\lambda_{1},\lambda^{\prime}_{1},\lambda_{p}$. The angular distribution is 
\begin{equation}
	\begin{aligned}
		\frac{d\Gamma}{d\Omega}\propto(A+D)+\alpha_{2}(A-D)\cos\theta_{2}+2\alpha_{2}\left[\mathcal{R}e(B)\cos\phi_{2}-\mathcal{I}m(B)\sin\phi_{2}\right]\sin\theta_{2}
	\end{aligned}
\end{equation}
where symbol $A,B,D$ are some abbreviations
\begin{equation}\label{AD}
	\begin{aligned}
		A+D&= 1+\alpha\alpha_{1}\cos\theta_{1}\\
		A-D&= \alpha_{1}+\alpha\cos\theta_{1}\\
		2\mathcal{R}e(B)&= -\gamma_{1}\alpha\sin\theta_{1}\\
		2\mathcal{I}m(B)&=- \beta_{1}\alpha\sin\theta_{1}\\
	\end{aligned}
\end{equation}
The definitions of the angles $\theta_{1},\theta_{2},\phi_{2}$ are similar to the corresponding angles in Fig.~\ref{Fig-4}.